\documentclass[showpacs,preprintnumbers,amsmath,amssymb]{revtex4}
\usepackage{graphicx}% Include figure files
\usepackage{dcolumn}% Align table columns on decimal point
\usepackage{bm}% bold math
\usepackage{epsfig}

\newcommand{\dhd}{{\textstyle d}
\lower.03ex\hbox{\kern-0.40em$^{\scriptstyle-}$}\kern-0.08em{}}  

\newcommand{\half}{{1\over 2}}
\newcommand{\bu}{{\bullet}}

\newcommand{\barz}{{\bar z}}
\newcommand{\barhats}{\bar{\hat S}}

\newcommand{\calo}{{\cal O}}

\newcommand{\calv}{{\cal V}} 
\newcommand{\calz}{{\cal Z}}

\newcommand{\halo}{\hat{\cal O}}  

\begin{document}

\preprint{JLAB-THY-09-961}

\title{
NLO evolution of color
dipoles in ${\cal N}$=4 SYM}

\author{Ian Balitsky and Giovanni A. Chirilli}
\affiliation{
Physics Dept., ODU, Norfolk VA 23529, \\
and \\
Theory Group, Jlab, 12000 Jefferson Ave, Newport News, VA 23606
}
\email{balitsky@jlab.org, chirilli@jlab.org}

\date{\today}

\begin{abstract}
High-energy behavior of amplitudes in a gauge theory can be reformulated in terms of the evolution 
of Wilson-line operators. In the leading logarithmic approximation it is given by the conformally invariant  BK
equation for the evolution of color dipoles.
 In QCD, the next-to-leading order BK equation has both conformal and non-conformal parts, the latter providing
the running of the coupling constant. To separate the conformally invariant effects from the running-coupling 
effects, we calculate the NLO evolution of the color dipoles in the conformal ${\cal N}$=4 SYM theory. 
We define the ``composite dipole operators'' with the rapidity cutoff preserving conformal invariance. 
The resulting M\"obius invariant kernel for these operators agrees with the forward NLO BFKL calculation of Ref. \cite{lipkot}.

\end{abstract}

\pacs{12.38.Bx,  12.38.Cy}

\maketitle

\section{\label{sec:in}Introduction }

The high-energy scattering in a gauge theory can be described in terms of Wilson lines - infinite gauge
 factors ordered along the straight lines (see e.g. the review \cite{mobzor}). 
Indeed, the fast particle moves along its straight-line classical trajectory and the only quantum effect is 
the eikonal phase factor acquired along this propagation path. In QCD, for fast quark or gluon scattering 
off some target, this eikonal phase factor is a Wilson line - an infinite gauge link ordered along the straight 
line collinear to particle's velocity $n^\mu$:
\begin{equation}
U^\eta(x_\perp)={\rm Pexp}\Big\{ig\int_{-\infty}^\infty\!\!  du ~n_\mu 
~A^\mu(un+x_\perp)\Big\},~~~~
\label{defU}
\end{equation}
Here $A_\mu$ is the gluon field of the target, $x_\perp$ is the transverse
position of the particle which remains unchanged throughout the collision, and the 
index $\eta$ labels the rapidity of the particle.

The high-energy behavior of QCD amplitudes can be studied in the 
framework of the evolution of color dipoles. Let us consider  the small-$x$ behavior of structure functions of deep inelastic scattering (DIS).  
At high energies the virtual photon decomposes into quark and antiquark
which propagate along the straight lines separated by transverse distance and form a color dipole - two-Wilson-line operator.
\begin{equation}
\hat{\cal U}^\eta(x_\perp,y_\perp)=1-{1\over N_c}
{\rm tr}\{\hat{U}^\eta(x_\perp)\hat{U}^{\dagger\eta}(y_\perp)\}
\label{fla1}
\end{equation}
The energy dependence of the structure function is then translated into the dependence of the color dipole on the rapidity $\eta$.
There are two ways to restrict the rapidity of Wilson lines: one can consider Wilson lines with the support line collinear
 to the velocity of the fast-moving particle or
one can take the light-like Wilson line and cut the rapidity integrals ``by hand''. While the former method appears to be more natural, 
it is technically simpler to get the conformal results with the latter method of ``rigid cutoff'' in the longitudinal direction. 

Thus, the  small-x behavior of the structure functions is  governed by the 
rapidity evolution of color dipoles \cite{mu94,nnn}. 
At relatively high energies and for sufficiently small dipoles we can use the leading logarithmic approximation (LLA)
where  $ \alpha_s\ll 1,~ \alpha_s\ln x_B\sim 1$ and get the non-linear BK evolution equation for the color
dipoles \cite{npb96,yura}:
\begin{eqnarray}
&&\hspace{-1mm}
{d\over d\eta}~\hat{\cal U}^\eta(z_1,z_2)~=~
{\alpha_sN_c\over 2\pi^2}\!\int\!d^2z_3~ {z_{12}^2\over z_{13}^2z_{23}^2}
[\hat{\cal U}^\eta(z_1,z_3)+\hat{\cal U}^\eta(z_3,z_2))-\hat{\cal U}^\eta(z_1,z_3)-\hat{\cal U}^\eta(z_1,z_3)\hat{\cal U}^\eta(z_3,z_2)]
\label{bk}
\end{eqnarray}
where $\eta=\ln{1\over x_B}$ and $z_{12}\equiv z_1-z_2$ etc. (As usual, we denote operators by ``hat'').
The first three terms correspond to the linear BFKL evolution \cite{bfkl} and describe the parton emission 
while the last term is responsible for the parton annihilation. For sufficiently low $x_B$ the parton emission 
balances the parton annihilation so the partons reach the state of saturation \cite{saturation} with
the characteristic transverse momentum $Q_s$ growing with energy $1/x_B$
(for a review, see \cite{satreviews})

It is easy to see that the BK equation (\ref{bk}) is conformally invariant in the two-dimensional space. This follows from the conformal   
invariance of the light-like Wilson lines. Indeed, the  Wilson line
%¡¡
\begin{equation}
U(x_\perp)~=~{\rm Pexp}~\Big\{ig\!\int_{-\infty}^\infty\!dx^+~A_+(x^+,x_\perp)\Big\}
\end{equation}
is invariant under the inversion $x^\mu\rightarrow x^\mu/x^2$ (with respect to the point with zero (-) component). 
Indeed, $(x^+,x_\perp)^2=-x_\perp^2$ so after the inversion $x_\perp\rightarrow x_\perp/x_\perp^2$ and $x^+\rightarrow x^+/x_\perp^2$ and
therefore
\begin{equation}
U(x_\perp)~\rightarrow~{\rm Pexp}~\Big\{ig\!\int_{-\infty}^\infty\!d{x^+\over x_\perp^2}~A_+({x^+\over x_\perp^2},x_\perp)\Big\}
~=~U(x_\perp/x_\perp^2)
\end{equation}
It is easy to check that the Wilson line operators lie in the standard representation of the conformal M\"obius group SL(2,C) 
with conformal spin 0 (see appendix A).

The NLO evolution of color dipole in QCD \cite{nlobk} is not expected to be M\"obius invariant due to the conformal anomaly leading 
to dimensional transmutation and running coupling constant. However, the NLO BK equation in QCD \cite{nlobk}  has an additional 
term violating M\"obius invariance and not related to the conformal anomaly. 
To understand the relation between the high-energy behavior of amplitudes and M\"obius invariance of Wilson lines, 
it is instructive to consider the conformally invariant  ${\cal N}=4$ super Yang-Mils theory. 
This theory was  intensively studied in recent years due to the fact that at large coupling constants 
it is dual to the IIB string theory in the AdS$_5$ background. In the light-cone limit,
the contribution of scalar operators to Maldacena-Wilson line \cite{mwline} vanishes so one has the usual Wilson line constructed from gauge fields and therefore the LLA evolution
equation for color dipoles in the ${\cal N}=4$ SYM has the same form as (\ref{bk}). At the NLO level, the contributions from gluino and scalar loops enter the picture.

As we mentioned above, formally the light-like Wilson lines are  M\"obius invariant.
Unfortunately, the light-like Wilson lines are divergent in the longitudinal direction and moreover,  it is exactly the evolution 
equation with respect to this longitudinal cutoff which governs the high-energy behavior of amplitudes. 
At present, it is not known how to find the conformally invariant cutoff in the longitudinal direction.  When we use the non-invariant cutoff 
we expect, as usual, the invariance to hold in the leading order but to be
violated in higher orders in perturbation theory. In our calculation we restrict the longitudinal momentum of the gluons composing Wilson lines, 
and with this non-invariant cutoff the NLO evolution equation in QCD has extra non-conformal parts not related to the running of coupling constant.
Similarly, there will be non-conformal parts coming from the longitudinal cutoff of Wilson lines in the ${\cal N}=4$ SYM equation.
We will demonstrate below that it is possible to construct the
``composite conformal dipole operator'' (order by order in perturbation theory) which mimics the conformal cutoff
in the longitudinal direction so the corresponding evolution equation has no extra non-conformal parts. This is similar 
to the construction of the composite renormalized local operator in the case when the UV cutoff  does not respect the 
symmetries of the bare operator - in this case the symmetry of the UV-regularized operator is preserved 
order by order in perturbation theory by subtraction of the symmetry-restoring counterterms.

Let us present our result for the  NLO evolution of the color dipole in the adjoint representation
(hereafter we use notations $z_{ij}\equiv z_i-z_j$ and $(T^a)_{bc}=-if^{abc}$)
\begin{eqnarray}
&&\hspace{-5mm}
{d\over d\eta}\big[{\rm Tr}\{\hat{U}^\eta_{z_1}\hat{U}^{\dagger\eta}_{z_2}\}\big]^{\rm conf}~
\label{nlobksymconf}\\
&&\hspace{-5mm}
=~{\alpha_s\over \pi^2}
\!\int\!d^2z_3~
{z_{12}^2\over z_{13}^2 z_{23}^2}\Big[1-
{\alpha_sN_c\over 4\pi}{\pi^2\over 3}\Big]\big[{\rm Tr}\{T^a\hat{U}^\eta_{z_1}\hat{U}^{\dagger\eta}_{z_3}T^a\hat{U}_{z_3}\hat{U}^{\dagger\eta}_{z_2}\} 
-N_c {\rm Tr}\{\hat{U}^\eta_{z_1}\hat{U}^{\dagger\eta}_{z_2}\}\big]^{\rm conf}
\nonumber\\
&&\hspace{-2mm} 
-~{\alpha_s^2\over 4\pi^4}
\int \!d^2 z_3 d^2 z_4 {z_{12}^2\over z_{13}^2z_{24}^2z_{34}^2}
\Big\{2\ln{z_{12}^2z_{34}^2\over z_{14}^2z_{23}^2}
+\Big[1+{z_{12}^2z_{34}^2\over z_{13}^2z_{24}^2-z_{14}^2z_{23}^2}\Big]\ln{z_{13}^2z_{24}^2\over z_{14}^2z_{23}^2}\Big\}
\nonumber\\ 
&&\hspace{-2mm}
\times~ {\rm Tr}\{[T^a,T^b]\hat{U}^\eta_{z_1}T^{a'}T^{b'}\hat{U}^{\dagger\eta}_{z_2}
 + T^bT^a\hat{U}^\eta_{z_1} [T^{b'},T^{a'}]\hat{U}^{\dagger\eta}_{z_2}\}
 [(\hat{U}^\eta_{z_3})^{aa'}(\hat{U}^\eta_{z_4})^{bb'}-(z_4\rightarrow z_3)]
\nonumber
\end{eqnarray}
where
\begin{eqnarray}
&&\hspace{-5mm}
[{\rm Tr}\{\hat{U}^\eta_{z_1}\hat{U}^{\dagger\eta}_{z_2}\}\big]^{\rm conf}~=~{\rm Tr}\{\hat{U}^\eta_{z_1}\hat{U}^{\dagger\eta}_{z_2}\}
+{\alpha_s\over 2\pi^2}\!\int\! d^2 z_3~{z_{12}^2\over z_{13}^2z_{23}^2}
[ {\rm Tr}\{T^n\hat{U}^\eta_{z_1}\hat{U}^{\dagger\eta}_{z_3}T^n\hat{U}^\eta_{z_3}\hat{U}^{\dagger\eta}_{z_2}\}
 -N_c {\rm Tr}\{\hat{U}^\eta_{z_1}\hat{U}^{\dagger\eta}_{z_2}\}]
 \ln {az_{12}^2\over z_{13}^2z_{23}^2}
\label{confodipole}
 \end{eqnarray}

is the ``composite dipole'' with the conformal longitudinal cutoff in the next-to-leading order and $a$ is an arbitrary dimensional constant.
(Similar expression 
for the conformal two-dipole operator in the r.h.s. of this equation is presented below, see Eq. (\ref{confoper4})).
In fact,  $a(\eta)=ae^\eta$ plays the same role for the rapidity evolution as $\mu^2$ for the usual DGLAP evolution: the derivative ${d\over da}$ gives the 
evolution equation (\ref{nlobksymconf}).
The kernel in the r.h.s. of Eq. (\ref{nlobksymconf})  is obviously M\"obius invariant since it depends on two four-point conformal 
ratios ${z_{13}^2z_{24}^2\over z_{14}^2z_{23}^2}$ and ${z_{12}^2z_{34}^2\over z_{13}^2 z_{24}^2}$. We will also demonstrate that Eq. (\ref{nlobksymconf}) 
agrees with forward NLO BFKL calculation of Ref. \cite{lipkot}.

The paper is organized as follows. In  Sect. \ref{sec:bk} we remind the derivation of the 
BK equation in the leading order in $\alpha_s$. In Sect. \ref{sec:nlobksym}, which is central to the paper,
we calculate the scalar and gluino  contributions to the small-$x$ evolution of color dipoles
and assemble the NLO BK kernel in ${\cal N}=4$ SYM. We  rewrite the NLO BK kernel in the conformal form (\ref{nlobksymconf})
in Sect. \ref{sec:nlobksymconf} and compare our results with the NLO BFKL calculations in ${\cal N}=4$ SYM in Sect. \ref{sec:compare} . In Sect \ref{sec:fund}  we derive
the NLO BK equation for the composite dipoles in the fundamental representation, both in ${\cal N}=4$ SYM and in QCD.
The Mobius group for the light-like Wilson lines is presented in Appendix A while in Appendix B we 
find the explicit form (\ref{confodipole}) of the composite conformal dipole operator by 
calculating the appropriate impact factor. In Appendix C, we find the leading-order evolution for the four-Wilson-line operator 
and Appendices D and E contain some technical calculations which may distract readers from main discussion.

%%%%%%%%%%%%%%%%%%%%%%%%%%%%%%%%%%%%%%%%%%%%%%%%%%%%%%%%%%
\section{Derivation of the BK equation \label{sec:bk}}
Before discussing the small-x evolution of color dipole in the next-to-leading approximation it is instructive to recall the derivation of the leading-order (BK)
evolution equation. 

 For the NLO calculation we use the lightcone gauge $p_2^\mu A_\mu=0$. In addition we find it convenient to 
 use the ``rigid cutoff'' prescription in the longitudinal direction.  
 We consider the light-like dipoles (in the $p_1$ direction) and impose the cutoff
 on the maximal $\alpha$ emitted by any gluon from the Wilson lines so
\begin{eqnarray}
&&\hspace{-2mm} 
 U^\eta_x~=~{\rm Pexp}\Big[ig\!\int_{-\infty}^\infty\!\! du p_1^\mu A^\eta_\mu(up_1+x_\perp)\Big]
 \nonumber\\ 
&&\hspace{-2mm}
A^\eta_\mu(x)~=~\int\!\dhd^4 k \theta(e^\eta-|\alpha_k|)e^{-ik\cdot x} A_\mu(k)
\label{cutoff}
\end{eqnarray}
(hereafter we use the $\hbar$-inspired notation  $\dhd^n p\equiv {d^np\over (2\pi)^n}$ for brevity).
Note that the cutoff (\ref{cutoff}) respects the unitarity of Wilson lines ($U^\eta U^{\dagger\eta}=1$).

The momenta $p_1$ and $p_2$ are the light-like
vectors such that $q=p_1-x_B p_2$ and  $p=p_2+{m^2\over s}p_1$ where
$p$ is the momentum of the target and $m$ is the mass. Throughout the paper, we use the 
Sudakov variables $p=\alpha p_1+\beta p_2 +p_\perp$ and the notations 
$x_\bullet\equiv x_\mu p_1^\mu$ and $x_\ast\equiv x_\mu p_2^\mu$ related to
the light-cone coordinates: $x_\ast=x^+\sqrt{s/2},~x_\bullet=x^-\sqrt{s/2}$. Our metric is (1,-1,-1,-1) 
so $p^2=\alpha\beta s-p_\perp^2$ and $x^2={4\over s}x_\bu x_\ast-x_\perp^2$.

\begin{figure}[h]
\centering
\includegraphics[width=1.0\textwidth]{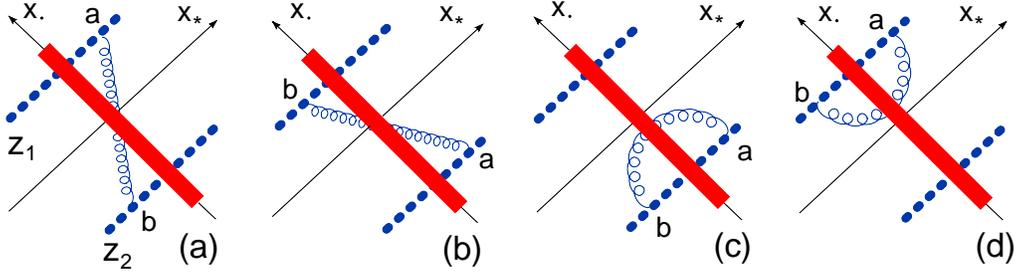}
\hspace{-4.6cm}
\vspace{-1cm}
\caption{Leading-order diagrams for the small-$x$ evolution of color dipole\label{bkevol}. Gauge links are denoted by dotted lines.}
\end{figure}

To find the evolution of the color dipole (\ref{fla1}) with respect to the slope of the 
Wilson lines in the leading log approximation
we consider the matrix element of the color dipole between (arbitrary) target states and 
integrate over the gluons with rapidities $\eta_1>\eta>\eta_2=\eta_1-\Delta\eta$ leaving the gluons with $\eta<\eta_2$ as
a background field (to be integrated over later).
In the frame of gluons with $\eta\sim\eta_1$ the fields with
$\eta<\eta_2$ shrink to a pancake and we obtain the four diagrams shown in Fig. 
\ref{bkevol}. Technically,  to find the kernel in the leading-ordrer approximation we 
write down the general form of the operator equation for the evolution of the color dipole 
\begin{eqnarray}
&&\hspace{-6mm}
{d \over d\eta}{\rm Tr}\{\hat{U}^\eta_{z_1}\hat{U}^{\dagger\eta}_{z_2}\}=
K_{\rm LO}{\rm Tr}\{\hat{U}^\eta_{z_1}\hat{U}^{\dagger\eta}_{z_2}\}+...
\label{eveq}
\end{eqnarray}
(where dots stand for the higher orders of the expansion) 
and calculate the l.h.s. of Eq. (\ref{eveq}) in the shock-wave background
\begin{eqnarray}
&&\hspace{-2mm}
{d \over d\eta}\langle{\rm Tr}\{\hat{U}^\eta_{z_1}\hat{U}^{\dagger\eta}_{z_2}\}\rangle_{\rm shockwave}=
\langle K_{\rm LO}{\rm Tr}\{\hat{U}^\eta_{z_1}\hat{U}^{\dagger\eta}_{z_2}\}\rangle_{\rm shockwave}
\label{eveqmaels}
\end{eqnarray}
In what follows we replace $\langle ...\rangle_{\rm shockwave}$ 
by $\langle ...\rangle$ for brevity.

%%%%%%%%%%%%%%FIGA%%%%%%%%%%%%%%%%%%%%%%%
%%%%%%%%%%%%%%%%%%%%%%%%%%%%%%%%%%%%%%%%%
With future NLO computation in view, we will perform the leading-order calculation in the lightcone gauge $~p_2^\mu A_\mu=0$. 
\vspace{-0.3cm}
The  gluon propagator in a shock-wave external field has the form\cite{prd99,balbel} 
\begin{eqnarray}
&&\hspace{-2mm}\langle \hat{A}^a_{\mu}(x)\hat{A}^b_{\nu}(y)\rangle ~=~
\theta(x_\ast y_\ast)\delta^{ab}
{s\over 2}\!\int\! \dhd\alpha\dhd\beta(x_\perp|{d_{\mu\nu}\over i(\alpha\beta s-p_\perp^2+i\epsilon)}|y_\perp)~
\label{gluprop}\\
&&\hspace{-2mm}
-~\theta(x_\ast)\theta(-y_\ast)
\!\int_0^\infty\!\dhd\alpha~ {e^{-i\alpha(x-y)_\bullet}\over 2\alpha}
(x_\perp|e^{-i{p_\perp^2\over\alpha s}x_\ast}
\Big[g^\perp_{\mu\xi}-{2\over \alpha s}(p^\perp_\mu p_{2\xi}+p_{2\mu}p^\perp_\xi)\Big]U^{ab} 
\Big[g^{\perp\xi}_\nu -{2\over \alpha s}(p_2^\xi p^\perp_\nu+p_{2\nu}p_\perp^\xi) \Big]
e^{i{p_\perp^2\over\alpha s}y_\ast}|y_\perp)
\nonumber \\
&&\hspace{-2mm}-~\theta(-x_\ast)\theta(y_\ast)
\!\int_0^\infty\!\dhd\alpha~ {e^{i\alpha(x-y)_\bullet}\over 2\alpha}
(x_\perp|e^{i{p_\perp^2\over\alpha s}x_\ast}
\Big[g^\perp_{\mu\xi}-{2\over \alpha s}(p^\perp_\mu p_{2\xi}+p_{2\mu}p^\perp_\xi)\Big]
U^{\dagger ab}
\Big[ g^{\perp\xi}_{~\nu}-{2\over \alpha s}(p_2^\xi p^\perp_\nu+p_{2\nu}p_\perp^\xi)\Big]
e^{-i{p_\perp^2\over\alpha s}y_\ast}|y_\perp)\dfrac{}{}
\nonumber
\end{eqnarray}
where 
\begin{equation}
d_{\mu\nu}(k)\equiv g^{\perp}_{\mu\nu}-
{2\over s\alpha}(k^{\perp}_{\mu}p_{2\nu}+k^{\perp}_{\nu}p_{2\mu})
-{4\beta\over s\alpha}p_{2\mu}p_{2\nu}
\label{demunu}
\end{equation}
Hereafter we use Schwinger's notations 
$(x_\perp|F(p_\perp)|y_\perp)\equiv \int\!\dhd p~e^{i(p,x-y)_\perp}F(p_\perp)$ 
(the scalar product of the four-dimensional vectors in our notations is 
$x\cdot y={2\over s}(x_\ast y_\bullet+x_\ast y_\bullet)-(x,y)_\perp$). 
Note that the interaction with the shock wave does not change the $\alpha$-component
of the gluon momentum.

We obtain
\begin{equation}
\hspace{-0mm}
g^2\!\int_0^\infty \! du\! \int^0_{-\infty} \! 
dv~\langle\hat{A}^{a,\eta_1}_\bu(up_1+x_\perp)
\hat{A}^{b,\eta_1}_\bu(vp_1+y_\perp)\rangle_{\rm Fig. \ref{bkevol}a}
~=~-4\alpha_s
\int_0^{e^{\eta_1}}\!{d\alpha\over\alpha}
(x_\perp|{p_i\over p_\perp^2-i\epsilon}
U^{ab}{p_i\over p_\perp^2-i\epsilon}|y_\perp)
\label{bk1}
\end{equation}

\begin{figure}[h]
\includegraphics[width=1.0\textwidth]{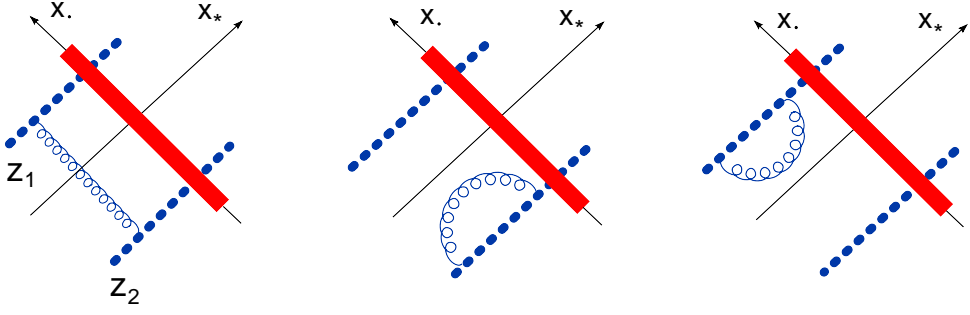}
\hspace{-4.8cm}
\vspace{-1cm}
\caption{Leading-order diagrams proportional to the original dipole. \label{virdiagrams}}
\end{figure}

Formally, the integral over $\alpha$ diverges at the lower limit, but since we integrate over the rapidities $\eta>\eta_2$ we get (in the LLA) 
\begin{eqnarray}
&&\hspace{-26mm}
g^2\!\int_0^\infty \! du \int^0_{-\infty} \! dv~\langle \hat{A}^{a,\eta_1}_\bu(up_1+x_\perp)
\hat{A}^{b,\eta_1}_\bu(vp_1+y_\perp)\rangle_{\rm Fig. \ref{bkevol}a}
~=~-4\alpha_s\Delta\eta
(x_\perp|{p_i\over p_\perp^2}U^{ab}{p_i\over p_\perp^2}|y_\perp)
\label{bk2}
\end{eqnarray}
and therefore
\begin{eqnarray}
&&\hspace{-2mm}\langle \hat{U}^\eta_{z_1}\otimes \hat{U}^{\dagger\eta}_{z_2}\rangle_{\rm Fig. \ref{bkevol}a}^{\eta_1}
~=~-{\alpha_s\over \pi^2}\Delta\eta~ (t^aU_{z_1}\otimes t^bU_{z_2}^\dagger)
\!\int\! d^2z_3 {(z_{13},z_{23})\over z_{13}^2z_{23}^2}
U_{z_3}^{ab}
\label{bk3}
\end{eqnarray}
The contribution of the diagram in Fig.  \ref{bkevol}b is obtained from Eq. (\ref{bk3})
by the replacement $t^aU_{z_1}\otimes t^bU_{z_2}^\dagger \rightarrow U_{z_1} t^b\otimes U_{z_2}^\dagger t^a$, $z_2\leftrightarrow z_1$ and the two remaining diagrams are obtained from
Eq. \ref{bk2} by taking $z_2=z_1$ (Fig. \ref{bkevol}c) and  $z_1=z_2$ (Fig. \ref{bkevol}d).
Finally, one obtains
\begin{eqnarray}
&&\hspace{-2mm}\langle \hat{U}^\eta_{z_1}\otimes \hat{U}^{\dagger\eta}_{z_2}\rangle_{\rm Fig. \ref{bkevol}}^{\eta_1}
~=~-{\alpha_s\Delta\eta\over \pi^2} (T^aU_{z_1}\otimes T^bU_{z_2}^\dagger
+ U_{z_1} T^b\otimes U_{z_2}^\dagger T^a)\!\int\! d^2z_\perp {(z_{13},z_{23})_\perp\over z_{13}^2z_{23}^2}
U_{z_3}^{ab}
\nonumber\\
&&\hspace{-2mm}
+~{\alpha_s\Delta\eta\over \pi^2}
 (T^aU_{z_1}T^b\otimes U_{z_2}^\dagger)
\!\int\! {d^2z_3 \over z_{13}^2}
U_{z_3}^{ab}
+{\alpha_s\Delta\eta\over \pi^2} (U_{z_1}\otimes T^bU_{z_2}^\dagger T^a)
\!\int\! {d^2z_3\over z_{23}^2}
U_{z_3}^{ab}
\label{bk4}
\end{eqnarray}
so
\begin{eqnarray}
&&\hspace{-2mm}
\langle{\rm Tr}\{\hat{U}^\eta_{z_1} \hat{U}^{\dagger\eta}_{z_2}\}\rangle_{\rm Fig. \ref{bkevol}}^{\eta_1}
~=~{\alpha_s\Delta\eta\over \pi^2} 
\!\int\! d^2z_3 {z_{12}^2\over z_{13}^2z_{23}^2}
[{\rm Tr}\{T^aU_{z_1} U^\dagger_{z_3}T^aU_{z_3} U_{z_2}^\dagger\}
-{1\over N_c}{\rm Tr}\{U_{z_1} U_{z_2}^\dagger\}]
\label{bk5}
\end{eqnarray}
There are also contributions coming from the diagrams shown in Fig. \ref{virdiagrams} (plus graphs obtained by reflection with respect to the shock wave).
%%%%%%%%%%%%%FIGA%%%%%%%%%%%%%%%%%%%%%%%
%%%%%%%%%%%%%%%%%%%%%%%%%%%%%%%%%%%%%%%%%

These diagrams are proportional to the original dipole ${\rm Tr}\{U_{z_1} U_{z_2}^\dagger\}$ 
and therefore the corresponding term can be derived from the contribution
of Fig. 1 graphs using the requirement that the r.h.s. of the evolution equation
should vanish in the absence  of the shock wave (when $U\equiv 1$).
It is easy to see that this requirement leads to 
\begin{eqnarray}
&&\hspace{-2mm}
\langle{\rm Tr}\{\hat{U}^{\eta_1}_{z_1} \hat{U}^{\dagger\eta_1}_{z_2}\}\rangle
~=~{\alpha_s\Delta\eta\over \pi^2} 
\!\int\! d^2z_3 {z_{12}^2\over z_{13}^2z_{23}^2}
[{\rm Tr}\{T^aU_{z_1} U^\dagger_{z_3}T^aU_{z_3} U_{z_2}^\dagger\}
-N_c{\rm Tr}\{U_{z_1} U_{z_2}^\dagger\}]
\nonumber
\end{eqnarray}
which gives  the BK equation for the evolution of the color dipole in the adjoint representation:.
\begin{eqnarray}
&&\hspace{-2mm}
{d\over d\eta}{\rm Tr}\{\hat{U}^\eta_{z_1} \hat{U}^{\dagger\eta}_{z_2}\}
~=~{\alpha_s\over \pi^2} 
\!\int\! d^2z_3 {z_{12}^2\over z_{13}^2z_{23}^2}
[{\rm Tr}\{T^aU_{z_1} U^\dagger_{z_3}T^aU_{z_3} U_{z_2}^\dagger\}
-N_c{\rm Tr}\{U_{z_1} U_{z_2}^\dagger\}]
\label{bk6}
\end{eqnarray}
%
%%%%%%%%%%%%%%%%%%%%%%%%%%%%%%%%%%%%%%%%%%%%%%%%%%%
\section{Calculation of the  NLO BK kernel in ${\cal N}=4$ SYM\label{sec:nlobksym}}

In the next-to-leading order the contributions to the kernel come from the one-loop diagrams 
for the color dipole in the shock-wave background. We will take the results for the gluon part of the NLO BK 
kernel from Ref. \cite{nlobk} and calculate the contribution of scalar and gluino loops. 
We use the ${\cal N}=4$ Lagrangian in the form (see e.g. Ref. \cite{bel03}):
\begin{eqnarray}
&&\mathcal{L}~=~
-{1\over 4}F^{\mu\nu}F_{\mu\nu} + {1\over 2}\big(D^\mu \Phi^a_I\big)\big(D_\mu \Phi^a_I\big)
-{1\over 4}g^2 f^{abc}f^{lmc} \Phi^a_I \Phi^b_J \Phi^l_I \Phi^m_J 
\nonumber\\
&&~~~~~~~~\,+\,\bar\lambda_{\dot\alpha A}^a
\sigma^{\dot\alpha \beta}_\mu {\cal D}^\mu\lambda^{aA}_\beta
\,-i\,\lambda^{\alpha A}_a\bar{\Sigma}^s_{AB}A^s_b\lambda^B_{\alpha k}f^{abc}
+\,i\,\bar{\lambda}^a_{\dot\alpha A}\Sigma^{s AB}A^s_b\bar{\lambda}^{\dot\alpha}_{B c}f^{abc}
\end{eqnarray}
Here $\Phi^a_I$ are scalars, $\lambda^{\alpha A}_a$ gluinos and $\Sigma^a_{IJ}=(\eta^i_{AB},i\bar{\eta}^i_{AB})$,
$\bar{\Sigma}^a_{IJ}=(\eta^i_{AB},-i\bar{\eta}^i_{AB})$  where 
$\eta^I_{AB}$ are standard `t Hooft symbols. The bare propagators are
\begin{eqnarray}
&&
\langle \Phi^a_I(x) \Phi^b_J(y)\rangle~=~i{\delta^{ab}\delta_{IJ}\!\int\!\dhd^4p {e^{-ip\cdot(x-y)}\over p^2+i\epsilon},~~~~~~
\langle\lambda^{aI}_\beta(x)\,\bar\lambda^{bJ}_{\dot\alpha}(y)\rangle =
\int\!\dhd^4p e^{-ip\cdot(x-y)}{ip_\mu\,\bar{\sigma}^\mu_{\beta\dot{\alpha}}\over p^2+i\epsilon},                         }
\end{eqnarray}
and the vertex of gluon emission in the momentum space is proportional to $(k_1-k_2)^\mu T^a\delta_{IJ}$ for the scalars and 
$\sigma^\mu T^a$ for gluinos.  (We do not need Yukawa or four-scalar vertices at this level). The diagrams in the shock-wave background 
are calculated similarly to the tree diagrams discussed in the previous Section.

%++++++++++++++++++++++++++++++++++++++++++++++++++++++++++++++
\subsection{Gluon contribution to NLO BK \label{subsect:gluonlobk}}
Let us start with the gluon contribution to the NLO evolution kernel. There is no difference between the gluon part of the kernel in QCD and in ${\cal N}=4$ SYM so we will just copy it from Ref. \cite{nlobk}) replacing ${\rm tr}\{t^aU_{z_1}t^bU_{z_2}^\dagger\}$ in the fundamental representation by 
${\rm Tr}\{T^aU_{z_1}T^bU_{z_2}^\dagger\}$ (throughout the paper we denote traces in the fundamental and the adjoint representations of color group by 
tr$\{...\}$ and Tr$\{...\}$, respectively)
\begin{eqnarray}
&&\hspace{-4mm}
{d\over d\eta}{\rm Tr}\{\hat{U}^\eta_{z_1} \hat{U}^{\dagger\eta}_{z_2}\}_{\rm gluon}~
=~{\alpha_s\over \pi^2}
\!\int\!d^2z_3~
{z_{12}^2\over z_{13}^2 z_{23}^2}\Big\{1+{\alpha_sN_c\over 4\pi}\Big[{11\over 3}\ln z_{12}^2\mu^2
-{11\over 3}{z_{13}^2-z_{23}^2\over  z_{12}^2}\ln{z_{13}^2\over z_{23}^2}+
{64 \over 9}-{\pi^2\over 3}
\label{gluon}\\
&&\hspace{62mm} 
-~
2\ln{z_{13}^2\over z_{12}^2}\ln{z_{23}^2\over z_{12}^2}\Big]\Big\}
~[{\rm Tr}\{T^a\hat{U}^\eta_{z_1} \hat{U}^{\dagger\eta}_{z_3}T^a\hat{U}_{z_3} \hat{U}^{\dagger\eta}_{z_2}\}
-N_c{\rm Tr}\{\hat{U}^\eta_{z_1} \hat{U}^{\dagger\eta}_{z_2}\}]   
\nonumber\\
&&\hspace{-4mm} 
+~{\alpha_s^2\over 8\pi^4}
~\!\int \!{d^2 z_3d^2 z_4\over z_{34}^4}~(\hat{U}^\eta_{z_3})^{aa'}
\Bigg[\Big\{2
\Big[2-{z_{13}^2z_{24}^2+z_{14}^2z_{23}^2-4 z_{12}^2z_{34}^2\over z_{13}^2z_{24}^2-z_{14}^2z_{23}^2}\ln{z_{13}^2z_{24}^2\over z_{14}^2z_{23}^2}
\Big]
{\rm Tr}\{[T^a,T^b]U_{z_1}[T^{a'},T^{b'}]U_{z_2}^\dagger\}~-~\Big[{ z_{12}^2z_{34}^2\over z_{13}^2 z_{24}^2}
\nonumber\\ 
&&\hspace{-4mm}
\times~\big(1+
{ z_{12}^2z_{34}^2\over z_{13}^2z_{24}^2-z_{23}^2z_{14}^2}\big)
\ln{z_{13}^2z_{24}^2\over z_{14}^2z_{23}^2}+z_3\leftrightarrow z_4\Big]
{\rm Tr}\{[T^a,T^b]U_{z_1}T^{a'}T^{b'}U_{z_2}^\dagger+T^bT^aU_{z_1}[T^{b'},T^{a'}]U_{z_2}^\dagger\}\Big\}(\hat{U}^\eta_{z_4}-\hat{U}^\eta_{z_3})^{ bb'}
\nonumber\\
&&\hspace{-4mm} 
-~\Big[{ z_{12}^2z_{34}^2\over z_{13}^2 z_{24}^2}
\big(1+
{ z_{12}^2z_{34}^2\over z_{13}^2z_{24}^2-z_{23}^2z_{14}^2}\big)
\ln{z_{13}^2z_{24}^2\over z_{14}^2z_{23}^2}-z_3\leftrightarrow z_4\Big]
{\rm Tr}\{[T^a,T^b]U_{z_1}T^{a'}T^{b'}U_{z_2}^\dagger+T^bT^aU_{z_1}[T^{b'},T^{a'}]U_{z_2}^\dagger\}U_{z_4}^{ bb'}\Bigg]
\nonumber
\end{eqnarray}
where $\mu$ is the normalization point in the $\overline{MS}$ scheme.
Our normalization fro Gell-Mann matrices is ${\rm tr}\{t^at^b\}=\half \delta^{ab}$. 

Note that the last term in r.h.s. is M\"obius invariant. 
The coefficient ${11\over 3}$ stands in front of the non-conformal terms 
coming from the running of the coupling constant and as discussed in the Introduction, there is an additional  non-conformal 
 term $\sim \ln { z_{12}^2\over z_{13}^2}\ln{ z_{12}^2\over z_{23}^2}$ coming from the non-invariance of the longitudinal cutoff (\ref{cutoff}).

It should be noted also that there is one small difference between QCD and ${\cal N}=4$ calculations of the gluon loop
due to the fact that in supersymmetic theories it is more natural to use the dimensional reduction scheme \cite{dimred} instead
of dimensional regularization. In dimensional reduction scheme the factor $g_{\mu\nu}^\perp g^{\perp\mu\nu}=d_\perp$ coming from the product of three-gluon vertices
should be replaced by 2: $g_{\mu\nu}^\perp g^{\perp\mu\nu}\rightarrow 2$. Making proper replacement in formulas 
in Sect. IV of Ref. \cite{nlobk} one gets the factor ${64\over 9}$ in the r.h.s. of the above equation. in place of ${67\over 9}$ in Eq. (5) in Ref. \cite{nlobk}).

%+++++++++++++++++++++++++++++++++++++++++++++++++++++
\subsection{Contribution of scalar particles \label{subsect:scalar}}
%-----------------------------------------------------------------------------------
\subsubsection{Diagrams with two scalar-shockwave intersections}

\begin{figure}[h]
%\centering
\vspace{-1cm}
\centering
\includegraphics[width=1.0\textwidth]{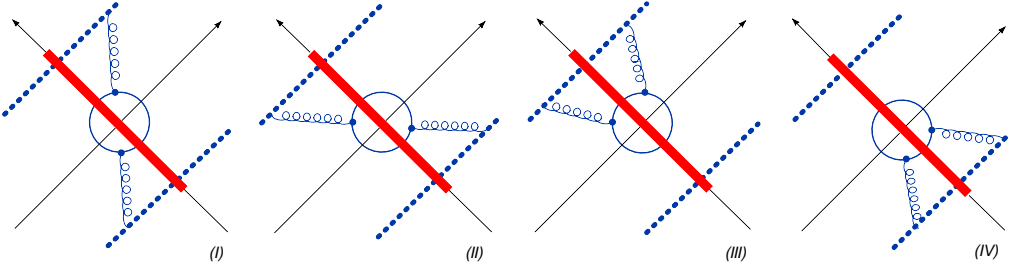}
\hspace{-3.2cm}
\vspace{-3.3cm}
%\hspace{-10cm}
%\vspace{-3.6cm}
\caption{Diagrams with the scalar loop bisected by the shock wave. \label{fig:scal}}
\end{figure}

First, we calculate the diagram with two scalar-shockwave intersections shown in Fig. \ref{fig:scal}.

\vspace{-9mm}
%%%%%%%%%%%%%FIGA%%%%%%%%%%%%%%%%%%%%%%%
%%%%%%%%%%%%%%%%%%%%%%%%%%%%%%%%%%%%%%%%
The scalar propagator in the shock-wave background has the form \cite{mobzor}:
\begin{eqnarray}
&&\hspace{-2mm}\langle \hat{\Phi}^a_I(x)\hat{\Phi}^b_J(y)\rangle ~=~
\theta(x_\ast y_\ast)\delta^{ab}
{s\over 2}\!\int\! \dhd\alpha\dhd\beta(x_\perp|{i\delta_{IJ}\over i(\alpha\beta s-p_\perp^2+i\epsilon)}|y_\perp)~
\label{scalprop}\\
&&\hspace{-2mm}
+~\delta_{IJ}
\!\int_0^\infty\!\dhd\alpha~ {e^{-i\alpha(x-y)_\bullet}\over 2\alpha}
\Big[\theta(x_\ast)\theta(-y_\ast)(x_\perp|e^{-i{p_\perp^2\over\alpha s}x_\ast}U^{ab} e^{i{p_\perp^2\over\alpha s}y_\ast}|y_\perp)
+\theta(-x_\ast)\theta(y_\ast)
(x_\perp|e^{i{p_\perp^2\over\alpha s}x_\ast}U^{\dagger ab}
e^{-i{p_\perp^2\over\alpha s}y_\ast}|y_\perp)\Big]
\nonumber
\end{eqnarray}

We start with the calculation of the Fig. \ref{fig:scal}a diagram.  Multiplying two propagators (\ref{scalprop}),  two 
scalar-gluon vertices and two bare gluon propagators we get

\begin{eqnarray}
&&\hspace{-6mm}
\langle\hat{U}^\eta_x\hat{U}^{\dagger\eta}_y\rangle_{\rm Fig. \ref{fig:scal}a}
~=~
g^2\int_0^{\infty}du\int^0_{-\infty}dv 
\langle \hat{A}^a_{\bullet}(up_1+z_1)\hat{A}^b_{\bullet}(vp_1+z_2)\rangle
\label{scalar1}\\
&&\hspace{-6mm}
=~
-6g^4{s^2\over 8}f^{anl}f^{bn'l'}\int
\dhd\alpha\dhd\alpha_1\dhd\beta\dhd\beta' \dhd\beta_1\dhd\beta'_1
\dhd\beta_2\dhd\beta'_2\!\int d^2z d^2z_4
\int \dhd^2q_1\dhd^2q_2\dhd^2k_1\dhd^2k_2~e^{i(q_1+q_2,z_1)_{\perp}
-i(k_1+k_2,z_2)_{\perp}}
\nonumber\\ 
&&\hspace{-6mm}
{4\alpha_1(\alpha-\alpha_1) 
U^{nn'}_{z_3}U^{ll'}_{z_4}e^{-i(q_1-k_1,z_3)_\perp-i(q_2-k_2,z_4)_\perp}
\over(\beta-\beta_1-\beta_2+i\epsilon)
(\beta'-\beta'_1-\beta'_2+i\epsilon)(\beta-i\epsilon)(\beta'-i\epsilon)}
~{d_{\bullet\lambda}(\alpha p_1+\beta p_2+q_{1\perp}+k_{1\perp})\over \alpha\beta
s-(q_1+q_2)_\perp^2+i\epsilon}
{d_{\lambda'\bullet}(\alpha p_1+\beta' p_2+q_{2\perp}+k_{2\perp})\over \alpha\beta'
s-(k_1+k_2)_\perp^2+i\epsilon}
\nonumber\\ 
&&\hspace{-6mm}
{[(2\alpha_1-\alpha)p_1+(q_1-q_2)_\perp]^\lambda
[(2\alpha_1-\alpha)p_1+(k_1-k_2)_\perp]^{\lambda'}
\over (\alpha_1\beta_1 s-q_{1\perp}^2+i\epsilon)
(\alpha_1\beta'_1 s-k_{1\perp}^2+i\epsilon)
[(\alpha-\alpha_1)\beta_2 s-q_{2\perp}^2+i\epsilon]
[(\alpha-\alpha_1)\beta'_2 s-k_{2\perp}^2+i\epsilon]}
\nonumber
\end{eqnarray}
Taking residues at $\beta=\beta'=0$ and $\beta_2=-\beta_1$, $\beta'_2=-\beta'_1$ 
we obtain
\begin{eqnarray}
&&\hspace{-6mm}g^2\!\int_0^{\infty}du\int^0_{-\infty}dv 
\langle \hat{A}^a_{\bullet}(up_1+z_1)\hat{A}^b_{\bullet}(vp_1+z_2)\rangle
\label{scalar2}\\
&&\hspace{-6mm}=~
-6g^4{s^2\over 8}f^{anl}f^{bn'l'}\!\int\!
\dhd\alpha\dhd\alpha_1\dhd\beta_1\dhd\beta'_1\!\int d^2z d^2z_4
\!\int \!\dhd^2q_1\dhd^2q_2\dhd^2k_1\dhd^2k_2
~e^{i(q_1+q_2,z_1)_{\perp}
-i(k_1+k_2,z_2)_{\perp}}
\nonumber\\ 
&&\hspace{-6mm}
\times~
4{\alpha_1(\alpha-\alpha_1)\over\alpha^2}
 U^{nn'}_{z_3}U^{ll'}_{z_4}e^{-i(q_1-k_1)z-i(q_2-k_2)z_4}
~{q_{1\perp}^2-q_{2\perp}^2\over (q_1+q_2)_\perp^2}
{k_{1\perp}^2-k_{2\perp}^2\over (k_1+k_2)_\perp^2}
\nonumber\\ 
&&\hspace{-6mm}
\times~
{1 \over (\alpha_1\beta_1 s-q_{1\perp}^2+i\epsilon)
(\alpha_1\beta'_1 s-k_{1\perp}^2+i\epsilon)
[-(\alpha-\alpha_1)\beta_1 s-q_{2\perp}^2+i\epsilon]
[-(\alpha-\alpha_1)\beta'_1 s-k_{2\perp}^2+i\epsilon]}
\nonumber
\end{eqnarray}

Finally, we can take  residues at $\beta_1={q_1^2\over\alpha_1 s}$ 
and $\beta'_1={k_1^2\over\alpha_1 s}$:
\begin{eqnarray}
&&\hspace{-6mm}
\langle\hat{U}^\eta_x\hat{U}^{\dagger\eta}_y\rangle_{\rm Fig. \ref{fig:scal}a}~=~-6{g^4\over 8\pi^2}
f^{anl}f^{bn'l'}\int_0^\sigma\!
{d\alpha\over\alpha}\!\int_0^1\! du~\bar{u} u\!\int d^2z d^2z_4
\!\int \!\dhd^2q_1\dhd^2q_2\dhd^2k_1\dhd^2k_2~e^{i(q_1+q_2,z_1)_\perp
-i(k_1+k_2,z_2)_\perp}
\nonumber\\ 
&&\hspace{-6mm}
\times~e^{-i(q_1-k_1,z_3)_\perp-i(q_2-k_2,z_4)_\perp}
{(q_{1\perp}^2-q_{2\perp}^2)(k_{1\perp}^2-k_{2\perp}^2)
\over (q_1+q_2)_\perp^2(k_1+k_2)_\perp^2}
~{1
\over (q_{1\perp}^2\bar{u}+q_{2\perp}^2 u)(k_{1\perp}^2\bar{u}+k_{2\perp}^2 u)}~ U^{nn'}_{z_3}U^{ll'}_{z_4}
\label{scalar3}
\end{eqnarray}
 where $u=\alpha_1/\alpha$ and $\bar{u}\equiv 1-u$. The contribution of the diagram in Fig. \ref{fig:scal}b
 is obtained by replacing $e^{i(q_1+q_2,z_1)_\perp}$ by $-e^{i(q_1+q_2,z_2)_\perp}$ and the two remaining diagrams in
 Fig. \ref{fig:scal}c and Fig. \ref{fig:scal}d are obtained by change $x\leftrightarrow y$. 
We get

\begin{eqnarray}
&&\hspace{-2mm}     
\langle {\rm Tr}\{\hat{U}^\eta_{z_1} \hat{U}^{\dagger\eta}_{z_2}\}
\rangle_{\rm Fig. \ref{fig:scal}}    
~=~ 6{g^4\over 8\pi^2}
f^{anl}f^{bn'l'}\int_0^\sigma\!
{d\alpha\over\alpha}\!\int_0^1\! du~\bar{u} u\!\int d^2z d^2z_4
\!\int \!\dhd^2q_1\dhd^2q_2\dhd^2k_1\dhd^2k_2~[e^{i(q_1+q_2,z_1)_\perp}-e^{i(q_1+q_2,z_2)_\perp}]
\nonumber\\
&&\hspace{-2mm}
\times~
[e^{i(k_1+k_2,z_1)_\perp}-e^{-i(k_1+k_2,z_2)_\perp}]
{(q_{1\perp}^2-q_{2\perp}^2)(k_{1\perp}^2-k_{2\perp}^2)
\over (q_1+q_2)_\perp^2(k_1+k_2)_\perp^2}
~{e^{-i(q_1-k_1,z_3)-i(q_2-k_2,z_4)}
\over (q_{1\perp}^2\bar{u}+q_{2\perp}^2 u)(k_{1\perp}^2\bar{u}+k_{2\perp}^2 u)}U^{nn'}_{z_3}U^{ll'}_{z_4}
\label{scalar4}
\end{eqnarray}
Performing the Fourier transformation
\begin{equation}
\hspace{-5mm}
\int\dhd^2 q_1\dhd^2 q_2~e^{i(q_1,x_1)+i(q_2,x_2)}~
{q_1^2-q_2^2\over (q_1+q_2)^2(q_1^2\bar{u}+q_2^2u)}
~=~
-{x_1^2-x_2^2\over 4\pi^2(x_1-x_2)^2(ux_1^2+\bar{u}x_2^2)}
\label{furie1}
\end{equation}
we obtain
\begin{eqnarray}
&&\hspace{-2mm}   
\langle {\rm Tr}\{\hat{U}^\eta_{z_1} \hat{U}^{\dagger\eta}_{z_2}\}
\rangle_{\rm Fig. \ref{fig:scal}}    
~=~
6{g^4\over 128\pi^6}f^{anl}f^{bn'l'}
\!\int_0^\sigma\!
{d\alpha\over\alpha}\!
\int_0^1\! du \int d^2 zd^2 z_4~
U^{nn'}(z)U^{ll'}(z_4)~{\bar{u}u\over z_{34}^4 }
\nonumber\\
&&\hspace{-6mm}
\times~
\Big[{z_{13}^2-z_{14}^2\over z_{13}^2u+z_{14}^2\bar{u}}-{z_{23}^2-z_{24}^2\over z_{23}^2u+z_{24}^2\bar{u}}\Big]
\Big[{z_{13}^2-z_{14}^2\over z_{13}^2u+z_{14}^2\bar{u}}-{z_{23}^2-z_{24}^2\over z_{23}^2u+z_{24}^2\bar{u}}\Big]
{\rm Tr}\{T^aU_{z_1}T^bU_{z_2}^\dagger\}
\label{scalar4a}
\end{eqnarray}
and therefore

\begin{eqnarray}
&&\hspace{-2mm}                     
\sigma{d\over d\sigma}\langle {\rm Tr}\{\hat{U}^\eta_{z_1} \hat{U}^{\dagger\eta}_{z_2}\}
\rangle_{\rm Fig. \ref{fig:scal}}    
~
\nonumber\\
&&\hspace{-2mm}
 =~
-{3\alpha^2\over 4\pi^4}\!\int\! {d^2 z_3d^2 z_4\over z_{34}^4}~\Big[-2
+{z_{13}^2z_{24}^2+z_{14}^2z_{23}^2\over z_{13}^2z_{24}^2-z_{14}^2z_{23}^2}\ln{z_{13}^2z_{24}^2\over z_{14}^2z_{23}^2}\Big]~
U_{z_3}^{aa'}U_{z_4}^{bb'}{\rm Tr}\{[T^a,T^b]U_{z_1}[T^{a'},T^{b'}]U_{z_2}^\dagger\}
\label{scalar5}
\end{eqnarray}

Following the method suggested in Refs. \cite{prd75,kw1,nlobk} we separate the UV-divergent part by adding and subtracting 
$z_4\rightarrow z_3$ contribution: $U_{z_3}^{aa'}U_{z_4}^{bb'}=(U_{z_3}^{aa'}U_{z_4}^{bb'}-z_4\rightarrow z_3)+U_{z_3}^{aa'}U_{z_3}^{bb'}$. We get
\begin{eqnarray}
&&\hspace{-2mm}                     
\sigma{d\over d\sigma}\langle {\rm Tr}\{\hat{U}^\eta_{z_1} \hat{U}^{\dagger\eta}_{z_2}\}
\rangle_{\rm Fig. \ref{fig:scal}}    
~ =~
-{3\alpha^2\over 4\pi^4}\!\int\! {d^2 z_3d^2 z_4\over z_{34}^4}~\Big[-2
+{z_{13}^2z_{24}^2+z_{14}^2z_{23}^2\over z_{13}^2z_{24}^2-z_{14}^2z_{23}^2}\ln{z_{13}^2z_{24}^2\over z_{14}^2z_{23}^2}\Big]
\nonumber\\
&&\hspace{22mm}
\times~(U_{z_3}^{aa'}U_{z_4}^{bb'}-z_4\rightarrow z_3){\rm Tr}\{[T^a,T^b]U_{z_1}[T^{a'},T^{b'}]U_{z_2}^\dagger\}
\nonumber\\
&&\hspace{-2mm}
+~{3\alpha^2N_c\over 4\pi^4}\!\int\! d^2 z
{\rm Tr}\{T^aU_{z_1}U^\dagger_{z_3}T^aU_{z_3}U_{z_2}^\dagger\}
\!\int\! {d^2 z_4\over z_{34}^4}~\Big[-2
+{z_{13}^2z_{24}^2+z_{14}^2z_{23}^2\over z_{13}^2z_{24}^2-z_{14}^2z_{23}^2}\ln{z_{13}^2z_{24}^2\over z_{14}^2z_{23}^2}\Big]
\label{scalar6}
\end{eqnarray}
The second (UV-divergent) part should be calculated at $d_\perp\neq 2$. As in the case of gluon loop, the Fourier transform (\ref{furie1}) at $d_\perp\neq 2$ is complicated so it is convenient to return to Eq. (\ref{scalar4}) in the momentum representation.
After replacing $U_{z_3}^{mm'}U_{z_4}^{nn'}$ by $U_{z_3}^{mm'}U_{z_3}^{nn'}$, integrating over $u$  
and changing variables to
$k_2=q_2=k'$, $p=q_1+q_2$, $l=q_1-k_1$ (so that
$q_1=p-k'$, $k_1=p-l-k'$ and $k_1+k_2=p-l$)  the Eq. (\ref{scalar4}) turns into  
\begin{eqnarray}
&&\hspace{-3mm}
\langle{\rm Tr}\{\hat{U}^\eta_{z_1} 
\hat{U}^{\dagger\eta}_{z_2}\}\rangle_{\rm Fig. \ref{fig:scal}~z_4\rightarrow z_3}
\label{scalar7}\\
&&\hspace{-3mm}=~
{3g^4N_c\over 4\pi^2}\!\int_0^\sigma\!{d\alpha\over\alpha}
\!\int d^2z_3~{\rm Tr}\{T^aU_{z_1}U^\dagger_{z_3}T^aU_{z_3}U_{z_2}^\dagger\}
\!\int\! \dhd^{2-\varepsilon} p\dhd^{2-\varepsilon}l~
~(e^{i(p,z_{13})}-e^{i(p,z_{23})})(e^{-i(p-l,z_{13})}-e^{-i(p-l,z_{23})})\Phi(p,l)
\nonumber
\end{eqnarray}
where
\begin{eqnarray}
&&\hspace{-2mm}
\Phi(p,l)~=~
\mu^{2\epsilon}\!\int\!\dhd^{2-\varepsilon}k'~
{1\over p^2(p-l)^2}~\Bigg[-2
-{(p-k')^2+(p-k'-l)^2\over (p-k')^2-(p-k'-l)^2}
\ln{(p-k')^2\over(p-k'-l)^2}
+{{k'}^2+(p-k')^2\over (p-k')^2-{k'}^2}
\ln{ (p-k')^2\over {k'}^2}
\nonumber\\
&&\hspace{-2mm}
+~{(p-k'-l)^2+{k'}^2\over (p-k'-l)^2-{k'}^2}
\ln{(p-k'-l)^2\over {k'}^2}\Bigg]~=~
{\Gamma^2(1-{\epsilon\over 2})\Gamma({\epsilon\over 2})\over 4\pi(3-\epsilon)\Gamma(2-\epsilon)}
{1\over p^2(p-l)^2}(p^{2-\epsilon}+|p-l|^{2-\epsilon}-l^{2-\epsilon})
\nonumber\\
&&\hspace{-2mm}
=~{(p,p-l)\over 2\pi p^2(p-l)^2}\Big({2\over\epsilon}-\ln{l^2\over\mu^2}+{8\over 3}\Big)
-{\ln p^2/l^2\over 3(p-l)^2}-{\ln(p-l)^2/l^2\over 3p^2}+O(\epsilon)
\label{scalar8}
\end{eqnarray}
Subtracting the pole in $\epsilon$ corresponding to counterterm 
(see the discussion in Refs \cite{prd75, nlobk}) we obtain
\begin{eqnarray}
&&\hspace{-2mm}
\sigma{d\over d\sigma}\langle{\rm Tr}\{\hat{U}^\eta_{z_1} 
\hat{U}^{\dagger\eta}_{z_2}\}\rangle_{\rm Fig. \ref{fig:scal}~z_4\rightarrow z_3}
=~
3{\alpha_s^2\over\pi}N_c
\!\int d^2z~
{\rm Tr}\{T^aU_{z_1}U^\dagger_{z_3}T^bU_{z_3}U_{z_2}^\dagger\}
\!\int\! \dhd^{2-\varepsilon} p~\dhd^{2-\varepsilon}l
\label{scalar9}\\
&&\hspace{-2mm}
\times~
~(e^{i(p,z_{13})}-e^{i(p,z_{23})})(e^{-i(p-l,z_{13})}-e^{-i(p-l,z_{23})})
\Big[{2(p,p-l)\over 3p^2(p-l)^2}\Big(-\ln{l^2\over\mu^2}+{8\over 3}\Big)
-{\ln p^2/l^2\over 3(p-l)^2}-{\ln(p-l)^2/l^2\over 3p^2}\Big]
\nonumber
\end{eqnarray}
Using the Fouriers integral from Appendix A to \cite{nlobk} we get
\begin{eqnarray}
&&\hspace{-2mm}
\int\! \dhd^2p\dhd^2l~e^{i(p,z_{13})-i(p-l,z_{23})}\Bigg[{2(p,p-l)\over 3p^2(p-l)^2}
\Big(-\ln {l^2\over\mu^2}+{8\over 3}\Big)-{\ln p^2/l^2\over 3(p-l)^2}-{\ln(p-l)^2/l^2\over 3p^2}\Bigg]
~
\nonumber\\ 
&&\hspace{-2mm}
=~{(z_{13},z_{23})\over 12\pi^2 z_{13}^2z_{23}^2}\Big[2\ln{z_{13}^2z_{23}^2\mu^2\over  z_{12}^2}+{16\over 3}\Big]
-{1\over 12\pi^2z_{23}^2}\ln{z_{13}^2\over  z_{12}^2}-{1\over 12\pi^2z_{13}^2}\ln{z_{23}^2\over  z_{12}^2}
\nonumber\\ 
&&\hspace{-2mm}
=~-{ z_{12}^2\over 12\pi^2 z_{13}^2z_{23}^2}\Big[\ln{z_{13}^2z_{23}^2\mu^2\over  z_{12}^2}+{8\over 3}\Big]
+{1\over 12\pi^2z_{13}^2}\ln{z_{13}^2\mu^2}+{1\over 12\pi^2z_{23}^2}\ln{z_{23}^2\mu^2}
\label{scalar10}
\end{eqnarray}
and therefore
\begin{eqnarray}
&&\hspace{-2mm}
\sigma{d\over d\sigma}\langle{\rm Tr}\{\hat{U}^\eta_{z_1} 
\hat{U}^{\dagger\eta}_{z_2}\}\rangle_{\rm Fig. \ref{fig:scal}~z_4\rightarrow z_3}~
=~{\alpha_s^2N_c\over 2\pi^3}
\!\int d^2z_3~{ z_{12}^2\over z_{13}^2z_{23}^2}\Big[\ln{z_{13}^2z_{23}^2\mu^2\over z_{12}^2}+{8\over 3}\Big]
{\rm Tr}\{T^aU_{z_1}U^\dagger_{z_3}T^bU_{z_3}U_{z_2}^\dagger\}
\label{scalar11}
\end{eqnarray}
Combining Eqs. (\ref{scalar6}) and (\ref{scalar11}) we obtain the full contribution of diagrams
 in Fig. \ref{fig:scal}
to the NLO kernel in the form
\begin{eqnarray}
&&\hspace{-2mm}                     
{d\over d\eta}( {\rm Tr}\{\hat{U}^\eta_{z_1} \hat{U}^{\dagger\eta}_{z_2}\})_{\rm scalars}    
~ 
\nonumber\\
&&\hspace{-2mm}
=~
-{3\alpha^2\over 4\pi^4}\!\int\! {d^2 z_3d^2 z_4\over z_{34}^4}~\Big[-2
+{z_{13}^2z_{24}^2+z_{14}^2z_{23}^2\over z_{13}^2z_{24}^2-z_{14}^2z_{23}^2}\ln{z_{13}^2z_{24}^2\over z_{14}^2z_{23}^2}\Big]
(\hat{U}_{z_3}^{aa'}\hat{U}_{z_4}^{bb'}
-z_4\rightarrow z_3){\rm Tr}\{[T^a,T^b]\hat{U}^\eta_{z_1}\hat[T^{a'},T^{b'}]\hat{U}_{z_2}\}
\nonumber\\
&&\hspace{-2mm}
+~{\alpha_s^2N_c\over 2\pi^3}
\!\int d^2z~{ z_{12}^2\over z_{13}^2z_{23}^2}\Big[\ln{z_{13}^2z_{23}^2\mu^2\over z_{12}^2}+{8\over 3}\Big]
{\rm Tr}\{T^a\hat{U}^\eta_{z_1}\hat{U}^{\dagger\eta}_{z_3}T^a\hat{U}_{z_3}\hat{U}^{\dagger\eta}_{z_2}\}
\label{scalar12}
\end{eqnarray}
where we have promoted the shock-wave Wilson lines to operators.
%-----------------------------------------------------------------------------------
\subsubsection{Scalar loop}
\begin{center}
\begin{figure}[h]
\centering
\includegraphics[width=\textwidth]{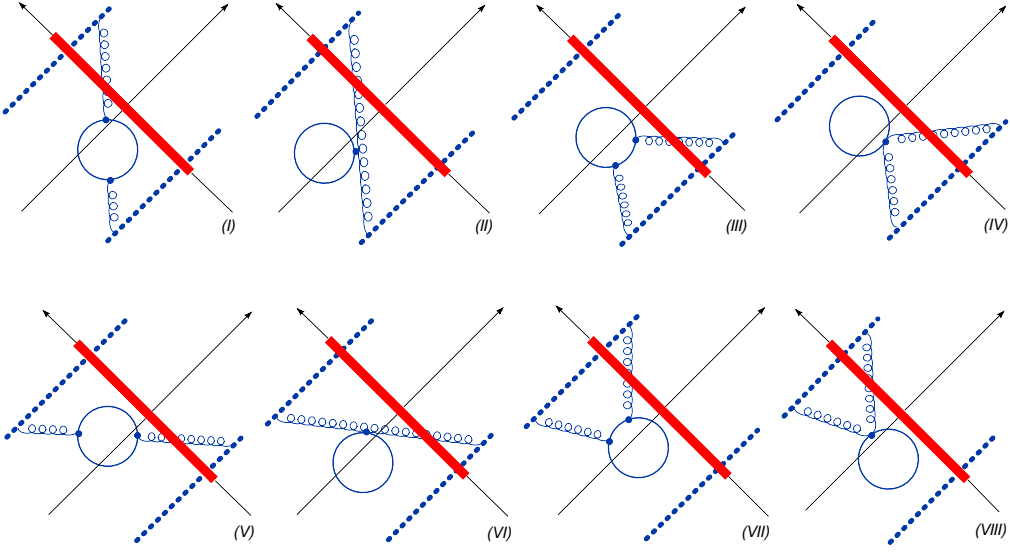}
\hspace{-3.8cm}
\vspace{-20mm}
\caption{Diagrams with bare scalar loop. \label{fig:scaloop}}
\end{figure}
\end{center}

Besides diagrams with the scalar loop bisected by the shock wave calculated above,   
there are diagrams with the ordinary scalar loop shown in Fig. \ref{fig:scaloop}.

%%%%%%%%%%%%%FIGA%%%%%%%%%%%%%%%%%%%%%%%
%%%%%%%%%%%%%%%%%%%%%%%%%%%%%%%%%%%%%%%%

The integral for the scalar loop has the form
\begin{equation}
\hspace{-1mm}
\!\int\! \dhd^{4-\epsilon}k'~{(2k'-k)_\mu(2k'-k)_\nu\over ({k'}^2+i\epsilon)[(k-k')^2+i\epsilon]}
~=~{2i\Gamma({\epsilon\over 2})\Gamma(1-{\epsilon\over 2})\Gamma(2-{\epsilon\over 2})
\over (4\pi)^{2-{\epsilon\over 2}}\Gamma(4-\epsilon) (-k^2)^{\epsilon\over 2} } 
(k_\mu k_\nu -g_{\mu\nu})
~\simeq~{i\over 48\pi^2}\Big[{2\over \epsilon}+\ln{1\over -k^2}+{8\over 3}\Big]
(k_\mu k_\nu -\delta_{\mu\nu})
\end{equation}
and therefore the contribution of the diagram shown in Fig. \ref{fig:scaloop} takes the form

\begin{eqnarray}
&&\hspace{-2mm}
{d\over d\eta}\langle{\rm Tr}\{\hat{U}^\eta_{z_1} \hat{U}^{\dagger\eta}_{z_2}\}
\rangle_{\rm Fig.~\ref{fig:scaloop}}
~=~-
{\alpha_s^2N_c\over \pi}\!\int\! d^2z_3
[{\rm Tr}\{T^aU_{z_1} U^\dagger_{z_3}T^aU_{z_3} U_{z_2}^\dagger\}
-{1\over N_c}{\rm Tr}\{U_{z_1} U_{z_2}^\dagger\}]\!
 \nonumber\\
&&\hspace{-2mm}                            
\times\int\!\dhd^2k\dhd^2k'\dhd^2q
~[e^{i(q,z_{13})}-e^{i(q,z_{23})}][e^{-i(k,z_{13})}-e^{-i(k,z_{23})}]{(q,k)\over k^2q^2}
\Big\{\ln{\mu^2\over k^2}+{8\over 3}\Big\}               
\label{vkladscal}
\end{eqnarray}
As usual we should add diagrams obtained by the reflection of diagrams shown in Fig. \ref{fig:scaloop}
with respect to the shock-wave line. Their contribution is obtained from Eq. (\ref{vkladscal}) by the replacement
 $q\leftrightarrow k$ in the logarithm so the final result for the sum of all 
 diagrams of Fig.~\ref{fig:scaloop} type has the form
\begin{eqnarray}
&&\hspace{-4mm}
{d\over d\eta}\langle {\rm Tr}\{\hat{U}^\eta_{z_1}\hat{U}^{\dagger\eta}_{z_2}\}\rangle_{\rm Fig.~\ref{fig:scaloop}~+refl.}
~=~
-{\alpha_s^2N_c\over \pi}\!\int\! d^2z_3
{\rm Tr}\{T^aU_{z_1} U^\dagger_{z_3}T^aU_{z_3} U_{z_2}^\dagger\}
\label{vklad1final}\\
&&\hspace{44mm}                            
\times\int\!\dhd^2k\dhd^2k'\dhd^2q
~[e^{i(q,z_{13})}-e^{i(q,z_{23})}][e^{-i(k,z_{13})}-e^{i(k,z_{23})}]{(q,k)\over k^2q^2}
\Big\{\ln{\mu^4\over q^2k^2}+{16\over 3}\Big\}    
 \nonumber\\
&&\hspace{-4mm}         
 =~
-{\alpha_s^2N_c\over 4\pi^3}\!\int\! d^2z_3
{\rm Tr}\{T^a\hat{U}^\eta_{z_1} \hat{U}^{\dagger\eta}_{z_3}T^a\hat{U}_{z_3} \hat{U}^{\dagger\eta}_{z_2}\}
\Big\{{ z_{12}^2\over z_{13}^2z_{23}^2} \Big[\ln {z_{13}^2z_{23}^2\over \mu^{-4}}+{16\over 3}\Big]
+\Big[{1\over z_{13}^2}-{1\over z_{23}^2}\Big]\ln{z_{13}^2\over z_{23}^2}\Big\}      
 \nonumber               
\end{eqnarray}
The total contribution of scalar particles to the NLO kernel from the diagrams in Fig. \ref{fig:scal} and  \ref{fig:scaloop} is a sum of Eqs. (\ref{vklad1final})
and (\ref{scalar12})
\begin{eqnarray}
&&\hspace{-2mm}                     
{d\over d\eta}\langle {\rm Tr}\{\hat{U}^\eta_{z_1} \hat{U}^{\dagger \eta}_{z_2}\}\rangle_{\rm scalars}    
~ =~
-{3\alpha^2\over 4\pi^4}\!\int\! {d^2 z_3d^2 z_4\over z_{34}^4}~\Big[-2
+{z_{13}^2z_{24}^2+z_{14}^2z_{23}^2\over z_{13}^2z_{24}^2-z_{14}^2z_{23}^2}\ln{z_{13}^2z_{24}^2\over z_{14}^2z_{23}^2}\Big]
\label{totalscalars}\\
&&\hspace{62mm}
\times~(U_{z_3}^{aa'}U_{z_4}^{bb'}-z_4\rightarrow z_3)
{\rm Tr}\{[T^a,T^b]U^\eta_{z_1}[T^{a'},T^{b'}]U^{\dagger\eta}_{z_2}\}
\nonumber\\
&&\hspace{-2mm}         
-~{\alpha_s^2N_c\over 4\pi^3}\!\int\! d^2z_3
{\rm Tr}\{T^aU_{z_1} U^\dagger_{z_3}T^a U_{z_3} U_{z_2}^\dagger\}
\Big\{{ z_{12}^2\over z_{13}^2z_{23}^2} \Big[\ln  z_{12}^2 \mu^2+{8\over 3}\Big]
+\Big[{1\over z_{13}^2}-{1\over z_{23}^2}\Big]\ln{z_{13}^2\over z_{23}^2}\Big\}      
\nonumber
\end{eqnarray}

Finally one needs to add the contribution of diagrams without scalar-shockwave intersection shown in Fig.\ref{fig:scalar-virtual}.
They are proportional to the ``parent dipole''  ${\rm Tr}\{U_{z_1}U_{z_2}^\dagger\}$, 
and their contribution can be found from Eq. (\ref{totalscalars}) 
using the requirement that the r.h.s. of the evolution equation must vanish as $z_1\rightarrow z_2$ 
(since $\lim_{z_1\rightarrow  z_2}\hat{U}^\eta_{z_1}\hat{U}^{\eta}\dagger_{z_2}=1$, see
the discussion below Eq. (\ref{cutoff})).
 It is easy to see that the following formula for the total contribution of scalar particle fulfils this requirement: 

\vspace{-11mm}
%%%%%%%%%%%%%FIGA%%%%%%%%%%%%%%%%%%%%%%%
\begin{center}
\begin{figure}[h]
\centering
\includegraphics[width=1\textwidth]{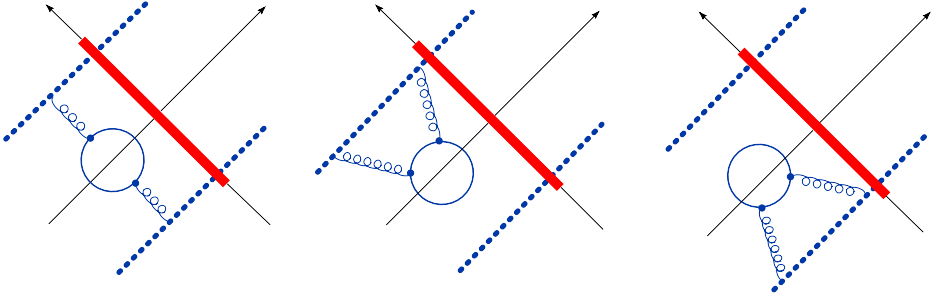}
\hspace{-3.8cm}
\vspace{-45mm}
\caption{Diagrams without scalar-shockwave intersection. \label{fig:scalar-virtual}}
\end{figure}
\end{center}
%%%%%%%%%%%%%%%%%%%%%%%%%%%%%%%%%%%%%%%%

\begin{eqnarray}
&&\hspace{-2mm}                     
{d\over d\eta}( {\rm Tr}\{\hat{U}^\eta_{z_1} \hat{U}^{\dagger\eta}_{z_2}\})_{\rm scalars}    
~ =~
-{3\alpha^2\over 4\pi^4}\!\int\! {d^2 z_3d^2 z_4\over z_{34}^4}~\Big[-2
+{z_{13}^2z_{24}^2+z_{14}^2z_{23}^2\over z_{13}^2z_{24}^2-z_{14}^2z_{23}^2}\ln{z_{13}^2z_{24}^2\over z_{14}^2z_{23}^2}\Big]
\label{scalars}\\
&&\hspace{62mm}
\times~(\hat{U}_{z_3}^{\eta,aa'}\hat{U}_{z_4}^{\eta,bb'}-z_4\rightarrow z_3)
{\rm Tr}\{[T^a,T^b]\hat{U}^\eta_{z_1}[T^{a'},T^{b'}]\hat{U}^{\dagger\eta}_{z_2}\}
\nonumber\\
&&\hspace{-2mm}         
-~{\alpha_s^2N_c\over 4\pi^3}\!\int\! d^2z_3
[{\rm Tr}\{T^a\hat{U}^\eta_{z_1} \hat{U}^{\dagger\eta}_{z_3}T^a\hat{U}_{z_3} \hat{U}^{\dagger\eta}_{z_2}\}
-N_c{\rm Tr}\{\hat{U}^\eta_{z_1} \hat{U}^{\dagger\eta}_{z_2}\}]
\Big\{{ z_{12}^2\over z_{13}^2z_{23}^2} \Big[\ln  z_{12}^2 \mu^2+{8\over 3}\Big]
+\Big[{1\over z_{13}^2}-{1\over z_{23}^2}\Big]\ln{z_{13}^2\over z_{23}^2}\Big\}      
\nonumber
\end{eqnarray}
Note that we have written this equation in the operator form by promoting the shock-wave fields in the r.h.s. of Eq. (\ref{totalscalars}) to operators.

%+++++++++++++++++++++++++++++++++++++++++++++++++++++
\subsection{Gluino contribution \label{sect:gluino}}
\begin{figure}
\includegraphics[width=1.0\textwidth]{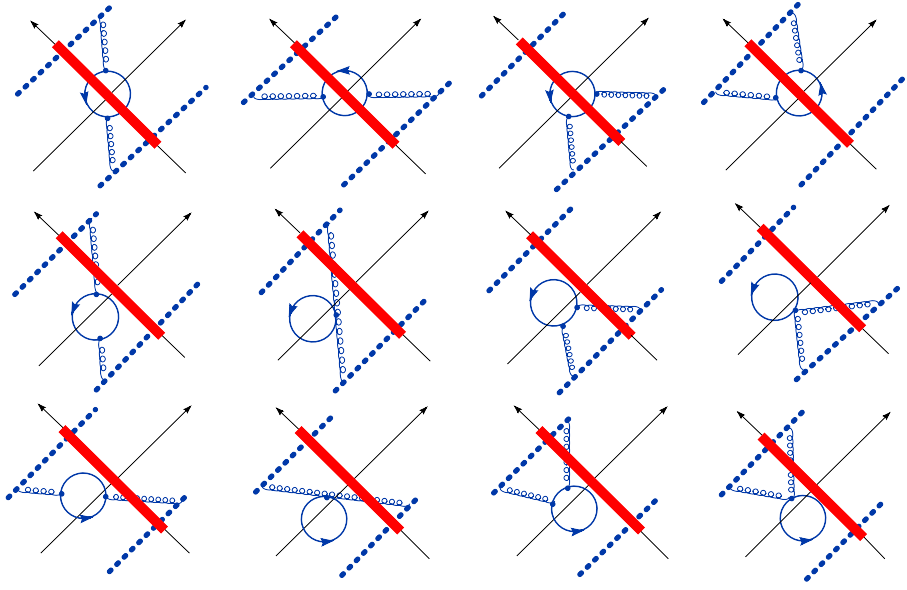}
\hspace{-54mm}

\vspace{-19mm}
\caption{Diagrams with the gluino loop. \label{fig:gluino}}
\end{figure}

%-----------------------------------------------------------------------------------
The diagrams for the gluino contribution to the NLO kernel are shown in Fig. \ref{fig:gluino}. 
The gluino propagator in the shock-wave background has the form
\begin{eqnarray}
&&\hspace{-2mm}
\langle{\rm T}\lambda^{aI}_\alpha(x)\,\bar\lambda^{bJ}_{\dot\alpha}(y)\rangle =
\theta(x_\ast y_\ast)\int \dhd^4 k\;e^{-ik\cdot(x-y)}\;
{ik_\mu\,\bar{\sigma}^\mu_{\alpha\dot{\alpha}}
\over k^2+i\epsilon}
\delta^{ab}\delta^{IJ}\nonumber
\label{gluinoprop}\\
&&\hspace{-2mm}
+~\delta_{IJ}\theta(x_\ast)\theta(-y_\ast)
\!\int_0^\infty\!\dhd\alpha~ {e^{-i\alpha(x-y)_\bullet}\over 2\alpha^2s}
(x_\perp|(\alpha \bar{p}_1+\bar{p}_\perp)e^{-i{p_\perp^2\over\alpha s}x_\ast}p_2U^{ab} e^{i{p_\perp^2\over\alpha s}y_\ast}
(\alpha \bar{p}_1+\bar{p}_\perp)|y_\perp)_{\alpha\dot{\alpha}}
\nonumber \\
&&\hspace{-2mm}+~\delta_{IJ}\theta(-x_\ast)\theta(y_\ast)
\!\int_0^\infty\!\dhd\alpha~ {e^{i\alpha(x-y)_\bullet}\over 2\alpha^2s}
(x_\perp|(\alpha \bar{p}_1+\bar{p}_\perp)e^{i{p_\perp^2\over\alpha s}x_\ast}p_2U^{\dagger ab}
e^{-i{p_\perp^2\over\alpha s}y_\ast}(\alpha \bar{p}_1+\bar{p}_\perp)|y_\perp)_{\alpha\dot{\alpha}}
\nonumber
\end{eqnarray}
where $\bar{p}_{\alpha\dot{\alpha}}\equiv p_\mu\bar{\sigma}^\mu_{\alpha\dot{\alpha}}$ and 
$p^{\dot{\alpha}\alpha}\equiv p^\mu\bar{\sigma}_\mu^{\dot{\alpha}\alpha}$.

\vspace{-0.5mm}
%%%%%%%%%%%%%FIGA%%%%%%%%%%%%%%%%%%%%%%%
%%%%%%%%%%%%%%%%%%%%%%%%%%%%%%%%%%%%%%%%

This propagator has the same form as the quark propagator in the shock-wave background in QCD so one can use the result for the quark 
part of the NLO BK kernel in QCD calculated in Refs. \cite{prd75,kw1}. We replace ${\rm tr}\{t^aU_{z_3}t^bU^\dagger_{z_4}\}$ for quarks in the fundamental representation 
by $2{\rm Tr}\{T^aU_{z_3}T^bU^\dagger_{z_4}\}$ for gluinos in the adjoint representation (and  ${\rm tr}\{t^aU_{z_1}t^bU_{z_2}^\dagger\}$ by 
${\rm Tr}\{T^aU_{z_1}T^bU_{z_2}^\dagger\}$ for Wilson loop as usual) and obtain 
\begin{eqnarray}
&&\hspace{-2mm}
{d\over d\eta}{\rm Tr}\{\hat{U}^\eta_{z_1} \hat{U}^{\dagger\eta}_{z_2}\}^{\rm gluino}~
\label{gluino}\\
&&\hspace{-2mm} 
=~{2\alpha_s^2N_c\over  3\pi^3}
\!\int\!d^2z_3~[{\rm Tr}\{T^a\hat{U}^\eta_{z_1} \hat{U}^{\dagger\eta}_{z_3}T^a\hat{U}^\eta_{z_3} \hat{U}^{\dagger\eta}_{z_2}\}
-N_c{\rm Tr}\{\hat{U}^\eta_{z_1} \hat{U}^{\dagger\eta}_{z_2}\}]
\Big[-{ z_{12}^2\over z_{13}^2 z_{23}^2}\big(\ln  z_{12}^2\mu^2+{5\over 3}\big)+{z_{13}^2-z_{23}^2\over z_{13}^2z_{23}^2}\ln{z_{13}^2\over z_{23}^2}\Big]
\nonumber\\
&&\hspace{-2mm}
-~{2\alpha^2_s\over \pi^4}
\!\int\!{d^2z_3 d^2z_4\over z_{34}^4}
\Big\{1
-{z_{14}^2z_{23}^2+z_{24}^2z_{13}^2- z_{12}^2z_{34}^2\over 2(z_{13}^2z_{24}^2-z_{14}^2z_{23}^2)}
\ln{z_{13}^2 z_{24}^2\over z_{14}^2z_{23}^2}\Big\}
(\hat{U}_{z_3}^\eta)^{aa'}(\hat{U}^\eta_{z_4}-\hat{U}^\eta_{z_3})^{bb'}
{\rm Tr}\{[T^a,T^b]\hat{U}^\eta_{z_1}[T^{a'},T^{b'}]\hat{U}^{\dagger\eta}_{z_2}\}
\nonumber
\end{eqnarray}
%

%+++++++++++++++++++++++++++++++++++++++++++++++++++++
\subsection{The N=4 kernel \label{subsect:N=4}}

Now we are in a position to assemble the NLO BK kernel in ${\cal N}=4$ SYM. Adding the gluon contribution (\ref{gluon}), 
the scalar part (\ref{scalars}), and the gluino term (\ref{gluino}) we obtain
\begin{eqnarray}
&&\hspace{-4mm}
{d\over d\eta}{\rm Tr}\{\hat{U}^\eta_{z_1} \hat{U}^{\dagger\eta}_{z_2}\}~
=~{\alpha_s\over \pi^2}
\!\int\!d^2z_3~
{z_{12}^2\over z_{13}^2 z_{23}^2}\Big[1-{\alpha_sN_c\over 4\pi}\Big({\pi^2\over 3}+2\ln{z_{12}^2\over z_{13}^2}\ln{z_{12}^2\over z_{23}^2}\Big)\Big]
~[{\rm Tr}\{T^a\hat{U}^\eta_{z_1} \hat{U}^{\dagger\eta}_{z_3}T^a\hat{U}_{z_3} \hat{U}^{\dagger\eta}_{z_2}\}
-N_c{\rm Tr}\{\hat{U}^\eta_{z_1} \hat{U}^{\dagger\eta}_{z_2}\}]   
\nonumber\\
&&\hspace{-4mm} 
-~{\alpha_s^2\over 8\pi^4}
~\!\int \!{d^2 z_3d^2 z_4\over z_{34}^4}~(\hat{U}^\eta_{z_3})^{aa'}
\Big\{
\Big[{ z_{12}^2z_{34}^2\over z_{13}^2 z_{24}^2}
\Big(1+
{ z_{12}^2z_{34}^2\over z_{13}^2z_{24}^2-z_{23}^2z_{14}^2}\Big)
\ln{z_{13}^2z_{24}^2\over z_{14}^2z_{23}^2}+z_3\leftrightarrow z_4\Big]
\label{nlobksyminus}\\
&&\hspace{54mm} 
\times~
{\rm Tr}\{[T^a,T^b]\hat{U}^\eta_{z_1}T^{a'}T^{b'}\hat{U}_{z_2}^{\dagger\eta}+T^bT^a\hat{U}^\eta_{z_1}[T^{b'},T^{a'}]
\hat{U}_{z_2}^{\dagger\eta}\}(\hat{U}^\eta_{z_4}-\hat{U}^\eta_{z_3})^{ bb'}
\nonumber\\
&&\hspace{-4mm} 
+~\Big[{ z_{12}^2z_{34}^2\over z_{13}^2 z_{24}^2}
\Big(1+
{ z_{12}^2z_{34}^2\over z_{13}^2z_{24}^2-z_{23}^2z_{14}^2}\Big)
\ln{z_{13}^2z_{24}^2\over z_{14}^2z_{23}^2}-z_3\leftrightarrow z_4\Big]
{\rm Tr}\{[T^a,T^b]\hat{U}^\eta_{z_1}T^{a'}T^{b'}\hat{U}_{z_2}^{\dagger\eta}
+T^bT^a\hat{U}^\eta_{z_1}[T^{b'},T^{a'}]\hat{U}_{z_2}^{\dagger\eta}\}(\hat{U}^\eta_{z_4})^{ bb'}\Big\}
\nonumber
\end{eqnarray}
Using Eq. (\ref{nashel}) from the Appendix E one can rewrite this equation as follows:
\begin{eqnarray}
&&\hspace{-2mm}
{d\over d\eta}{\rm Tr}\{\hat{U}^\eta_{z_1} \hat{U}^{\dagger\eta}_{z_2}\}~
\nonumber\\
&&\hspace{-2mm} =~{\alpha_s\over \pi^2}
\!\int\!d^2z_3~
{ z_{12}^2\over z_{13}^2 z_{23}^2}\Big\{1-{\alpha_sN_c\over 4\pi}\Big[{\pi^2\over 3}
+2\ln{z_{13}^2\over z_{12}^2}\ln{z_{23}^2\over z_{12}^2}\Big]\Big\}
~[{\rm Tr}\{T^a\hat{U}^\eta_{z_1} \hat{U}^{\dagger\eta}_{z_3}T^a\hat{U}^\eta_{z_3} \hat{U}^{\dagger\eta}_{z_2}\}-N_c{\rm Tr}\{\hat{U}^\eta_{z_1} \hat{U}^{\dagger\eta}_{z_2}\}]
\nonumber\\
&&\hspace{-2mm} 
-~{\alpha_s^2\over 4\pi^4}
~\!\int \!{d^2 z_3d^2 z_4\over z_{34}^4}~
{ z_{12}^2z_{34}^2\over z_{13}^2 z_{24}^2}
\Big[1+
{ z_{12}^2z_{34}^2\over z_{13}^2z_{24}^2-z_{23}^2z_{14}^2}\Big]
\ln{z_{13}^2z_{24}^2\over z_{14}^2z_{23}^2}
\nonumber\\
&&\hspace{32mm} 
\times~
{\rm Tr}\{[T^a,T^b]\hat{U}^\eta_{z_1}T^{a'}T^{b'}\hat{U}_{z_2}^{\dagger\eta}+T^bT^a\hat{U}^\eta_{z_1}[T^{b'},T^{a'}]
\hat{U}_{z_2}^{\dagger\eta}\}(\hat{U}^\eta_{z_3})^{aa'}(\hat{U}^\eta_{z_4}-\hat{U}^\eta_{z_3})^{ bb'}
\label{nlobksym}
\end{eqnarray}
All terms in the r.h.s. of this equation are M\"obius invariant except the double-log term 
proportional to $\ln{z_{13}^2\over z_{12}^2}\ln{z_{23}^2\over z_{12}^2}$. As we discussed in the Introduction, the reason
for this non-invariance is the cutoff in the longitudinal direction which violates the formal
invariance of the non-cut Wilson lines.

It is worth noting that conformal and non-conformal terms come from graphs with different topology: the conformal terms come from 1$\rightarrow$3 dipoles diagrams (see Fig. 6 in Ref. \cite{nlobk}) which describe the dipole creation while the non-conformal double-log term comes from the 
1$\rightarrow$2 dipole transitions (Fig.9 in Ref. \cite{nlobk})) which can be regarded as a combination of dipole creation and dipole recombination.

%%%%%%%%%%%%%%%%%%%%%%%%%%%%%%%%%%%%%%%%%%%%%%%%%%
\section{Conformal dipole and conformal NLO kernel \label{sec:nlobksymconf}}
A we discussed in the Introduction, it is possible to define 
the composite conformal dipole operator order by order in perturbation theory in such a way that the evolution equation for this operator
would be M\"obius invariant. The form of this operator can be guessed from the expression 
(\ref{ope1}) (see the discussion in the Appendix B)
\begin{eqnarray}
&&\hspace{-5mm}
[{\rm Tr}\{\hat{U}^\eta_{z_1}\hat{U}^{\dagger\eta}_{z_2}\}\big]^{\rm conf}~=~{\rm Tr}\{\hat{U}^\eta_{z_1}\hat{U}^{\dagger\eta}_{z_2}\}
+{\alpha_s\over 2\pi^2}\!\int\! d^2 z_3{z_{12}^2\over z_{13}^2z_{23}^2}
[ {\rm Tr}\{T^n\hat{U}^\eta_{z_1}\hat{U}^{\dagger\eta}_{z_3}T^n\hat{U}^\eta_{z_3}\hat{U}^{\dagger\eta}_{z_2}\}
 -N_c {\rm Tr}\{\hat{U}^\eta_{z_1}\hat{U}^{\dagger\eta}_{z_2}\}]
 \ln {a z_{12}^2\over z_{13}^2z_{23}^2}
\label{confodipole1}
 \end{eqnarray}
Let us find the NLO evolution kernel for this operator and demonstrate that it is conformal. 

For the evolution of the composite operator (\ref{confodipole1}) we get
\begin{eqnarray}
&&\hspace{-5mm}
{d\over d\eta}[{\rm Tr}\{\hat{U}^\eta_{z_1}\hat{U}^{\dagger\eta}_{z_2}\}\big]^{\rm conf}~
\nonumber\\
&&\hspace{-5mm}=~{d\over d\eta}{\rm Tr}\{\hat{U}^\eta_{z_1}\hat{U}^{\dagger\eta}_{z_2}\}
+{\alpha_s\over 2\pi^2}\!\int\! d^2 z_3{z_{12}^2\over z_{13}^2z_{23}^2}
[{d\over d\eta} {\rm Tr}\{T^a\hat{U}^\eta_{z_1}\hat{U}^{\dagger\eta}_{z_3}T^a\hat{U}^\eta_{z_3}\hat{U}^{\dagger\eta}_{z_2}\}
 -N_c {d\over d\eta}{\rm Tr}\{\hat{U}^\eta_{z_1}\hat{U}^{\dagger\eta}_{z_2}\}]
 \ln {a z_{12}^2\over z_{13}^2z_{23}^2}
\label{confevol1}
 \end{eqnarray}
Writing down the evolution of the four-Wilson-line operator (\ref{4wlinevolad}) calculated in Appendix C  we obtain 
\begin{eqnarray}
&&\hspace{-3mm}
{d\over d\eta}[{\rm Tr}\{\hat{U}^\eta_{z_1}\hat{U}^{\dagger\eta}_{z_2}\}\big]^{\rm conf}
\nonumber\\
&&\hspace{-1mm}
=~{\alpha_s\over \pi^2}
\!\int\!d^2z_3~
{z_{12}^2\over z_{13}^2 z_{23}^2}\Big\{1+{\alpha_sN_c\over 4\pi}\Big[-{\pi^2\over 3}
-2\ln{z_{13}^2\over z_{12}^2}\ln{z_{23}^2\over z_{12}^2}\Big]\Big\}
~[{\rm Tr}\{T^a\hat{U}^\eta_{z_1} \hat{U}^{\dagger\eta}_{z_3}T^a\hat{U}^\eta_{z_3} \hat{U}^{\dagger\eta}_{z_2}\}
-N_c{\rm Tr}\{\hat{U}^\eta_{z_1} \hat{U}^{\dagger\eta}_{z_2}\}]
\nonumber\\
&&\hspace{-3mm}
+~{\alpha_s^2\over 8\pi^4}\!\int\! d^2 z_3d^2z_4
(\hat{U}^\eta_{z_3})^{aa'}
 \Big[{z_{12}^2\over z_{14}^2z_{24}^2}
 {\rm Tr}\{T^aT^b\hat{U}^\eta_{z_1}T^{a'}T^{b'}\hat{U}^{\dagger\eta}_{z_2}
 + T^bT^a\hat{U}^\eta_{z_1}T^{b'}T^{a'}\hat{U}^{\dagger\eta}_{z_2}\}
 (2\hat{U}^\eta_{z_4}-\hat{U}^\eta_{z_1}-\hat{U}^\eta_{z_2})^{bb'}
 \nonumber\\
&&\hspace{-3mm}
-~ {z_{13}^2\over z_{14}^2z_{34}^2}
 {\rm Tr}\{T^aT^b\hat{U}^\eta_{z_1}[T^{a'},T^{b'}]\hat{U}^{\dagger\eta}_{z_2}
 +[T^b,T^a]\hat{U}^\eta_{z_1}T^{b'}T^{a'}\hat{U}^{\dagger\eta}_{z_2}\}
 (2\hat{U}^\eta_{z_4} -\hat{U}^\eta_{z_1}-\hat{U}^\eta_{z_3})^{bb'}
 \nonumber\\
&&\hspace{-3mm}
-~{z_{23}^2\over z_{24}^2z_{34}^2}
 {\rm Tr}\{[T^a,T^b]\hat{U}^\eta_{z_1}T^{a'}T^{b'}\hat{U}^{\dagger\eta}_{z_2}
 + T^bT^a\hat{U}^\eta_{z_1}[T^{b'},T^{a'}]\hat{U}^{\dagger\eta}_{z_2}\}
 (2\hat{U}^\eta_{z_4}-\hat{U}^\eta_{z_2}-\hat{U}^\eta_{z_3})^{bb'} \Big]
  {z_{12}^2\over z_{13}^2z_{23}^2}
 \ln{az_{12}^2\over z_{13}^2z_{23}^2}  
 \nonumber\\
&&\hspace{-1mm}
-~{\alpha_s^2N_c\over 2\pi^4}\!\int\! d^2z_3d^2z_4
({\rm Tr}\{T^a\hat{U}^\eta_{z_1}\hat{U}^{\dagger\eta}_{z_4} T^a \hat{U}^\eta_{z_4}\hat{U}^{\dagger\eta}_{z_2}\}
-N_c{\rm Tr}\{U_{z_1}U_{z_2}^\dagger\}){z_{12}^2\over z_{13}^2z_{23}^2}{z_{12}^2\over z_{14}^2z_{24}^2}
\ln{az_{12}^2\over z_{13}^2z_{23}^2}
~+~...
\label{confevol2}
 \end{eqnarray}
where the dots stand for the last (conformal) term in the r.h.s. of Eq. (\ref{nlobksym}).

Next we need the ``counterterms'' converting the four-Wilson-line operator 
${\rm Tr}\{T^a\hat{U}^\eta_{z_1} \hat{U}^{\dagger\eta}_{z_3}T^a\hat{U}^\eta_{z_3} \hat{U}^{\dagger\eta}_{z_2}\}$
into the conformal operator. In principle, this should be done similarly to obtaining the ``conformal dipole'' (\ref{confodipole}) in Appendix C: 
one should expand the T-product of conformal operators in the next ($\alpha_s^2$) order in perturbation theory and rearrange 
the 6-Wilson-line operators in such a way that the coefficient in front of the combination
 $[{\rm Tr}\{T^a\hat{U}^\eta_{z_1} \hat{U}^{\dagger\eta}_{z_3}T^a\hat{U}^\eta_{z_3} \hat{U}^{\dagger\eta}_{z_2}\}]^{\rm conf}$ 
 is conformal. Since it means the calculation of the NNLO impact factor which is a formidable task, we will use 
 another method to get the four-Wilson-line conformal operator. We will make a guess 
\begin{eqnarray}
&&\hspace{-1mm}
[{\rm Tr}\{T^a\hat{U}^\eta_{z_1}\hat{U}^{\dagger\eta}_{z_3}T^a\hat{U}^\eta_{z_3}\hat{U}^{\dagger\eta}_{z_2}\}
-N_c{\rm Tr}\{\hat{U}^\eta_{z_1}\hat{U}^{\dagger\eta}_{z_2}\}]^{\rm conf}
\nonumber\\
&&\hspace{-1mm}
 =~{\rm Tr}\{T^a\hat{U}^\eta_{z_1}\hat{U}^{\dagger\eta}_{z_3}T^a\hat{U}^\eta_{z_3}\hat{U}^{\dagger\eta}_{z_2}\}
-N_c{\rm Tr}\{\hat{U}^\eta_{z_1}\hat{U}^{\dagger\eta}_{z_2}\}~\nonumber\\
&&\hspace{-1mm}
+~{\alpha_s\over 8\pi^2}\!\int\! d^2z_4(\hat{U}^\eta_{z_3})^{aa'}
 \Big\{
 {\rm Tr}\{T^aT^b\hat{U}^\eta_{z_1}T^{a'}T^{b'}\hat{U}^{\dagger\eta}_{z_2}
 + T^bT^a\hat{U}^\eta_{z_1}T^{b'}T^{a'}\hat{U}^{\dagger\eta}_{z_2}\}
 (2\hat{U}^\eta_{z_4}-\hat{U}^\eta_{z_1}-\hat{U}^\eta_{z_2})^{bb'}{z_{12}^2\over z_{14}^2z_{24}^2}\ln\big({az_{12}^2\over z_{14}^2z_{24}^2}\big)
  \nonumber\\
&&\hspace{-1mm}
 -~
 {\rm Tr}\{T^aT^b\hat{U}^\eta_{z_1}[T^{a'},T^{b'}]\hat{U}^{\dagger\eta}_{z_2}
 +[T^b,T^a]\hat{U}^\eta_{z_1}T^{b'}T^{a'}\hat{U}^{\dagger\eta}_{z_2}\}
 (2\hat{U}^\eta_{z_4}-\hat{U}^\eta_{z_1}-\hat{U}^\eta_{z_3})^{bb'}{z_{13}^2\over z_{14}^2z_{34}^2}\ln\big({az_{13}^2\over z_{14}^2z_{34}^2}\big)
 \nonumber\\
&&\hspace{-1mm}
-~
 {\rm Tr}\{[T^a,T^b]\hat{U}^\eta_{z_1}T^{a'}T^{b'}\hat{U}^{\dagger\eta}_{z_2}
 + T^bT^a\hat{U}^\eta_{z_1}[T^{b'},T^{a'}]\hat{U}^{\dagger\eta}_{z_2}\}
 (2\hat{U}^\eta_{z_4}-\hat{U}^\eta_{z_2}-\hat{U}^\eta_{z_3})^{bb'} 
 {z_{23}^2\over z_{24}^2z_{34}^2}\ln\big({az_{23}^2\over z_{24}^2z_{34}^2}\big)
 \Big\}
  \nonumber\\
&&\hspace{-1mm}
-~{\alpha_sN_c\over 2\pi^2}\!\int\! d^2z_4{z_{12}^2\over z_{14}^2z_{24}^2}
({\rm Tr}\{T^aU_{z_1}U_{z_4}^\dagger T^a U_{z_4}U_{z_2}^\dagger\}
-N_c{\rm Tr}\{U_{z_1}U_{z_2}^\dagger\})\ln{a z_{12}^2\over z_{14}^2z_{24}^2}
\label{confoper4}
 \end{eqnarray}
and check that it leads to the conformal evolution equation (\ref{nlobksymconf}).  

Rewriting the evolution equation (\ref{confevol2}) in terms of conformal 
operators (\ref{confodipole1}) and (\ref{confoper4}) we obtain 

\begin{eqnarray}
&&\hspace{-1mm}
{d\over d\eta}[{\rm Tr}\{\hat{U}^\eta_{z_1}\hat{U}^{\dagger\eta}_{z_2}\}\big]^{\rm conf}
\nonumber\\
&&\hspace{-1mm}
 =~{\alpha_s\over \pi^2}\!\int\!d^2z_3~{z_{12}^2\over z_{13}^2 z_{23}^2}\Big\{1+{\alpha_sN_c\over 4\pi}\Big[-{\pi^2\over 3}
-2\ln{z_{12}^2\over z_{13}^2}\ln{z_{12}^2\over z_{23}^2}\Big]\Big\}
~[{\rm Tr}\{T^a\hat{U}^\eta_{z_1} \hat{U}^{\dagger\eta}_{z_3}T^a\hat{U}^\eta_{z_3} \hat{U}^{\dagger\eta}_{z_2}\}
-N_c{\rm Tr}\{\hat{U}^\eta_{z_1} \hat{U}^{\dagger\eta}_{z_2}\}]^{\rm conf}
\nonumber\\
&&\hspace{-3mm}
+~{\alpha_s^2\over 8\pi^4}\!\int\! d^2 z_3d^2z_4{z_{12}^2\over z_{13}^2z_{23}^2}
 \Big\{
(\hat{U}^\eta_{z_3})^{aa'} {\rm Tr}\{T^aT^b\hat{U}^\eta_{z_1}T^{a'}T^{b'}\hat{U}^{\dagger\eta}_{z_2}
 + T^bT^a\hat{U}^\eta_{z_1}T^{b'}T^{a'}\hat{U}^{\dagger\eta}_{z_2}\}
 (\hat{U}^\eta_{z_1}+\hat{U}^\eta_{z_2})^{bb'}{z_{12}^2\over z_{14}^2z_{24}^2}
 \ln\big({z_{13}^2z_{23}^2\over z_{14}^2z_{24}^2}\big)
  \nonumber\\
&&\hspace{-1mm}
 +~4(\hat{U}^\eta_{z_3})^{aa'}(\hat{U}^\eta_{z_4}-\hat{U}^\eta_{z_3})^{bb'}
 {\rm Tr}\{[T^a,T^b]\hat{U}^\eta_{z_1}T^{a'}T^{b'}\hat{U}^{\dagger\eta}_{z_2}
 + T^bT^a\hat{U}^\eta_{z_1}[T^{b'},T^{a'}]\hat{U}^{\dagger\eta}_{z_2}\}\
~{z_{23}^2\over z_{24}^2z_{34}^2}
 \ln\big({z_{14}^2z_{23}^2\over z_{12}^2z_{34}^2}\big)
   \nonumber\\
&&\hspace{-1mm}
 +~4(\hat{U}^\eta_{z_3})^{aa'}(\hat{U}^\eta_{z_3})^{bb'}
 {\rm Tr}\{[T^a,T^b]\hat{U}^\eta_{z_1}T^{a'}T^{b'}\hat{U}^{\dagger\eta}_{z_2}
 + T^bT^a\hat{U}^\eta_{z_1}[T^{b'},T^{a'}]\hat{U}^{\dagger\eta}_{z_2}\}
~{z_{23}^2\over z_{24}^2z_{34}^2}
 \ln\big({z_{14}^2z_{23}^2\over z_{12}^2z_{34}^2}\big)
  \nonumber\\
&&\hspace{-1mm}
-~ (\hat{U}^\eta_{z_1}+\hat{U}^\eta_{z_3})^{aa'}(\hat{U}^\eta_{z_3})^{bb'}
 {\rm Tr}\{[T^a,T^b]\hat{U}^\eta_{z_1}T^{a'}T^{b'}\hat{U}^{\dagger\eta}_{z_2}
 + T^bT^a\hat{U}^\eta_{z_1}[T^{b'},T^{a'}]\hat{U}^{\dagger\eta}_{z_2}\}\
{z_{13}^2\over z_{14}^2z_{34}^2}
 \ln\big({z_{13}^4z_{23}^2\over z_{12}^2z_{14}^2z_{34}^2}\big)
 \nonumber\\
&&\hspace{-1mm}
-~
 (\hat{U}^\eta_{z_3})^{aa'} (\hat{U}^\eta_{z_2}+\hat{U}^\eta_{z_3})^{bb'} {\rm Tr}\{[T^a,T^b]\hat{U}^\eta_{z_1}T^{a'}T^{b'}\hat{U}^{\dagger\eta}_{z_2}
 + T^bT^a\hat{U}^\eta_{z_1}[T^{b'},T^{a'}]\hat{U}^{\dagger\eta}_{z_2}\}
 {z_{23}^2\over z_{24}^2z_{34}^2}\ln\big({z_{23}^4z_{13}^2\over z_{12}^2z_{24}^2z_{34}^2}\big)\Big\}
 \nonumber\\
&&\hspace{-1mm}
+~{\alpha_s^2N_c\over 2\pi^4}\!\int\! d^2 z_3d^2z_4{z_{12}^2\over z_{13}^2z_{23}^2}{z_{12}^2\over z_{14}^2z_{24}^2}
[{\rm Tr}\{T^a\hat{U}_{z_1}\hat{U}_{z_3}^\dagger T^a \hat{U}_{z_3}\hat{U}_{z_2}^\dagger\}-N_c{\rm Tr}\{\hat{U}_{z_1}\hat{U}_{z_2}^\dagger\}]
\ln{z_{14}^2z_{24}^2\over z_{13}^2z_{23}^2}+...
\label{work1}
 \end{eqnarray}
Note that with our accuracy we do not need to specify the form of  ``counterterms''  for the conformal composite operators in the 
$\alpha_s^2$ term in the r.h.s. of this equation.

Using Eq. (\ref{traces}) and the integral (\ref{integral1}) from Appendix we get
\begin{eqnarray}
&&\hspace{-1mm}
{d\over d\eta}[{\rm Tr}\{\hat{U}^\eta_{z_1}\hat{U}^{\dagger\eta}_{z_2}\}\big]^{\rm conf}
\label{corr1}\\
&&\hspace{-1mm}
 =~{\alpha_s\over \pi^2}\!\int\!d^2z_3~{z_{12}^2\over z_{13}^2 z_{23}^2}\Big\{1+{\alpha_sN_c\over 4\pi}\Big[-{\pi^2\over 3}
-2\ln{z_{12}^2\over z_{13}^2}\ln{z_{12}^2\over z_{23}^2}\Big]\Big\}
~[{\rm Tr}\{T^a\hat{U}^\eta_{z_1} \hat{U}^{\dagger\eta}_{z_3}T^a\hat{U}^\eta_{z_3} \hat{U}^{\dagger\eta}_{z_2}\}
-N_c{\rm Tr}\{\hat{U}^\eta_{z_1} \hat{U}^{\dagger\eta}_{z_2}\}]^{\rm conf}
\nonumber\\
&&\hspace{-3mm}
+~{\alpha_s\over 2\pi^4}\!\int\! d^2 z_3d^2z_4{z_{12}^2\over z_{13}^2 z_{24}^2z_{34}^2}
(\hat{U}^\eta_{z_3})^{aa'}(\hat{U}^\eta_{z_4}-\hat{U}^\eta_{z_3})^{bb'}
 {\rm Tr}\{[T^a,T^b]\hat{U}^\eta_{z_1}T^{a'}T^{b'}\hat{U}^{\dagger\eta}_{z_2}
 + T^bT^a\hat{U}^\eta_{z_1}[T^{b'},T^{a'}]\hat{U}^{\dagger\eta}_{z_2}\}
 \ln {z_{14}^2z_{23}^2\over z_{12}^2z_{34}^2}
\Big\}
 \nonumber\\
&&\hspace{-1mm}
+~{\alpha_sN_c\over 2\pi^3}\!\int\! \!d^2z_3~{z_{12}^2\over z_{13}^2 z_{23}^2}\ln{z_{12}^2\over z_{13}^2}\ln{z_{12}^2\over z_{23}^2}
{\rm Tr}\{T^a\hat{U}^\eta_{z_1} \hat{U}^{\dagger\eta}_{z_3}T^a\hat{U}^\eta_{z_3} \hat{U}^{\dagger\eta}_{z_2}\}
-{\alpha_sN_c^2\over \pi^3}2\zeta(3){\rm Tr}\{\hat{U}^\eta_{z_1}\hat{U}^{\dagger\eta}_{z_2}\}+...
 \nonumber\\
&&\hspace{-1mm}
 =~{\alpha_s\over \pi^2}\!\int\!d^2z_3~{z_{12}^2\over z_{13}^2 z_{23}^2}\Big\{1-{\alpha_sN_c\over 4\pi}{\pi^2\over 3}\Big\}
~[{\rm Tr}\{T^a\hat{U}^\eta_{z_1} \hat{U}^{\dagger\eta}_{z_3}T^a\hat{U}^\eta_{z_3} \hat{U}^{\dagger\eta}_{z_2}\}
-N_c{\rm Tr}\{\hat{U}^\eta_{z_1} \hat{U}^{\dagger\eta}_{z_2}\}]^{\rm conf}
+~{\alpha_s\over 2\pi^4}\!\int\! d^2 z_3d^2z_4{z_{12}^2\over z_{13}^2 z_{24}^2z_{34}^2}
\nonumber\\
&&\hspace{-3mm}
\times~
(\hat{U}^\eta_{z_3})^{aa'}(\hat{U}^\eta_{z_4}-\hat{U}^\eta_{z_3})^{bb'}
 {\rm Tr}\{[T^a,T^b]\hat{U}^\eta_{z_1}T^{a'}T^{b'}\hat{U}^{\dagger\eta}_{z_2}
 + T^bT^a\hat{U}^\eta_{z_1}[T^{b'},T^{a'}]\hat{U}^{\dagger\eta}_{z_2}\}
 \ln {z_{14}^2z_{23}^2\over z_{12}^2z_{34}^2}
+...
\nonumber
 \end{eqnarray}
We see that the non-conformal term double-log term $\sim\ln{z_{12}^2\over z_{13}^2}\ln{z_{12}^2\over z_{23}^2}$ is canceled 
with the correction coming from the subtitution of the dipole by the  composite operator (\ref{confodipole}). This confirms our expression
(\ref{confodipole}) for the conformal dipole and 
justifies our guess for the conformal  composite 4-Wilson-line operator (\ref{confoper4}).

Substituting the dots in the r.h.s. of this equation for the last (conformal) term in Eq. (\ref{nlobksym}) we get the final 
evolution equation for the conformal composite operator cited in the Introduction (\ref{nlobksymconf}): 
\begin{eqnarray}
&&\hspace{-5mm}
{d\over d\eta}\big[{\rm Tr}\{\hat{U}^\eta_{z_1}\hat{U}^{\dagger\eta}_{z_2}\}\big]^{\rm conf}~
\label{nlobksymconf1}\\
&&\hspace{-5mm}
=~{\alpha_s\over \pi^2}
\!\int\!d^2z_3~
{z_{12}^2\over z_{13}^2 z_{23}^2}\Big[1-
{\alpha_sN_c\over 4\pi}{\pi^2\over 3}\Big]\big[{\rm Tr}\{T^a\hat{U}^\eta_{z_1}\hat{U}^{\dagger\eta}_{z_3}T^a\hat{U}_{z_3}\hat{U}^{\dagger\eta}_{z_2}\} 
-N_c {\rm Tr}\{\hat{U}^\eta_{z_1}\hat{U}^{\dagger\eta}_{z_2}\}\big]^{\rm conf}
\nonumber\\
&&\hspace{-2mm} 
-~{\alpha_s^2\over 4\pi^4}
\int \!d^2 z_3 d^2 z_4 {z_{12}^2\over z_{13}^2z_{24}^2z_{34}^2}
\Big\{2\ln{z_{12}^2z_{34}^2\over z_{14}^2z_{23}^2}
+\Big[1+{z_{12}^2z_{34}^2\over z_{13}^2z_{24}^2-z_{14}^2z_{23}^2}\Big]\ln{z_{13}^2z_{24}^2\over z_{14}^2z_{23}^2}\Big\}
\nonumber\\ 
&&\hspace{-2mm}
\times~ {\rm Tr}\{[T^a,T^b]\hat{U}^\eta_{z_1}T^{a'}T^{b'}\hat{U}^{\dagger\eta}_{z_2}
 + T^bT^a\hat{U}^\eta_{z_1} [T^{b'},T^{a'}]\hat{U}^{\dagger\eta}_{z_2}\}
 [(\hat{U}^\eta_{z_3})^{aa'}(\hat{U}^\eta_{z_4})^{bb'}-(z_4\rightarrow z_3)]
\nonumber
\end{eqnarray}

At this point we would like to discuss the gauge invariance of our evolution equation (\ref{nlobksymconf1}). The Wilson-line operator is gauge invariant up
to gauge rotations at $\pm\infty$. As it was discussed in Refs. \cite{npb96, mobzor}, the evolution equation should be reformulated in terms of gauge-invariant Wilson loops. 
In particular,   ${\rm Tr}\{\hat{U}^\eta_{z_1}\hat{U}^{\dagger\eta}_{z_2}\}$ in the l.h.s. of this equation should be promoted to
\begin{eqnarray}
{\rm Tr}\{\hat{U}^\eta_{z_1}\hat{U}^{\dagger\eta}_{z_2}\}~=~{\rm Tr}\{[\infty p_1+z_1,-\infty p_1+z_1][z_1,z_2]_{-\infty}
[-\infty p_1+z_2,\infty p_1+z_2][z_2,z_1]_{\infty}\}
\end{eqnarray}
where we use the notation $[z_1,z_2]_{\pm \infty}\equiv [z_1\pm\infty p_1,z_2\pm\infty p_1]$ and the precise form of contours connecting these points  does
not matter since the fields at infinity are pure gauges.
We do not have a simple way to introduce these gauge links at infinity to the r.h.s. of Eq. (\ref{nlobksymconf1})  in the adjoint representation, but 
it can be easily done if one rewrites the adjoint traces in terms of traces in the fundamental representation using Eq. (\ref{trace}) from the 
Appendix D.
For example, ${\rm tr}\{U_{z_4}U_{z_1}^\dagger \}{\rm tr}\{ U_{z_2} U_{z_3}^\dagger U_{z_1}U_{z_2}^\dagger U_{z_3}U_{z_4}^\dagger\}$ 
should be replaced by
\begin{eqnarray}
&&\hspace{-3mm}
{\rm Tr}\{[\infty p_1+z_4,-\infty p_1+z_4][z_4,z_1]_{-\infty}
[-\infty p_1+z_1,\infty p_1+z_1][z_1,z_4]_{\infty}\}
\nonumber\\
&&\hspace{-3mm}
\times~{\rm Tr}\{[\infty p_1+z_2,-\infty p_1+z_2][z_2,z_3]_{-\infty}\}[-\infty p_1+z_3,\infty p_1+z_3][z_3,z_1]_{\infty}[z_1,z_2]_{-\infty}
[-\infty p_1+z_2,\infty p_1+z_2]
\nonumber\\
&&\hspace{-3mm}
\times~[z_2,z_3]_{\infty}[\infty p_1+z_3,-\infty p_1+z_3][z_3,z_4]_{-\infty}[-\infty p_1+z_4,\infty p_1+z_4][z_4,z_2]_{\infty}\}
\nonumber
\end{eqnarray}
and similarly for other traces in the r.h.s. of Eq. (\ref{traces}). With this replacement, the evolution equation (\ref{nlobksymconf1}) is gauge invariant.

%%%%%%%%%%%%%%%%%%%%%%%%%%%%%%%%%%%%%%%%%%%%%%
\section{Comparison to NLO BFKL  \label{sec:compare}}

In this section we compare our kernel with the forward NLO BFKL results for ${\cal N}=4$ SYM \cite{lipkot}. To compare to 
 the BFKL amplitude of gluon-gluon scattering at high energies we need to expand our Wilson lines 
 up to two-gluon accuracy.
 We define the analog of Eq. (1) in the adjoint approximation:
 \begin{equation}
 \hat{\cal V}^\eta(x,y)~=~1-{1\over N_c^2-1}{\rm Tr}\{ \hat{U}^\eta_{x} \hat{U}^{\dagger\eta}_y\}=~{N_c^2\over N_c^2-1}[\hat{\cal U}(x,y)+\hat{\cal U}(y,x)
 -\hat{\cal U}(x,y)\hat{\cal U}(x,y)]~\simeq~{N_c^2\over N_c^2-1}[\hat{\cal U}(x,y)+\hat{\cal U}(y,x)]
 \end{equation}
The corresponding conformal dipole operator in the BFKL approximation has the form
 \begin{equation}
 \hat{\cal V}_{\rm conf}^\eta(z_1,z_2)~=~ \hat{\cal V}^\eta(z_1,z_2)+{\alpha_sN_c\over 4\pi^2}\!\int\! d^2z{z_{12}^2\over z_{13}^2 z_{23}^2}\ln {az_{12}^2\over z_{13}^2 z_{23}^2}
[\hat{\cal V}^\eta(z_1,z_3)+\hat{\cal V}^\eta(z_2,z_3)-\hat{\cal V}^\eta(z_1,z_2)]
\label{calv}
 \end{equation}
 Using color traces (\ref{tracelin}), (\ref{tracelintriv}), and (\ref{tracelin2})  from Appendix D it is possible to demonstrate that the conformal 4-Wilson-line operator (\ref{confoper4}) reduces to the sum of three conformal dipoles:
\begin{equation}
\hspace{-0mm}  
[{\rm Tr}\{T^a\hat{U}^\eta_{z_1}\hat{U}^{\dagger\eta}_{z_3}T^a\hat{U}^\eta_{z_3}\hat{U}^{\dagger\eta}_{z_2}\}-N_c{\rm Tr}\{\hat{U}^\eta_{z_1}\hat{U}^{\dagger\eta}_{z_2}\}]^{\rm conf}~
=~-{N_c\over 2}(N_c^2-1)\big[\hat{\cal V}_{\rm conf}^\eta(z_1,z_3)+\hat{\cal V}_{\rm conf}^\eta(z_2,z_3)-\hat{\cal V}_{\rm conf}^\eta(z_1,z_2)\big]
\label{fla70}
\end{equation}
and therefore the evolution equation (\ref{nlobksymconf1}) turns into
\begin{eqnarray}
&&\hspace{-5mm}
{d\over d\eta}\hat{\cal V}^\eta_{\rm conf}(z_1,z_2)~
~
=~{\alpha_sN_c\over 2\pi^2}
\!\int\!d^2z_3~
{z_{12}^2\over z_{13}^2 z_{23}^2}\Big[1-{\alpha_sN_c\over 4\pi}{\pi^2\over 3}\Big]
[\hat{\cal V}^\eta_{\rm conf}(z_1,z_3)+\hat{\cal V}^\eta_{\rm conf}(z_2,z_3)-\hat{\cal V}^\eta_{\rm conf}(z_1,z_2)]
\label{nlosymlin}\\
&&\hspace{-5mm} 
+~{\alpha_s^2N_c^2\over 8\pi^4}\!
\int \!{d^2 z_3d^2 z_4\over z_{34}^4}~{z_{12}^2z_{34}^2\over z_{13}^2z_{24}^2}[\hat{\cal V}^\eta_{\rm conf}(z_3,z_4)
+\hat{\cal V}^\eta_{\rm conf}(z_2,z_4)-\hat{\cal V}^\eta_{\rm conf}(z_2,z_3)]
\Big\{2\ln{z_{12}^2z_{34}^2\over z_{14}^2z_{23}^2}+
\Big[1+{ z_{12}^2z_{34}^2\over z_{13}^2z_{24}^2- z_{14}^2z_{23}^2}\Big]\ln{z_{13}^2z_{24}^2\over z_{14}^2z_{23}^2}\Big\}
\nonumber
\end{eqnarray}
It is convenient to change  $z_4\leftrightarrow z_3$ in the second term in square brackets and to perform the integral over $z_4$ 
for the second and the third terms. One obtains:
\begin{eqnarray}
&&\hspace{-5mm}
\int \!{d^2 z_3d^2 z_4\over z_{34}^4}~{z_{12}^2z_{34}^2\over z_{13}^2z_{24}^2}
[\hat{\cal V}^\eta_{\rm conf}(z_2,z_4)-\hat{\cal V}^\eta_{\rm conf}(z_2,z_3)]
\Big\{2\ln{z_{12}^2z_{34}^2\over z_{14}^2z_{23}^2}+
\Big[1+{ z_{12}^2z_{34}^2\over z_{13}^2z_{24}^2- z_{14}^2z_{23}^2}\Big]\ln{z_{13}^2z_{24}^2\over z_{14}^2z_{23}^2}\Big\}
\nonumber\\
&&\hspace{-5mm}
=\int \!d^2 z_3~\hat{\cal V}^\eta_{\rm conf}(z_2,z_3){z_{12}^2\over z_{13}^2 z_{23}^2} \!\int \!d^2 z_4~
\Big\{2\Big[{z_{13}^2\over z_{14}^2z_{34}^2}\ln {z_{12}^2z_{34}^2\over z_{13}^2z_{24}^2}
-{z_{23}^2\over z_{24}^2z_{34}^2}\ln {z_{12}^2z_{34}^2\over z_{14}^2z_{23}^2}\Big]
-\Big[{z_{13}^2\over z_{14}^2 z_{34}^2}+{z_{23}^2\over z_{24}^2 z_{34}^2}-{z_{12}^2\over z_{14}^2 z_{24}^2}\Big]
\ln{z_{13}^2z_{24}^2\over z_{14}^2z_{23}^2}\Big\}
\nonumber\\
&&\hspace{-5mm}
=~12\pi\zeta(3)\hat{\cal V}^\eta_{\rm conf}(z_1,z_2)
\label{zeterm}
\end{eqnarray}
because
\begin{eqnarray}
&&\hspace{-5mm}
{z_{12}^2\over z_{13}^2 z_{23}^2} \!\int \!d^2 z_4~
\Big\{2\Big[{z_{13}^2\over z_{14}^2z_{34}^2}\ln {z_{12}^2z_{34}^2\over z_{13}^2z_{24}^2}
-{z_{23}^2\over z_{24}^2z_{34}^2}\ln {z_{12}^2z_{34}^2\over z_{14}^2z_{23}^2}\Big]
-\Big[{z_{13}^2\over z_{14}^2 z_{34}^2}+{z_{23}^2\over z_{24}^2 z_{34}^2}-{z_{12}^2\over z_{14}^2 z_{24}^2}\Big]
\ln{z_{13}^2z_{24}^2\over z_{14}^2z_{23}^2}\Big\}
\nonumber\\
&&\hspace{-5mm}
=~12\pi\zeta(3)[\delta(z_{13}-\delta(z_{23})]
\nonumber
\end{eqnarray}
as it is easily seen from Eqs. (\ref{nashel}) and (\ref{integral2}) from Appendix E.

Using Eq. (\ref{zeterm}) one can rewrite the linearized equation (\ref{nlosymlin}) in the following way:
\begin{eqnarray}
&&\hspace{-2mm}
{d\over d\eta}\hat{\cal V}^\eta_{\rm conf}(z_1,z_2)~
~
=~{\alpha_sN_c\over 2\pi^2}
\!\int\!d^2z_3~
{z_{12}^2\over z_{13}^2 z_{23}^2}\Big[1-{\alpha_sN_c\over 4\pi}{\pi^2\over 3}\Big]
[\hat{\cal V}^\eta_{\rm conf}(z_1,z_3)+\hat{\cal V}^\eta_{\rm conf}(z_2,z_3)-\hat{\cal V}^\eta_{\rm conf}(z_1,z_2)]
\label{nlobfklsym}\\
&&\hspace{-2mm} 
+~{\alpha_s^2N_c^2\over 8\pi^4}\!
\int \!{d^2 z_3d^2 z_4\over z_{34}^4}~{z_{12}^2z_{34}^2\over z_{13}^2z_{24}^2}
\Big\{2\ln{z_{12}^2z_{34}^2\over z_{14}^2z_{23}^2}+
\Big[1+{ z_{12}^2z_{34}^2\over z_{13}^2z_{24}^2- z_{14}^2z_{23}^2}\Big]
\ln{z_{13}^2z_{24}^2\over z_{14}^2z_{23}^2}\Big\}\hat{\cal V}^\eta_{\rm conf}(z_3,z_4)
+{3\alpha_s^2N_c^2\over 2\pi^3}\zeta(3)\hat{\cal V}^\eta_{\rm conf}(z_1,z_2)
\nonumber
\end{eqnarray}
In this form the evolution equation (\ref{nlobfklsym}) coincides with the conformal NLO BFKL equation 
restored in Ref. \cite{nlobfklconf} from the eigenvalues calculated in Ref. \cite{lipkot}.

%%%%%%%%%%%%%%%%%%%%%%%%%%%%%%%%%%%%%%%%%%%%%%%%%%%%%%%%%
\section{Evolution equation in the fundamental representation \label{sec:fund}}

%++++++++++++++++++++++++++++++++++++++++++++++++++++++++++++++++++++++++++++
\subsection{In ${\cal N}=4$ SYM}
For comparison with QCD let us calculate  the evolution equation for color dipoles in the fundamental representation. 
The gluon part can be taken from Ref. \cite{nlobk}
\begin{eqnarray}
&&\hspace{-2mm}
\big({d\over d\eta}{\rm tr}\{\hat{U}^\eta_{z_1} \hat{U}^{\dagger\eta}_{z_2}\}\big)_{\rm gluon}~
\label{gluonfund}\\
&&\hspace{-2mm}
=~{\alpha_s\over 2\pi^2}
\!\int\!d^2z_3~
{ z_{12}^2\over z_{13}^2 z_{23}^2}\Big\{1+{\alpha_sN_c\over 4\pi}\Big[{11\over 3}\ln z_{12}^2\mu^2
-{11\over 3}{z_{13}^2-z_{23}^2\over  z_{12}^2}\ln{z_{13}^2\over z_{23}^2}+
{64 \over 9}-{\pi^2\over 3}
\nonumber\\
&&\hspace{52mm} 
-~
2\ln{z_{13}^2\over z_{12}^2}\ln{z_{23}^2\over z_{12}^2}\Big]\Big\}
~[{\rm tr}\{\hat{U}^\eta_{z_1} \hat{U}^{\dagger\eta}_{z_3}\}{\rm tr}\{\hat{U}_{z_3} \hat{U}^{\dagger\eta}_{z_2}\}
-N_c{\rm tr}\{\hat{U}^\eta_{z_1} \hat{U}^{\dagger\eta}_{z_2}\}]   
\nonumber\\
&&\hspace{-2mm} 
+~{\alpha_s^2\over 16\pi^4}
\int \!d^2 z_3d^2 z_4
\Bigg[
\Big\{-{4\over z_{34}^4}
+2{z_{13}^2z_{24}^2+z_{14}^2z_{23}^2-4 z_{12}^2z_{34}^2\over  z_{34}^4[z_{13}^2z_{24}^2-z_{14}^2z_{23}^2]}
\ln{z_{13}^2z_{24}^2\over z_{14}^2z_{23}^2}
+\Big[\Big(1+{z_{12}^2z_{34}^2\over z_{13}^2z_{24}^2-z_{14}^2z_{23}^2}
\Big)
\nonumber\\
&&\hspace{-2mm} 
\times~
{ z_{12}^2\over z_{13}^2z_{24}^2 z_{34}^2}\ln{z_{13}^2z_{24}^2\over z_{14}^2z_{23}^2}+z_3\leftrightarrow z_4\Big]\Big\}
\times~[{\rm tr}\{\hat{U}^\eta_{z_1}\hat{U}^{\dagger\eta}_{z_3}\}
{\rm tr}\{\hat{U}_{z_3}\hat{U}^{\dagger\eta}_{z_4}\}{\rm tr}\{\hat{U}_{z_4}\hat{U}^{\dagger\eta}_{z_2}\}
-{\rm tr}\{\hat{U}^\eta_{z_1}\hat{U}^{\dagger\eta}_{z_3} \hat{U}_{z_4}
U_{z_2}^\dagger\hat{U}_{z_3}\hat{U}^{\dagger\eta}_{z_4}\}-(z_4\rightarrow z_3)]
\nonumber\\ 
&&\hspace{-2mm}
+~\Big\{{ z_{12}^2\over z_{34}^2 }\Big[{1\over z_{13}^2z_{24}^2}+{1\over z_{23}^2z_{14}^2}\Big]
-{z_{12}^4\over  z_{13}^2z_{24}^2z_{14}^2z_{23}^2}\Big\}\ln{z_{13}^2z_{24}^2\over z_{14}^2z_{23}^2}
~{\rm tr}\{\hat{U}^\eta_{z_1} \hat{U}^{\dagger\eta}_{z_3}\}{\rm tr}\{\hat{U}_{z_3}\hat{U}^{\dagger\eta}_{z_4}\}
{\rm tr}\{\hat{U}_{z_4}\hat{U}^{\dagger\eta}_{z_2}\}
\Bigg]
\nonumber
\end{eqnarray}
The scalar part can be obtained from Eq. (\ref{scalars}) by the replacement $T^a\rightarrow t^a$ and Tr $\rightarrow$ tr:
\begin{eqnarray}
&&\hspace{-2mm}                     
\big({d\over d\eta} {\rm tr}\{\hat{U}^\eta_{z_1} \hat{U}^{\dagger\eta}_{z_2}\}\big)_{\rm scalars}    
~ =~
{6\alpha^2\over 16\pi^4}\!\int\! {d^2 z_3d^2 z_4\over z_{34}^4}~\Big[-2
+{z_{13}^2z_{24}^2+z_{14}^2z_{23}^2\over z_{13}^2z_{24}^2-z_{14}^2z_{23}^2}\ln{z_{13}^2z_{24}^2\over z_{14}^2z_{23}^2}\Big]
\nonumber\\
&&\hspace{22mm}
\times~
[{\rm tr}\{\hat{U}^\eta_{z_1}\hat{U}^{\dagger\eta}_{z_3}\}
{\rm tr}\{\hat{U}^\eta_{z_3}\hat{U}^{\dagger\eta}_{z_4}\}
{\rm tr}\{\hat{U}^\eta_{z_4}\hat{U}^{\dagger\eta}_{z_2}\}
-{\rm tr}\{\hat{U}^\eta_{z_1}\hat{U}^{\dagger\eta}_{z_3} \hat{U}^\eta_{z_4}\hat{U}^{\dagger\eta}_{z_2}\hat{U}^\eta_{z_3}\hat{U}^{\dagger\eta}_{z_4}\}-(z_4\rightarrow z_3)]
\nonumber\\
&&\hspace{-2mm}         
-~{\alpha_s^2N_c\over 8\pi^3}\!\int\! d^2z_3
[{\rm tr}\{\hat{U}^\eta_{z_1} \hat{U}^{\dagger\eta}_{z_3}\}{\rm tr}\{\hat{U}_{z_3} \hat{U}^{\dagger\eta}_{z_2}\}
-{1\over N_c}{\rm tr}\{\hat{U}^\eta_{z_1} \hat{U}^{\dagger\eta}_{z_2}\}]\!
~\Big\{{ z_{12}^2\over z_{13}^2z_{23}^2} \Big[\ln  z_{12}^2 \mu^2+{8\over 3}\Big]
+\Big[{1\over z_{13}^2}-{1\over z_{23}^2}\Big]\ln{z_{13}^2\over z_{23}^2}\Big\}      
\label{scalarfund}
\end{eqnarray}

Similarly, for gluino contribution one gets from Eq. (\ref{gluino})
\begin{eqnarray}
&&\hspace{-6mm}
\big({d\over d\eta} {\rm tr}\{\hat{U}^\eta_{z_1} \hat{U}^{\dagger\eta}_{z_2}\}\big)_{\rm gluino}    
~\nonumber\\
&&\hspace{-6mm}
=~{\alpha^2_sN_c\over 3\pi^3}
\!\int\!d^2z_3~[{\rm tr}\{\hat{U}^\eta_{z_1}\hat{U}^{\dagger\eta}_{z_3}\}{\rm tr}\{\hat{U}^\eta_{z_3}\hat{U}^{\dagger\eta}_{z_2}\}
-{1\over N_c}{\rm tr}\{\hat{U}^\eta_{z_1}\hat{U}^{\dagger\eta}_{z_2}\}]
\Big[-{ z_{12}^2\over z_{13}^2 z_{23}^2}[\ln  z_{12}^2\mu^2+{5\over 3}]+
{z_{13}^2-z_{23}^2\over z_{13}^2z_{23}^2}\ln{z_{13}^2\over z_{23}^2}\Big]
\nonumber\\
&&\hspace{-6mm}
+~{\alpha^2_s\over \pi^4}
\!\int\!d^2z_3 d^2z_4~
~{1\over z_{34}^4}
\Big\{1
+{z_{14}^2z_{23}^2+z_{24}^2z_{13}^2- z_{12}^2z_{34}^2\over 2(z_{14}^2z_{23}^2-z_{24}^2z_{13}^2)}
\ln{z_{24}^2z_{13}^2 \over z_{14}^2z_{23}^2}\Big\}
\nonumber\\
&&\hspace{26mm}
\times~[{\rm tr}\{\hat{U}^\eta_{z_1}\hat{U}^{\dagger\eta}_{z_3}\}{\rm tr}\{\hat{U}^\eta_{z_3}\hat{U}^{\dagger\eta}_{z_4}\}
{\rm tr}\{\hat{U}^\eta_{z_4}\hat{U}^{\dagger\eta}_{z_2}\}
-~{\rm tr}\{\hat{U}^\eta_{z_1}\hat{U}^{\dagger\eta}_{z_3} \hat{U}^\eta_{z_4}\hat{U}^{\dagger\eta}_{z_2}
\hat{U}^\eta_{z_3}\hat{U}^{\dagger\eta}_{z_4}\}-z_4\rightarrow z_3]
\label{gluinofund}
\end{eqnarray}
Adding together Eq. (\ref{gluonfund}), (\ref{scalarfund}) and (\ref{gluinofund}) we obtain the evolution equation
for the color dipole with longitudinqal cutoff (\ref{cutoff}) in the fundamental representation:
\begin{eqnarray}
&&\hspace{-2mm}
{d\over d\eta}{\rm tr}\{\hat{U}^\eta_{z_1} \hat{U}^{\dagger\eta}_{z_2}\}~
\label{nlobksymfundrigid}\\
&&\hspace{-2mm}
=~{\alpha_s\over 2\pi^2}
\!\int\!d^2z_3~
{ z_{12}^2\over z_{13}^2 z_{23}^2}\Big\{1+{\alpha_sN_c\over 4\pi}\Big[-{\pi^2\over 3}
-2\ln{z_{13}^2\over z_{12}^2}\ln{z_{23}^2\over z_{12}^2}\Big]\Big\}
~[{\rm tr}\{\hat{U}^\eta_{z_1} \hat{U}^{\dagger\eta}_{z_3}\}{\rm tr}\{\hat{U}_{z_3} \hat{U}^{\dagger\eta}_{z_2}\}
-N_c{\rm tr}\{\hat{U}^\eta_{z_1} \hat{U}^{\dagger\eta}_{z_2}\}]   
\nonumber\\ 
&&\hspace{-2mm}
+~{\alpha_s^2\over 16\pi^4}
\int \!d^2 z_3d^2 z_4 ~\Bigg[\Big\{{z_{12}^2\over z_{13}^2z_{24}^2z_{34}^2}\Big(1
+{ z_{12}^2 z_{34}^2\over z_{13}^2z_{24}^2-z_{14}^2z_{23}^2}\Big)
\ln{z_{13}^2z_{24}^2\over z_{14}^2z_{23}^2}+z_3\leftrightarrow z_4\Big\}
\nonumber\\ 
&&\hspace{32mm}
\times~[{\rm tr}\{\hat{U}^\eta_{z_1}\hat{U}^{\dagger\eta}_{z_3}\}
{\rm tr}\{\hat{U}_{z_3}\hat{U}^{\dagger\eta}_{z_4}\}{\rm tr}\{\hat{U}_{z_4}\hat{U}^{\dagger\eta}_{z_2}\}
-{\rm tr}\{\hat{U}^\eta_{z_1}\hat{U}^{\dagger\eta}_{z_3} \hat{U}_{z_4}
U_{z_2}^\dagger\hat{U}_{z_3}\hat{U}^{\dagger\eta}_{z_4}\}-(z_4\rightarrow z_3)]
\nonumber\\ 
&&\hspace{-2mm}
+~\Big\{{ z_{12}^2\over z_{34}^2 }\Big[{1\over z_{13}^2z_{24}^2}+{1\over z_{23}^2z_{14}^2}\Big]
-{z_{12}^4\over  z_{13}^2z_{24}^2z_{14}^2z_{23}^2}\Big\}\ln{z_{13}^2z_{24}^2\over z_{14}^2z_{23}^2}
~{\rm tr}\{\hat{U}^\eta_{z_1} \hat{U}^{\dagger\eta}_{z_3}\}{\rm tr}\{\hat{U}_{z_3}\hat{U}^{\dagger\eta}_{z_4}\}
{\rm tr}\{\hat{U}_{z_4}\hat{U}^{\dagger\eta}_{z_2}\}
\Bigg]
\nonumber
\end{eqnarray}
which can be rewritten as 
\begin{eqnarray}
&&\hspace{-2mm}
{d\over d\eta}{\rm tr}\{\hat{U}^\eta_{z_1} \hat{U}^{\dagger\eta}_{z_2}\}~
=~{\alpha_s\over 2\pi^2}
\!\int\!d^2z_3~
{ z_{12}^2\over z_{13}^2 z_{23}^2}\Big\{1-{\alpha_sN_c\over 4\pi}\Big[{\pi^2\over 3}
+2\ln{z_{13}^2\over z_{12}^2}\ln{z_{23}^2\over z_{12}^2}\Big]\Big\}
~[{\rm tr}\{\hat{U}^\eta_{z_1} \hat{U}^{\dagger\eta}_{z_3}\}{\rm tr}\{\hat{U}_{z_3} \hat{U}^{\dagger\eta}_{z_2}\}
-N_c{\rm tr}\{\hat{U}^\eta_{z_1} \hat{U}^{\dagger\eta}_{z_2}\}]   
\nonumber\\ 
&&\hspace{-2mm}
+~{\alpha_s^2\over 16\pi^4}
\int \!d^2 z_3d^2 z_4 ~{z_{12}^2\over z_{13}^2z_{24}^2z_{34}^2}\Big(1
+{ z_{12}^2 z_{34}^2\over z_{13}^2z_{24}^2-z_{14}^2z_{23}^2}\Big)
\ln{z_{13}^2z_{24}^2\over z_{14}^2z_{23}^2}
\nonumber\\ 
&&\hspace{-2mm}
\times~[2{\rm tr}\{\hat{U}^\eta_{z_1}\hat{U}^{\dagger\eta}_{z_3}\}
{\rm tr}\{\hat{U}_{z_3}\hat{U}^{\dagger\eta}_{z_4}\}{\rm tr}\{\hat{U}_{z_4}\hat{U}^{\dagger\eta}_{z_2}\}
-{\rm tr}\{\hat{U}^\eta_{z_1}\hat{U}^{\dagger\eta}_{z_3} \hat{U}_{z_4}
U_{z_2}^\dagger\hat{U}_{z_3}\hat{U}^{\dagger\eta}_{z_4}\}
-{\rm tr}\{\hat{U}^\eta_{z_1}\hat{U}^{\dagger\eta}_{z_4} \hat{U}_{z_3}
U_{z_2}^\dagger\hat{U}_{z_4}\hat{U}^{\dagger\eta}_{z_3}\}-(z_4\rightarrow z_3)]
\label{nlobksymfund}
\end{eqnarray}
due to Eq. (\ref{nashel}) from Appendix D.

The composite conformal dipole in the fundamental representation can be obtained from Eq. (\ref{confodipole})) by the usual substitution $T^a\rightarrow t^a$ and Tr $\rightarrow$ tr:
\begin{eqnarray}
&&\hspace{-2mm}
[{\rm tr}\{\hat{U}^\eta_{z_1} \hat{U}^{\dagger\eta}_{z_2}\}]^{\rm conf}
~=~{\rm tr}\{\hat{U}^\eta_{z_1} \hat{U}^{\dagger\eta}_{z_2}\}-{\alpha_s\over 4\pi^2}\!\int\! d^2z_3{z_{12}^2\over z_{13}^2 z_{23}^2}\ln {az_{12}^2\over z_{13}^2 z_{23}^2}
[{\rm tr}\{\hat{U}^\eta_{z_1} \hat{U}^{\dagger\eta}_{z_3}\}{\rm tr}\{\hat{U}_{z_3} \hat{U}^{\dagger\eta}_{z_2}\}
-N_c{\rm tr}\{\hat{U}^\eta_{z_1} \hat{U}^{\dagger\eta}_{z_2}\}]   
\label{confodipolefund}
\end{eqnarray}
Similarly, the conformal 4-Wilson-line operator (\ref{confoper4}) turns to
\begin{eqnarray}
&&\hspace{-5mm}
[{\rm tr}\{\hat{U}^\eta_{z_1}\hat{U}_{z_3}^{\dagger\eta}\}{\rm tr}\{U_{z_3}\hat{U}_{z_2}^{\dagger\eta}\} 
-N_c {\rm tr}\{\hat{U}^\eta_{z_1}\hat{U}_{z_2}^{\dagger\eta}\}]^{\rm conf}~=~
[{\rm tr}\{\hat{U}^\eta_{z_1}\hat{U}_{z_3}^{\dagger\eta}\}{\rm tr}\{\hat{U}^\eta_{z_3}\hat{U}_{z_2}^{\dagger\eta}\} 
-N_c {\rm tr}\{\hat{U}^\eta_{z_1}\hat{U}_{z_2}^{\dagger\eta}\}]
\label{conf4fund}\\
&&\hspace{-5mm}
+~{\alpha_s^2\over 4\pi^2}\!\int\!d^2z_4
\Big\{{z_{13}^2\over z_{14}^2z_{34}^2}
[{\rm tr}\{\hat{U}^\eta_{z_1} \hat{U}^{\dagger\eta}_{z_4}\}{\rm tr}\{\hat{U}^\eta_{z_4}\hat{U}^{\dagger\eta}_{z_3}\} 
-N_c {\rm tr}\{\hat{U}^\eta_{z_1}\hat{U}^{\dagger\eta}_{z_3}\}]{\rm tr}\{\hat{U}^\eta_{z_3}\hat{U}^{\dagger\eta}_{z_2}\}
 \ln {az_{13}^2\over z_{14}^2z_{34}^2 }
\nonumber\\
&&\hspace{-5mm}
+~{z_{23}^2\over z_{34}^2z_{24}^2}{\rm tr}\{\hat{U}^\eta_{z_1}\hat{U}^{\dagger\eta}_{z_3}\}
[{\rm tr}\{\hat{U}^\eta_{z_3}\hat{U}^{\dagger\eta}_{z_4}\}{\rm tr}\{\hat{U}^\eta_{z_4}\hat{U}^{\dagger\eta}_{z_2}\} 
-N_c {\rm tr}\{\hat{U}^\eta_{z_3}\hat{U}^{\dagger\eta}_{z_2}\}] \ln {az_{23}^2\over z_{24}^2z_{34}^2}
\nonumber\\
&&\hspace{75mm}
-~{ z_{12}^2N_c\over z_{14}^2z_{24}^2}
[ {\rm tr}\{\hat{U}^\eta_{z_1}\hat{U}^{\dagger\eta}_{z_4}\} 
{\rm tr}\{\hat{U}^\eta_{z_4}\hat{U}^{\dagger\eta}_{z_2}\}
 -N_c{\rm tr}\{\hat{U}^\eta_{z_1}\hat{U}^{\dagger\eta}_{z_2}\}] \ln {a z_{12}^2\over z_{14}^2z_{24}^2}
 \nonumber\\
&&\hspace{-5mm}
+~{1\over 2}\Big[
{z_{13}^2\over z_{14}^2z_{34}^2}\ln{az_{13}^2\over z_{14}^2z_{34}^2}
+{z_{23}^2\over z_{24}^2z_{34}^2}\ln{az_{23}^2\over z_{34}^2z_{24}^2}
-{z_{12}^2\over z_{14}^2z_{24}^2}\ln{az_{12}^2\over z_{14}^2z_{24}^2}\Big]
\big[{\rm tr}\{\hat{U}^\eta_{z_1} \hat{U}^{\dagger\eta}_{z_3} 
\hat{U}^\eta_{z_4}\hat{U}^{\dagger\eta}_{z_2}\hat{U}^\eta_{z_3}\hat{U}^{\dagger\eta}_{z_4}\} 
+(z_3\leftrightarrow z_4)-2{\rm tr}\{\hat{U}^\eta_{z_1} \hat{U}^{\dagger\eta}_{z_2}\}\big]\Big\}
\nonumber
 \end{eqnarray}
Repeating the steps which lead us to the Eq. (\ref{nlobksymconf1}) in Sect. \ref{sec:nlobksymconf} we obtain the conformal evolution equation in the fundamental representation 
%OK
\begin{eqnarray}
&&\hspace{-2mm}
{d\over d\eta}[{\rm tr}\{\hat{U}^\eta_{z_1} \hat{U}^{\dagger\eta}_{z_2}\}]^{\rm conf}~
=~{\alpha_s\over 2\pi^2}
\!\int\!d^2z_3~
{z_{12}^2\over z_{13}^2 z_{23}^2}\Big[1-
{\alpha_sN_c\over 4\pi}{\pi^2\over 3}\Big]
~[{\rm tr}\{\hat{U}^\eta_{z_1} \hat{U}^{\dagger\eta}_{z_3}\}{\rm tr}\{\hat{U}^\eta_{z_3} \hat{U}^{\dagger\eta}_{z_2}\}
-N_c{\rm tr}\{\hat{U}^\eta_{z_1} \hat{U}^{\dagger\eta}_{z_2}\}]^{\rm conf}
\label{nlobksymconfund}\\
&&\hspace{-2mm} 
+~{\alpha_s^2\over 16\pi^4}
\int \!d^2 z_3d^2 z_4~{z_{12}^2\over z_{13}^2z_{34}^2z_{24}^2}\Big\{2\ln{z_{12}^2z_{34}^2\over z_{14}^2z_{23}^2} +
\Big[1+{z_{12}^2z_{34}^2\over z_{13}^2z_{24}^2-z_{14}^2z_{23}^2}
\Big]
\ln{z_{13}^2z_{24}^2\over z_{14}^2z_{23}^2}
\nonumber\\ 
&&\hspace{-2mm}
\times~[2{\rm tr}\{\hat{U}^\eta_{z_1}\hat{U}^{\dagger\eta}_{z_3}\}{\rm tr}\{\hat{U}^\eta_{z_3}\hat{U}^{\dagger\eta}_{z_4}\}{\rm tr}\{\hat{U}^\eta_{z_4}\hat{U}^{\dagger\eta}_{z_2}\}
-{\rm tr}\{\hat{U}^\eta_{z_1}\hat{U}^{\dagger\eta}_{z_3} \hat{U}^\eta_{z_4}U_{z_2}^{\dagger\eta}\hat{U}^\eta_{z_3}\hat{U}^{\dagger\eta}_{z_4}\}
-{\rm tr}\{\hat{U}^\eta_{z_1}\hat{U}^{\dagger\eta}_{z_4} \hat{U}^\eta_{z_3}
U_{z_2}^{\dagger\eta}\hat{U}^\eta_{z_4}\hat{U}^{\dagger\eta}_{z_3}\}-(z_4\rightarrow z_3)]
\nonumber
\end{eqnarray}
Note that it can be obtained from the equation in the adjoint representation (\ref{nlobksymconf1}) by same 
replacement $T^a\rightarrow t^a$ , Tr $\rightarrow$ tr.
%++++++++++++++++++++++++++++++++++++++++++++++
\subsection{In QCD}

It is instructive also to present the evolution equation for composite operator (\ref{confodipolefund}) in QCD. 
The resulting equation will not be M\"obius invariant because of the running coupling constant so composite 
operators (\ref{confodipolefund}) and (\ref{conf4fund}) are not strictly speaking conformal. We will, however,
keep the notation $[...]^{\rm conf}$ as a reminder that these operators were conformal in ${\cal N}=4$ SYM.

To get the evolution equation for ``conformal'' composite operators (\ref{confodipolefund}) in QCD we subtract the scalar (\ref{scalarfund}) and gluino (\ref{gluinofund}) contributions from  Eq. (\ref{nlobksymconfund}) and add the quark contribution calculated in Refs. \cite{prd75,kw1}. We obtain
\begin{eqnarray}
&&\hspace{-5mm}
{d\over d\eta}\big[{\rm tr}\{ \hat{U}^\eta_{z_1} \hat{U}^{\dagger\eta}_{z_2}\}\big]^{\rm conf}~
=~{\alpha_s\over 2\pi^2}
\!\int\!d^2z_3~
{z_{12}^2\over z_{13}^2z_{23}^2}\Big[1+
\nonumber\\ 
&&\hspace{-2mm}
+~{\alpha_s\over 4\pi}\Big[b\ln z_{12}^2\mu^2
-b{z_{13}^2-z_{23}^2\over  z_{12}^2}\ln{z_{13}^2\over z_{23}^2}+
\big({67 \over 9}-{\pi^2\over 3}\big)N_c-{10\over 9}n_f\Big]
\big[{\rm tr}\{\hat{U}^\eta_{z_1}\hat{U}^{\dagger\eta}_{z_3}\}{\rm tr}\{\hat{U}^\eta_{z_3}\hat{U}^{\dagger\eta}_{z_2}\}
-N_c {\rm tr}\{\hat{U}^\eta_{z_1}\hat{U}^{\dagger\eta}_{z_2}\}\big]^{\rm conf}
\nonumber\\
&&\hspace{-2mm} 
+~{\alpha_s^2\over 16\pi^4}
\int \!{d^2 z_3d^2 z_4\over z_{34}^4}\Bigg[ \Big\{\Big(
-2+2{z_{12}^2z_{34}^2\over z_{13}^2z_{24}^2}\ln{z_{12}^2z_{34}^2\over z_{14}^2z_{23}^2}
+
\Big[{z_{12}^2z_{34}^2\over z_{13}^2z_{24}^2}
\big(1+{z_{12}^2z_{34}^2\over z_{13}^2z_{24}^2-z_{14}^2z_{23}^2}\big)
+{2z_{13}^2z_{24}^2-4z_{12}^2z_{34}^2\over z_{13}^2z_{24}^2-z_{14}^2z_{23}^2}
\Big]\ln{z_{13}^2z_{24}^2\over z_{14}^2z_{23}^2}\Big)
\nonumber\\ 
&&\hspace{-2mm}
+~\big(z_3\leftrightarrow z_4\big)\Big\}~\big[\big({\rm tr}\{\hat{U}^\eta_{z_1}\hat{U}^{\dagger\eta}_{z_3}\}{\rm tr}\{\hat{U}^\eta_{z_3}\hat{U}^{\dagger\eta}_{z_4}\}
{\rm tr}\{\hat{U}^\eta_{z_4}\hat{U}^{\dagger\eta}_{z_2}\}
-{\rm tr}\{\hat{U}^\eta_{z_1}\hat{U}^{\dagger\eta}_{z_3}\hat{U}^\eta_{z_4}
\hat{U}^{\dagger\eta}_{z_2}\hat{U}^\eta_{z_3}\hat{U}^{\dagger\eta}_{z_4}\}
\big)
-(z_4\rightarrow z_3)\big]
\nonumber\\
&&\hspace{-2mm} 
+~{z_{12}^2z_{34}^2\over z_{13}^2z_{24}^2}
\Big\{2\ln{z_{12}^2z_{34}^2\over z_{14}^2z_{23}^2}
+\Big[1+{z_{12}^2z_{34}^2\over z_{13}^2z_{24}^2-z_{14}^2z_{23}^2}\Big]\ln{z_{13}^2z_{24}^2\over z_{14}^2z_{23}^2}\Big\}
\big({\rm tr}\{\hat{U}^\eta_{z_1}\hat{U}^{\dagger\eta}_{z_3}\}{\rm tr}\{\hat{U}^\eta_{z_3}\hat{U}^{\dagger\eta}_{z_4}\}
{\rm tr}\{\hat{U}^\eta_{z_4}\hat{U}^{\dagger\eta}_{z_2}\}
-z_3\leftrightarrow z_4\big)\Bigg]
\nonumber\\
&&\hspace{-2mm} 
+~{\alpha^2_sn_f \over 2\pi^4}\!\int\!{d^2z_3 d^2z_4\over z_{34}^4}
\Big\{2-{z_{13}^2z_{24}^2+z_{23}^2z_{14}^2- z_{12}^2z_{34}^2\over z_{13}^2z_{24}^2-z_{14}^2z_{23}^2}
\ln{z_{13}^2 z_{24}^2\over z_{14}^2z_{23}^2}\Big\}
{\rm tr}\{t^a\hat{U}^\eta_{z_1}t^b\hat{U}^{\dagger\eta}_{z_2}\}
{\rm tr}\{t^a\hat{U}^\eta_{z_3}t^b(\hat{U}^{\dagger\eta}_{z_4}-\hat{U}^\eta_{z_3})\}
\nonumber\\
&&\hspace{26mm}
\label{nlobksymqcd}
\end{eqnarray}
where $b={11\over 3}N_c-{2\over 3}n_f$ and we have ${67\over 9}$ instead of ${64\over 9}$ because in QCD we use dimensional regularization rather than dimensional reduction scheme.
Following the analysis of Ref. \cite{nlobk} we will outline how the above kernel reproduces the NLO BFKL eigenvalues \cite{nlobfkl}.  

In the two-gluon approximation
the conformal dipoles (\ref{confodipolefund}) reduces to
 \begin{equation}
 \hat{\cal U}_{\rm conf}^\eta(z_1,z_2)~=~ \hat{\cal U}^\eta(z_1,z_2)-{\alpha_sN_c\over 4\pi^2}\!\int\! d^2z_3{z_{12}^2\over z_{13}^2 z_{23}^2}\ln {az_{12}^2\over z_{13}^2 z_{23}^2}
[\hat{\cal U}^\eta(z_1,z_3)+\hat{\cal U}^\eta(z_2,z_3)-\hat{\cal U}^\eta(z_1,z_2)]
\label{confodipolin}
 \end{equation}
Using Eq. (\ref{tracelin1}) it is easy to demonstrate that (cf. Eq. (\ref{fla70}))
\begin{equation}
\hspace{-2mm}  
[{\rm tr}\{\hat{U}^\eta_{z_1}\hat{U}^{\dagger\eta}_{z_3}\}{\rm tr}\{\hat{U}^\eta_{z_3}\hat{U}^{\dagger\eta}_{z_2}\}-N_c{\rm tr}\{\hat{U}^\eta_{z_1}\hat{U}^{\dagger\eta}_{z_2}\}]^{\rm conf}~
=~-N_c\big[\hat{\cal U}_{\rm conf}^\eta(z_1,z_3)+\hat{\cal U}_{\rm conf}^\eta(z_2,z_3)-\hat{\cal U}_{\rm conf}^\eta(z_1,z_1)\big]
\label{confoper4lin}
\end{equation}
and therefore the evolution equation (\ref{nlobksymqcd}) turns into
\begin{eqnarray}
&&\hspace{-5mm}
{d\over d\eta}\hat{\cal U}^\eta_{\rm conf}(z_1,z_2)~
=~{\alpha_sN_c\over 2\pi^2}
\!\int\!d^2z_3~
{z_{12}^2\over z_{13}^2z_{23}^2}\Big[1
\nonumber\\ 
&&\hspace{-2mm}
+~{\alpha_s\over 4\pi}\Big[b\ln z_{12}^2\mu^2
-b{z_{13}^2-z_{23}^2\over  z_{12}^2}\ln{z_{13}^2\over z_{23}^2}+
\big({67 \over 9}-{\pi^2\over 3}\big)N_c-{10\over 9}n_f\Big]
\big[\hat{\cal U}^\eta_{\rm conf}(z_1,z_3)+\hat{\cal U}^\eta_{\rm conf}(z_2,z_3)-\hat{\cal U}^\eta_{\rm conf}(z_1,z_2)\big]
\nonumber\\
&&\hspace{-2mm} 
+~{\alpha_s^2N_c^2\over 8\pi^4}
\int \!{d^2 z_3d^2 z_4\over z_{34}^2} \Big\{
2{z_{12}^2z_{34}^2\over z_{13}^2z_{24}^2}\ln{z_{12}^2z_{34}^2\over z_{14}^2z_{23}^2}
+
{z_{12}^2z_{34}^2\over z_{13}^2z_{24}^2}
\Big(1+{z_{12}^2z_{34}^2\over z_{13}^2z_{24}^2-z_{14}^2z_{23}^2}\Big)\ln{z_{13}^2z_{24}^2\over z_{14}^2z_{23}^2}
-{3z_{12}^2z_{34}^2\over z_{13}^2z_{24}^2-z_{14}^2z_{23}^2}
\ln{z_{13}^2z_{24}^2\over z_{14}^2z_{23}^2}
\nonumber\\ 
&&\hspace{-2mm}
+~\big(1+{n_f\over N_c^3}\big)\Big({z_{13}^2z_{24}^2+z_{14}^2z_{23}^2-z_{12}^2z_{34}^2\over z_{13}^2z_{24}^2-z_{14}^2z_{23}^2}
\ln{z_{13}^2z_{24}^2\over z_{14}^2z_{23}^2}-2\Big)
\Big\}
\hat{\cal U}^\eta_{\rm conf}(z_3,z_4)+{3\alpha_s^2N_c^2\over 2\pi^2}\zeta(3)\hat{\cal U}^\eta(z_1,z_2)
\label{nlolin}
\end{eqnarray}
where we used formula
\begin{eqnarray}
&&\hspace{-5mm}
\int \!d^2 z_4~\Big\{{z_{12}^2\over z_{13}^2z_{24}^2z_{34}^2}
\Big(2\ln{z_{12}^2z_{34}^2\over z_{14}^2z_{23}^2}
+\Big[1+{z_{12}^2z_{34}^2\over z_{13}^2z_{24}^2-z_{14}^2z_{23}^2}\Big]\ln{z_{13}^2z_{24}^2\over z_{14}^2z_{23}^2}\Big)
-z_3\leftrightarrow z_4\Big\}
~=~12\pi\zeta(3)[\delta(z_{23})-\delta(z_{13})]
\label{formula85}
\end{eqnarray}
following from integrals (\ref{nashel}) and (\ref{integral2}) from Appendix E.

For the case of forward scattering $\langle\hat{\cal U}(x,y)\rangle={\cal U}(x-y)$ and the linearized 
equation (\ref{nlolin}) can be reduced to an integral equation with respect to one variable $z\equiv z_{12}$. Using
integrals (104)-(106) from Ref. \cite{nlobk} and the integral
\begin{eqnarray}
&&\hspace{-5mm}
\int\! d\tilde{z}{1\over \tilde{z}^2(z-z'-\tilde{z})^2}\ln{z^2{z'}^2\over (z-\tilde{z}^2)(z'-\tilde{z}^2)}~=~-{\pi\over (z-z')^2}\ln^2{z^2\over {z'}^2}
\nonumber
\end{eqnarray}
we obtain
\begin{eqnarray}
&&\hspace{-2mm}
{d\over d\eta}\hat{\cal U}(z)~
~
=~{\alpha_sN_c\over 2\pi^2}
\!\int\!d^2z~
{z^2\over (z-z')^2 {z'}^2}\Big\{1+{\alpha_s\over 4\pi}\Big[b\ln z^2\mu^2
-b{(z-z')^2-{z'}^2\over z^2}\ln{(z-z')^2\over {z'}^2}+
({67\over 9}-{\pi^2\over 3})N_c-{10\over 9}n_f\Big]
\nonumber\\
&&\hspace{62mm} 
\times~[\langle\hat{\cal U}(z-z')\rangle+\langle\hat{\cal U}(z')\rangle-\langle\hat{\cal U}(z)\rangle]
\nonumber\\
&&\hspace{-2mm} 
+~{\alpha_s^2N_c^2\over 4\pi^3}\!\int\! d^2z'~ {z^2\over {z'}^2}
\Big[-{1\over (z-z')^2}\ln^2{z^2\over {z'}^2}+F(z,z')+\Phi(z,z')\Big]~{\cal U}(z') +3{\alpha_s^2N_c^2\over 2\pi^2}\zeta(3){\cal U}(z)
\label{nlobfkernel}
\end{eqnarray}
where 
\begin{eqnarray}
&&\hspace{-2mm}
F(z,z')~=~\Big(1+{n_f\over N_c^3}\Big){3(z,z')^2-2z^2{z'}^2\over16z^2{z'}^2}
\Big({2\over z^2}+{2\over {z'}^2}+{z^2-{z'}^2\over z^2{z'}^2}\ln{z^2\over {z'}^2}\Big)
\nonumber\\
&&\hspace{-1mm}
-~\Big[3+\Big(1+{n_f\over N_c^3}\Big)\Big(1-{(z^2+{z'}^2)^2\over 8z^2{z'}^2}
+{3z^4+3{z'}^4-2z^2{z'}^2\over 16z^4{z'}^4}(z,z')^2\Big)\Big]\!\int_0^\infty\!dt{1\over z^2+t^2{z'}^2}\ln{1+t\over |1-t|} 
\nonumber
\end{eqnarray}
and 
\begin{eqnarray}
&&\hspace{-2mm}
\Phi(z,z')~=~{(z^2-{z'}^2)\over (z-z')^2(z+z')^2}
\Big[\ln{z^2\over {z'}^2}\ln{z^2{z'}^2(z-z')^4\over (z^2+{z'}^2)^4}
+2{\rm Li_2}\Big(-{{z'}^2\over z^2}\Big)-2{\rm Li_2}\Big(-{z^2\over {z'}^2}\Big)\Big]
\nonumber\\
&&\hspace{-1mm}
-~\Big(1-{(z^2-{z'}^2)^2\over (z-z')^2(z+z')^2}\Big)\Big[\!\int_0^1-\int_1^\infty\Big]
{du\over (z-z'u)^2}\ln{u^2{z'}^2\over z^2}
\nonumber
\end{eqnarray}

The function $-{1\over (q-q')^2}\ln^2{q^2\over {q'}^2}+F(q,q')+\Phi(q,q')$ enters the NLO BFKL equation in the momentum space \cite{nlobfkl}
and since the eigenfunctions of the forward BFKL equation are powers both in the coordinate and momentum space, 
it is clear that the corresponding eigenvalues coincide. As to the first term in the r.h.s. of Eq. (\ref{nlobfkernel}), one can 
demonstrate using the analysis carried out in Ref. \cite{nlobk} that this term also agrees with the eigenvalues of Ref. \cite{nlobfkl}. 
However, this analysis would lead us away from the main topic of this paper so we defer it until our next publicaton.
It should be also mentioned that the statement in our previous paper \cite{nlobk} that our equation (\ref{gluonfund}) disagrees with 
NLO BFKL was due to erroneous calculation of the integral (\ref{nashel}) which was assumed to be zero. After taking into account the 
$\delta$-function contributions in the r.h.s. of eq. (\ref{nashel}) the disagreement disappears.

 %%%%%%%%%%%%%%%%%%%%%%%%%%%%%%%%%%%%%%%%%%%%%%%%
\section{Conclusion}

The amplitudes in $N=4$ SYM are conformally invariant and therefore the Regge limit (\ref{reggelimit}) of these amplitudes must
be invariant with respect to M\"obius transformations of the transverse plane. If we want to use the operator expansion to find this amplitude,
it is better to expand in operators which are  M\"obius invariant. As we demonstrate in Appendix A, the light-like Wilson lines
are formally invariant. However,  they are divergent in the longitudinal direction, and at present the regularization of this
rapidity divergence which respects the conformal invariance is not known. We manage to circumvent this problem by using the non-invariant
``rigid cutoff'' (\ref{cutoff}) and restoring the conformal invariance order by order in perturbation theory by subtracting 
the proper counterterms (made again of Wilson lines). The resulting NLO evolution equation for ``composite conformal dipoles'' 
is M\"obius invariant and agrees with forward NLO BFKL calculations in ${\cal N}=4$ SYM.

 Let us comment on the non-conformal result of the calculation of NLO BFKL kernel in ${\cal N}=4$ SYM 
carried out in Ref. \cite{fadinlo}. 
We think that the difference between our kernel and that of Ref. \cite{fadinlo} is due to different cutoffs for 
longitudinal integrations.
The authors of Ref. \cite{fadinlo} propose that the transformation of their kernel of the type 
$\hat{K}_{\rm NLO}\rightarrow \hat{K}_{\rm NLO}-\alpha_s[\hat{K}_{\rm LO},\hat{\calo}]$ with some suitable operator $\hat{\calo}$ 
will restore conformal invariance. This is exactly what happens in our case of the kernel (\ref{nlobksym}) with 
the ``rigid cutoff'' (\ref{cutoff}) of the rapidity divergence. Let us discuss the transformation proposed in  Ref. \cite{fadinlo} in our language.
If we define ${\tilde{\hat{\cal V}}}(z)$ as
\begin{eqnarray}
&&\hspace{-2mm}
{\hat{\tilde{\cal V}}^\eta}(z)~
=~\hat{\cal V}^\eta(z)+\alpha_s\!\int\! d^2z' f(z,z')\hat{\cal V}^\eta(z')
\label{vtilda}
\end{eqnarray}
we see that the (linear) evolution equations for the operators $\hat{\cal V}^\eta$ and $\hat{\tilde{\cal V}}^\eta$ are related by
\begin{equation}
\hat{\tilde{K}}_{\rm NLO}~=~\hat{K}_{\rm NLO}-\alpha_s[\hat{K}_{\rm LO}, \hat{f}]
\label{kommu}
\end{equation}
Indeed, differentiating the operator (\ref{vtilda}) with respect to $\eta$ we obtain
\begin{eqnarray}
&&\hspace{-2mm}
{d\over d\eta}{\hat{\tilde{\cal V}}^\eta}(z)~=~{d\over d\eta}\hat{\cal V}^\eta(z)+\alpha_s\!\int\! d^2z'\! f(z,z'){d\over d\eta}\hat{\cal V}^\eta(z')
\nonumber\\
&&\hspace{-2mm}
=~\!\int\! dz' K_{\rm LO}(z,z')\hat{\cal V}^\eta(z')+\alpha_s \!\int\! dz' K_{\rm NLO}(z,z')\hat{\cal V}^\eta(z')+
\alpha_s \!\int\! dz' dz'' f(z,z')K_{\rm LO}(z',z'')\hat{\cal V}^\eta(z')
\nonumber\\
&&\hspace{-2mm}
=~\!\int\! dz' K_{\rm LO}(z,z')\hat{\tilde{{\cal V}}}^\eta(z)
\nonumber\\
&&\hspace{-2mm}
+~\alpha_s \!\int\! dz' K_{\rm NLO}(z,z')\hat{\tilde{{\cal V}}}^\eta(z')+
\alpha_s \!\int\! dz' dz''[f(z,z')K_{\rm LO}(z',z'')-K_{\rm LO}(z,z')F(z',z'')]{\hat{\tilde{\cal V}}^\eta}(z)+O(\alpha_s^3)
\label{kommut}
\end{eqnarray}
which corresponds to Eq. (\ref{kommu}). Our transition between ${\hat{\tilde{\cal V}}^\eta}$ and ${\hat{\tilde{\cal V}}_{\rm conf}^\eta}$ 
is of the type (\ref{vtilda}) so it is not surprising that the kernels (\ref{nlobksym}) and (\ref{nlobksymconf}) are different. 
We think that one can recover the conformal kernel (\ref{nlobksymconf}) from the kernel of Ref. \cite{fadinlo} as long as one finds
the appropriate $f(z,z')$.
It should be also mentioned that the transformation (\ref{vtilda})
with both $\hat{\cal V}^\eta$ and $f(z,z')$ conformally invariant does not change $K_{\rm NLO}$ as can be easily seen from
Eq. (\ref{kommut}). Thus, the form of the conformal composite dipole is not unique (our Eq. (\ref{confodipole}) is one particular choice)
but the conformal kernel $\hat{K}_{\rm NLO}$ is unique.

In conclusion let us discuss possible generalizations of our method. The operator expansion of the type (\ref{opeq}) is relevant for 
processes like deep inelastic scattering where the strong gluon fields come from the nucleon (or nucleus) target and 
the spectator (virtual photon) is a weak source of the gluon field. For the processes like heavy-ion collisions, the projectile-target symmetric
language of 2+1 - dimensional effective action seems more adequate (here 2 is the number of transverse dimensions and 1 stands for rapidity). 
There are many attempts in the literature to find comprehensive effective action \cite{effaction}
but the answer for the ultimate high-energy effective action eludes us so far.  It is possible that considering this problem in ${\cal N}=4$ SYM where we have the additional requirement of conformal (M\"obius) symmetry to restrict the effective action will help us to find the correct effective action at high energies.

\section*{Acknowledgments}
The authors are grateful to L.N. Lipatov and J. Penedones for valuable discussions. 
This work was supported by contract
 DE-AC05-06OR23177 under which the Jefferson Science Associates, LLC operate the Thomas Jefferson National Accelerator Facility.

%%%%%%%%%%%%%%%%%%%%%%%%%%%%%%%%%%%%%%%%%%%%%%%%%%%%%%%
\section{Appendix}
%+++++++++++++++++++++++++++++++++++++++++++++++
\subsection{Conformal properties of the light-like Wilson lines \label{sect:appa}}
%+++++++++++++++++++++++++++++++++++++++++++++++
In this Section we demonstrate that the light-like Wilson lines  are invariant under the conformal (M\"obius) group SL(2,C).
It is easy to demonstrate that the set of transverse-space operators 
\begin{eqnarray}
&&\hspace{-2mm} 
\hat{S}_-\equiv{i\over 2}(K^1+iK^2), ~~~\hat{S}_0\equiv{i\over 2}(D+iM^{12}),
 ~~~\hat{S}_+\equiv {i\over 2}(P^1-iP^2)
 \nonumber\\ 
&&\hspace{-2mm}
\barhats_-\equiv {i\over 2}(K^1-iK^2), ~~~\barhats_0\equiv {i\over 2}(D-iM^{12}),
 ~~~\barhats_+\equiv {i\over 2}(P^1+iP^2)
\label{Ss}
\end{eqnarray}
form an SL(2,C) algebra:
\begin{eqnarray}
&&\hspace{-2mm} 
[\hat{S}_0,\hat{S}_\pm]=\pm \hat{S}_\pm,~~~[\hat{S}_+,\hat{S}_-]=2\hat{S}_0,~~~~~
[\barhats_0,\barhats_\pm]=\pm\barhats_\pm,~~~ [\barhats_+,\barhats_-]=2\barhats_0
\label{sl2c}
\end{eqnarray}
Here we use standard textbook definitions of the momentum operator $\hat{P}$, angular mpmentum operator $\hat{M}$,
dilatation operator $\hat{D}$, and special conformal generator $\hat{K}$.
Using the conventional commutators of these generators with the gluon field
\begin{eqnarray}
&&\hspace{-2mm} 
i[\hat{P}^\mu, \hat{A}^\alpha] =\partial^\mu \hat{A}^\alpha,~~~i[\hat{D}, \hat{A}^\alpha] =(x_\mu\partial^\mu +1)\hat{A}^\alpha,
\nonumber\\ 
&&\hspace{-2mm}
i[\hat{M}^{\mu\nu},\hat{A}^\alpha]=(x^\mu\partial^\nu-x^\nu\partial^\mu)\hat{A}^\alpha
-(g^{\nu\alpha}\hat{A}^\mu-g^{\mu\alpha}\hat{A}^\nu)
\nonumber\\
&&\hspace{-2mm}
i[K^\mu,A^\alpha]~=~(2x^\mu x_\nu\partial^\nu-x^2\partial^\mu+2x^\mu )A^\alpha
-2x_\nu(g^{\nu\alpha} A^\mu-g^{\mu\alpha} A^\nu)
\end{eqnarray}
 one can easily obtain the action of these operators on the light-like Wilson lines. In complex notations
\begin{eqnarray}
&&\hspace{-2mm} 
z\equiv z^1+iz^2,~~~\barz\equiv z^1-iz^2, ~~~~
{\partial\over\partial z}= {1\over 2}\Big({\partial\over\partial z^1}-i{\partial\over\partial z^2}\Big), 
~~~~
{\partial\over\partial \barz}= {1\over 2}\Big({\partial\over\partial z^1}+i{\partial\over\partial z^2}\Big)
\label{znotations}
\end{eqnarray}
these commutators take the form
\begin{eqnarray}
&&\hspace{-2mm} 
[\hat{S}_-,\hat{U}(z,\barz)]=z^2\partial_z\hat{U}(z,\barz),
~~~[\hat{S}_0,\hat{U}(z,\barz)]=z\partial_z\hat{U}(z,\barz),
~~~[\hat{S}_+,\hat{U}(z,\barz)]=-\partial_z\hat{U}(z,\barz)
 \nonumber\\ 
&&\hspace{-2mm}
[\barhats_-, \hat{U}(z,\barz)]=\barz^2\partial_\barz\hat{U}(z,\barz),~~~
[\barhats_0, \hat{U}(z,\barz)]=\barz\partial_\barz\hat{U}(z,\barz),~~~
[\barhats_+, \hat{U}(z,\barz)]=-\partial_\barz\hat{U}(z,\barz)
\nonumber
\end{eqnarray}
These equations mean that the operators $U(z,\barz)$ lie in the standard representation of  conformal group 
$SL(2,C)$ with the conformal spin 0.

%%%%%%%%%%%%%%%%%%%%%%%%%%%%%%%%%%%%%%%%%%%%%%%%%%%%%%%%%
\subsection{NLO impact factor \label{sect:appenb}}

As we demonstrated in the Appendix A the light-like Wilson lines $U(x_\perp)$ are formally M\"obius invariant and this is why 
the leading-order BK equation is conformal.  However, because our cutoff of the rapidity divergence is not invariant, the NLO evolution kernel
(\ref{nlobksym})
has the non-invariant double-log term. To illustrate the non-invariance of the dipole with the cutoff (\ref{cutoff}) let us consider the high-energy 
operator expansion \cite{npb96}
of two BPS-protected currents ${\calo}\equiv{4\pi^2\sqrt{2}\over \sqrt{N_c^2-1}}
{\rm Tr\{Z^2\}}$ ($Z={1\over\sqrt{2}}(\phi_1+i\phi_2)$) in color
dipoles.  The Regge limit of the amplitude 
\begin{equation}
\hspace{-1mm}
A(x,y;x',y')~=~(x-y)^4(x'-y')^4\langle\calo(x)\calo(y)\calo(x')\calo(y')\rangle
\label{amplituda}
\end{equation}
can be achieved by the rescaling \cite{npb96}
\begin{eqnarray}
&&\hspace{-1mm}
x~\rightarrow~\lambda x_\ast p_1+{1\over\lambda}x_\bu p_2+x_\perp,~~~~~~~ y~\rightarrow~\lambda y_\ast +{1\over\lambda}y_\bu p_2+y_\perp
\label{reggelimit}
\end{eqnarray}
with $\lambda\rightarrow\infty$. In this regime, the T-product of the currents $\calo(x)$ and $\calo(y)$ 
can be expanded in color dipoles as follows:
\begin{eqnarray}
&&\hspace{-1mm}
 T\{\halo(x)\halo(y)\}~=~\int\! d^2z_1d^2z_2~I^{\rm LO}(z_1,z_2)
 {\rm Tr}\{\hat{U}^\eta_{z_1}\hat{U}^{\dagger\eta}_{z_2}\}
 \nonumber\\
&&\hspace{-1mm}
+\int\! d^2z_1d^2z_2d^2z_3~I^{\rm NLO}(z_1,z_2,z_3)
[ {\rm Tr}\{T^n\hat{U}^\eta_{z_1}\hat{U}^{\dagger\eta}_{z_3}T^n\hat{U}^\eta_{z_3}\hat{U}^{\dagger\eta}_{z_2}\}
 -N_c {\rm Tr}\{\hat{U}^\eta_{z_1}\hat{U}^{\dagger\eta}_{z_2}\}]
 \label{opeq}
 \end{eqnarray}
(structure of the NLO contribution is clear from the topology of diagrams in the shock-wave background, see Fig. \ref{fig:nloif} below).

Let us  calculate the impact factor taking $x_\bu=y_\bu=0$ for simplicity. The leading-order impact factor is proportional to the product of two propagators 
(\ref{scalar-sw1}) 
\begin{eqnarray}
&&
\langle T\{\halo(x)\halo(y)\}\rangle^{\rm LO}_A~
\label{loif}\\
&&=~{16\pi^4\over (N_c^2-1)}
\!\int\! d^4 z_1d^4z_2~\delta(z_{1\ast})\delta(z_{2\ast})
~U^{ab}_{z_1}U^{ab}_{z_2}
\Big({1\over4\pi^2[(x-z_1)^2-i\epsilon]}(2i)\partial_\ast^{(z_1)}
{1\over 4\pi^2[(y-z_1)^2-i\epsilon]}\Big)
\nonumber\\
&&
\times~
\Big({1\over 4\pi^2[(x-z_2)^2-i\epsilon]}(2i)\partial_\ast^{(z_2)}
{1\over4\pi^2[(y-z_2)^2-i\epsilon]}\Big)~
=~{1\over \pi^2(N_c^2-1)}\!\int\! d^2 z_{1\perp}d^2 z_{2\perp}
{(x_\ast y_\ast)^{-2}\over \calz_1^2\calz_2^2}~{\rm Tr}\{U_{z_1}U^\dagger_{z_2}\}
\nonumber
\end{eqnarray}
where $\calz_i\equiv {(x-z_i)_\perp^2\over x_\ast}- {(y-z_i)_\perp^2\over y_\ast}$ and the color trace is taken in the adjoint representation. This expressions coincides
with the result of Ref. \cite{penecostalba}. 
 It is easy to see that under the inversion $x_\perp\rightarrow x_\perp/x_\perp^2, x_\ast\rightarrow x_\ast/x_\perp^2$
\begin{equation}
x_\perp\rightarrow x_\perp/x_\perp^2, x_\ast\rightarrow x_\ast/x_\perp^2,~~~~~~y_\perp\rightarrow y_\perp/x_\perp^2, y_\ast\rightarrow y_\ast/y_\perp^2,
~~~~~~z_{i\perp}\rightarrow z_{i\perp}/z_{i\perp}^2
\label{inversion}
\end{equation}
$\calz_i$ is transformed as $\calz_i\rightarrow z_i^{-4}\calz_i$ so the leading-order impact factor is invariant.

The NLO impact factor for two $Z^2$ currents is given by the two diagrams shown in Fig. \ref{fig:nloif}.
%%%%%%%%%%%%%FIGA%%%%%%%%%%%%%%%%%%%%%%%
\begin{figure}[h]
\includegraphics[width=0.7\textwidth]{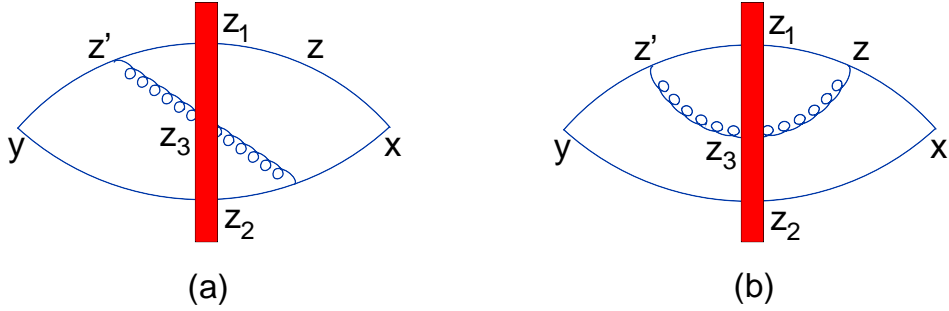}
\caption{Diagrams for the NLO impact factor. \label{fig:nloif}}
\end{figure}
%%%%%%%%%%%%%%%%%%%%%%%%%%%%%%%%%%%%%%%%
To calculate them we will use the following representation of scalar and gluon propagators in the shock-wave background (at $x_\ast>0>y_\ast$ )
\begin{eqnarray}
&&\hspace{-4mm}
\langle \hat{\Phi}_I(x)\hat{\Phi}_J(y)\rangle_A
=~2i\delta^{IJ}\!\int d^4 z\delta(z_\ast) {1\over 4\pi^2[(x-z)^2-i\epsilon]}~U^{ab}_{z_\perp}~
{\partial^{(z)}_\ast}{1\over 4\pi^2[(z-y)^2-i\epsilon]}
\nonumber \\
&&\hspace{-2mm}
=~
-{\delta_{IJ}\over 4\pi^3}\!\int\! d^2z_\perp{x_\ast y_\ast U^{ab}_{z_3}\over 
\Big[{4\over s}(x-y)_\bu x_\ast y_\ast-(x-z)_\perp^2y_\ast+(y-z)_\perp^2x_\ast+i\epsilon\Big]^2}
\label{scalar-sw1}
\end{eqnarray}
and
\begin{eqnarray}
&&\hspace{-4mm}
\langle \hat{A}^a_\mu(x)\hat{A}^b_\nu(y)\rangle_A
~=~-{i\over 2}\int d^4 z~\delta(z_\ast)~
{x_\ast g^\perp_{\mu\xi}-p_{2\mu}(x-z)^\perp_\xi \over \pi^2[(x-z)^2-i\epsilon]^2}\;U^{ab}_{z_\perp}
{1\over\partial_\ast^{(z)}}~{y_*\delta^{\perp\xi}_\nu
 - p_{2\nu}(y-z)_\perp^\xi\over \pi^2[(z-y)^2-i\epsilon]^2}
\label{gluon-sw1}
\end{eqnarray}
where ${1\over \partial_\ast}$ can be either ${1\over \partial_\ast+i\epsilon }$ or 
${1\over \partial_\ast-i\epsilon }$ which leads to the same result after subtraction of the leading-order contribution (see below).

To calculate the next-to-leading impact factor we need the three-point scalar-scalar-gluon vertex Green function (vertex with tails):
\begin{equation}
\hspace{-1mm}
\int\! \!d^4z~\theta(z_\ast)\Big[{1\over 4\pi^2[(x-z)^2-i\epsilon]}
\stackrel{\leftrightarrow}{\partial_\mu^z}{1\over 4\pi^2[(z-z_1)^2-i\epsilon]}\Big]
{z_\ast \delta_\mu^{\perp \xi} -p_{2\mu}(z-z_3)_\perp^\xi\over 2\pi^2[(z-z_3)^2-i\epsilon]^2}
~=~{ix_\ast z_{13}^{\perp\xi}\over 8\pi^4 z_{13}^2[(x-z_1)^2-i\epsilon][(x-z_3)^2-i\epsilon]}
\label{block}
\end{equation}
Since $z_{1\ast}=z_\ast=0$ one can easily check that $\theta(z_\ast)$ in the l.h.s. of this integral 
can be erased since the contribution from $z_\ast<0$ vanishes. 
After that, one easily gets Eq. (\ref{block}) from the conformal integral
\begin{equation}
\hspace{-2mm}
\int\! d^4z
~\Big[{1\over (x-z)^2}\stackrel{\leftrightarrow}{\partial_\mu}
{1\over(z-y)^2}\Big]{z_\nu\over z^4}-\mu\leftrightarrow\nu~=~4i\pi^2{x_\mu y_\nu-x_\nu y_\mu\over x^2y^2(x-y)^2}
\label{conformalinteg}
\end{equation}
which is easily calculated by inversion $z_\mu\rightarrow z_\mu/z^2$.

Now we are in position to calculate the contribution of the diagram in Fig. \ref{fig:nloif}a:
\begin{eqnarray}
&&\hspace{-1mm}
\langle T\{\halo(x)\halo(y)\}\rangle_A^{\rm Fig. \ref{fig:nloif}a}
\nonumber\\
&&\hspace{-1mm}
=~
g^2{32\pi^4\over N_c^2-1}\int d^4 z_1 d^4z_2 d^4 z_3~\delta(z_{3\ast})
{\rm Tr}\{T^n U_{z_1}T^{n'}U^\dagger_{z_2}\} U_{z_3}^{nn'}
\Big(2i\,\partial _{1*}{\delta(z_{1*})\over 4\pi^2[(y-z_1)^2-i\epsilon]}\Big)
\Big(2i\,\partial_{2*}{\delta(z_2*)\over 4\pi^2[(x-z_2)^2-i\epsilon]}\Big)
\nonumber\\
&&\hspace{-1mm}
\times~\int d^4z\Big({1\over 4\pi^2[(x-z)^2-i\epsilon]}
\stackrel{\leftrightarrow}{\partial_\mu^z} {\delta(z_1*)\over 4\pi^2[(z-z_1)^2-i\epsilon]}\Big)
{z_*\theta(z_*)\over 2\pi^2[(z-z_3)^2-i\epsilon]^2}\Big[g^\perp_{\mu\rho}-p_{2\mu}{(z-z_3)_\perp^\rho\over z_*}\Big]\;
{2i\over\partial_{3*}}\times
\nonumber\\
&&\hspace{-1mm}
\times\int d^4z'\Big({\delta(z_2*)\over 4\pi^2[(z_2-z')^2-i\epsilon]}
\stackrel{\leftrightarrow}{\partial_\mu^{z'}}{1\over 4\pi^2[(y-z')^2-i\epsilon]}\Big)
{z'_\ast\theta(-z'_\ast)\over 2\pi^2[(z_3-z')^2-i\epsilon]^2}\Big[g^{\perp\rho}_\nu - p_{2\nu}{(z'-z_3)_\perp^\rho\over z'_\ast}\Big]
\end{eqnarray}
Using the Eq. (\ref{block}) one can reduce this equation to 
\begin{eqnarray}
&&\hspace{-1mm}
\langle T\{\halo(x)\halo(y)\}\rangle_A^{\rm Fig. \ref{fig:nloif}a}
\nonumber\\
&&\hspace{-1mm}
=~
-g^2{32i\pi^4\over N^2_c-1}\int d^4 z_1 d^4z_2 d^4 z_3~\delta(z_{3\ast}){\rm Tr}\{T^n U_{z_1}T^{n'}U^\dagger_{z_2}\} U_{z_3}^{nn'}
{1\over 4\pi^2[(x-z_1)^2-i\epsilon]}\Big(2i\,\partial _{1*}{\delta(z_{1*})\over 4\pi^2[(y-z_1)^2-i\epsilon]}\Big)
\nonumber\\
&&\hspace{-1mm}
\Big(2i\,\partial_{2*}{\delta(z_2*)\over 4\pi^2[(x-z_2)^2-i\epsilon]}\Big)
{1\over 4\pi^2[(y-z_2)^2-i\epsilon]}
~{(z_{13},z_{23})_\perp\over z_{13}^2z_{23}^2}
{x_\ast\over 2\pi^2 [(x-z_3)^2-i\epsilon]}
{2\over\partial_{3*}}~
{y_\ast \over 2\pi^2 [(y-z_3)^2-i\epsilon]}
\end{eqnarray}
Performing integration with respect to $z_{\bu i}$ we get
\begin{equation}
\hspace{-0mm}
\langle T\{\halo(x)\halo(y)\}\rangle_A^{\rm Fig. \ref{fig:nloif}a}
~=~                     
{2\alpha_s\over (N^2_c-1)\pi^4x_\ast^2 y_\ast^2}\!\int\! {d^2z_1d^2z_2 \over\calz_1^2\calz_2^2}\!\int\! d^2z_3
{(z_{13},z_{23})_\perp\over z_{13}^2z_{23}^2}
{\rm Tr}\{T^n U_{z_1}U^\dagger_{z_3}T^nU_{z_3}U^\dagger_{z_2}\} 
\!\int_0^\infty\!{d\alpha\over\alpha}~e^{i\alpha {s\over 4}\calz_3}         
\label{ot5a}            
\end{equation}
(recall that $\calz_i={(x-z_i)_\perp^2\over x_\ast}- {(y-z_i)_\perp^2\over y_\ast}$). 
Similarly, one can demonstrate that the  diagram shown in Fig. \ref{fig:nloif}b yields
\begin{eqnarray}
&&\hspace{-1mm}
\langle T\{\halo(x)\halo(y)\}\rangle_A^{\rm Fig. \ref{fig:nloif}b}
~=~                     
{2\alpha_s\over (N^2_c-1)\pi^4x_\ast^2 y_\ast^2}\!\int\! {d^2z_1d^2z_2 \over\calz_1^2\calz_2^2}\!\int\! d^2z_3
{1\over z_{13}^2}
{\rm Tr}\{T^n U_{z_1}U^\dagger_{z_3}T^nU_{z_3}U^\dagger_{z_2}\} 
\!\int_0^\infty\!{d\alpha\over\alpha}~e^{i\alpha {s\over 4}\calz_3}         
\label{ot5b}           
\end{eqnarray}
Using the $z_1\leftrightarrow z_2$ symmetry in this equation one can write down the sum of 
Eq. (\ref{ot5a}) and Eq. (\ref{ot5b}) as
\begin{equation}
\hspace{1mm}
\langle T\{\halo(x)\halo(y)\}\rangle_A^{\rm Fig. \ref{fig:nloif}}
~=~                     
{\alpha_s\over (N^2_c-1)\pi^4x_\ast^2 y_\ast^2}\!\int\! {d^2z_1d^2z_2 \over\calz_1^2\calz_2^2}\!\int\! d^2z_3
{z_{12}^2\over z_{13}^2z_{23}^2}
{\rm Tr}\{T^n U_{z_1}U^\dagger_{z_3}T^nU_{z_3}U^\dagger_{z_2}\} 
\!\int_0^\infty\!{d\alpha\over\alpha}~e^{i\alpha {s\over 4}\calz_3}         
\label{ot5}           
\end{equation}
It is convenient to rewrite this in the following form
\begin{eqnarray}
&&\hspace{-1mm}
\langle T\{\halo(x)\halo(y)\}\rangle_A^{\rm Fig. \ref{fig:nloif}}
~=~                     
{\alpha_s\over (N^2_c-1)\pi^4x_\ast^2 y_\ast^2}\!\int\! {d^2z_1d^2z_2 \over\calz_1^2\calz_2^2}\!\int\! d^2z_3
{z_{12}^2\over z_{13}^2z_{23}^2}
{\rm Tr}\{T^n U_{z_1}U^\dagger_{z_3}T^nU_{z_3}U^\dagger_{z_2}\} 
\nonumber\\
&&\hspace{-1mm}
-~N_c{\rm Tr}\{U_{z_1}U^\dagger_{z_2}\} 
\!\int_0^\infty\!{d\alpha\over\alpha}~e^{i\alpha {s\over 4}\calz_3}         
+~{2\alpha_sN_c\over (N^2_c-1)\pi^4x_\ast^2 y_\ast^2}\!\int\! {d^2z_1d^2z_2 \over\calz_1^2\calz_2^2}
\!\int\! d^2z_3
{z_{12}^2\over z_{13}^2z_{23}^2}
{\rm Tr}\{U_{z_1}U^\dagger_{z_2}\} 
\!\int_0^\infty\!{d\alpha\over\alpha}~e^{i\alpha {s\over 4}\calz_3} 
\label{ot5fu}           
\end{eqnarray}
%
%%%%%%%%%%%%%FIGA%%%%%%%%%%%%%%%%%%%%%%%
\begin{figure}[h]
\includegraphics[width=0.7\textwidth]{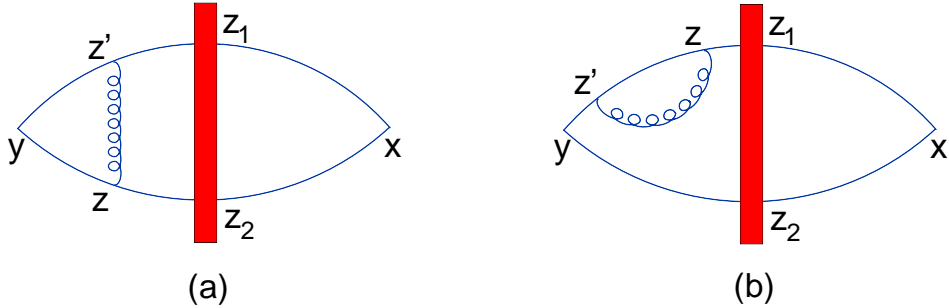}
\caption{Diagrams for the NLO impact factor without gluon-shockwave intersection. \label{fig:nloif1}}
\end{figure}
%%%%%%%%%%%%%%%%%%%%%%%%%%%%%%%%%%%%%%%%
Let us discuss now the contribution of Fig. \ref{fig:nloif1} diagrams. Since this contribution is proportional to 
${\rm Tr}\{U_{z_1}U^\dagger_{z_2}\} $ it can be restored from the comparison of Eq. (\ref{ot5fu}) 
with the pure perturbative series for the correlator $\langle T\{\halo(x)\halo(y)\}\rangle$. Indeed, if we 
switch off the shock wave the contribution of the Fig. \ref{fig:nloif} diagrams is given by the second term in 
Eq. (\ref{ot5fu}) (with $U,U^\dagger$ repaced by 1). On the other hand, perturbative series for the 
correlator $\langle T\{\halo(x)\halo(y)\}\rangle$ vanishes \cite{vanish} and therefore the contribution of the 
Fig. \ref{fig:nloif1} diagrams should be equal to the second term in 
the r.h.s. Eq. (\ref{ot5fu}) with opposite sign. 
Thus, the first term in the r.h.s. Eq. (\ref{ot5fu}) gives the total contribution to the impact factor:
\begin{eqnarray}
&&\hspace{-1mm}
\langle T\{\halo(x)\halo(y)\}\rangle_A^{\rm Fig. \ref{fig:nloif}+Fig. \ref{fig:nloif1}}
\nonumber\\
&&\hspace{-1mm}=~                     
{\alpha_s\over (N^2_c-1)\pi^4x_\ast^2 y_\ast^2}\!\int\! {d^2z_1d^2z_2 \over\calz_1^2\calz_2^2}\!\int\! d^2z_3
{z_{12}^2\over z_{13}^2z_{23}^2}
{\rm Tr}\{T^n U_{z_1}U^\dagger_{z_3}T^nU_{z_3}U^\dagger_{z_2}\} 
-N_c{\rm Tr}\{U_{z_1}U^\dagger_{z_2}\} 
\!\int_0^\infty\!{d\alpha\over\alpha}~e^{i\alpha {s\over 4}\calz_3}   
\label{ot5and6}           
\end{eqnarray}

The integral over $\alpha$ in the r.h.s. of Eq. (\ref{ot5and6}) diverges. This divergence reflects the fact that the r.h.s.  of Eq. (\ref{ot5and6}) is not exactly the NLO impact factor since we must subtract the 
matrix element of the leading-order contribution. Indeed,  the NLO impact factor is a coefficient function defined according to Eq. (\ref{opeq}).
To find the NLO impact factor, we consider the operator equation (\ref{opeq}) in 
the shock-wave background (in the leading order $\langle\hat{U}_{z_3}\rangle_A=U_{z_3}$):
\begin{eqnarray}
&&\hspace{-1mm}
 \langle T\{\halo(x)\halo(y)\}\rangle_A~-\int\! d^2z_1d^2z_2~I^{\rm LO}(x,y;z_1,z_2)
  \langle{\rm Tr}\{\hat{U}^\eta_{z_1}\hat{U}^{\dagger\eta}_{z_2}\}\rangle_A
 \nonumber\\
&&\hspace{-1mm}
=~\int\! d^2z_1d^2z_2d^2z_3~I^{\rm NLO}(x,y;z_1,z_2,z_3;\eta)
[ {\rm Tr}\{T^nU_{z_1}U^\dagger_{z_3}T^nU_{z_3}U^\dagger_{z_2}\}
 -N_c {\rm Tr}\{U_{z_1}U^\dagger_{z_2}\}]
 \label{opeq1}
 \end{eqnarray}
The NLO matrix element $ \langle T\{\halo(x)\halo(y)\}\rangle_A$ is given by Eq. (\ref{ot5and6}) 
while 
\begin{eqnarray}
&&\hspace{-1mm}
\int\! d^2z_1d^2z_2~I^{\rm LO}(x,y;z_1,z_2)
  \langle{\rm Tr}\{\hat{U}^\eta_{z_1}\hat{U}^{\dagger\eta}_{z_2}\}\rangle_A
 \nonumber\\
&&\hspace{-1mm}
=~{(x_\ast y_\ast)^{-2}\over \pi^2(N_c^2-1)}\!\int\! d^2z_1d^2z_2~
{1\over\calz_1^2\calz_2^2}{\alpha_s\over\pi^2}\!\int_0^\sigma\!{d\alpha\over\alpha}\!\int\! d^2z_3
{z_{12}^2\over z_{13}^2z_{23}^2}
[ {\rm Tr}\{T^nU_{z_1}U^\dagger_{z_3}T^nU_{z_3}U^\dagger_{z_2}\}
 -N_c {\rm Tr}\{U_{z_1}U^\dagger_{z_2}\}]
 \label{mael1}
 \end{eqnarray}
as follows from Eq. (\ref{bk4}).
The $\alpha$ integration is cut above by $\sigma=e^\eta$ in accordance with the definition of operators $\hat{U}^\eta$  (\ref{cutoff}). Subtracting  (\ref{mael1}) from  Eq. (\ref{ot5and6}) we get
\begin{eqnarray}
&&\hspace{-1mm}
I^{\rm NLO}(x,y;z_1,z_2,z_3;\eta)~=~{\alpha_s(x_\ast y_\ast)^{-2}\over\pi^4(N_c^2-1)}
{z_{13}^2\over z_{12}^2z_{23}^2\calz_1^2\calz_2^2}
\Big[\!\int_0^\infty\!{d\alpha\over\alpha}~e^{i\alpha {s\over 4}\calz_3}   
-\!\int_0^\sigma\!{d\alpha\over\alpha}\Big]
 \nonumber\\
&&\hspace{-1mm}
=~-{\alpha_s(x_\ast y_\ast)^{-2}\over\pi^4(N_c^2-1)}
{z_{13}^2\over z_{12}^2z_{23}^2\calz_1^2\calz_2^2}
\Big[\ln{\sigma s\over 4}\calz_3-{i\pi\over 2}+C\Big]
 \label{nloif}
 \end{eqnarray}

Let us rewrite the operator expansion (\ref{opeq}) in the explicit form
\begin{eqnarray}
&&\hspace{-5mm}
(x-y)^4 T\{\halo(x)\halo(y)\}~
=~{(x-y)^4\over \pi^2(N_c^2-1)}\!\int\! d^2 z_{1\perp}d^2 z_{2\perp}
{(x_\ast y_\ast)^{-2}\over \calz_1^2\calz_2^2}~{\rm Tr}\{\hat{U}^\eta_{z_1}\hat{U}^{\dagger\eta}_{z_2}\}
 \label{ope1}\\
&&\hspace{-5mm}
-~{\alpha_s(x-y)^4\over\pi^4(N_c^2-1)}\!\int\! d^2 z_{1\perp}d^2 z_{2\perp} d^2 z_3
{z_{12}^2(x_\ast y_\ast)^{-2}\over z_{13}^2z_{23}^2\calz_1^2\calz_2^2}
\Big[\ln{\sigma s\over 4}\calz_3-{i\pi\over 2}+C\Big]
[ {\rm Tr}\{T^n\hat{U}^\eta_{z_1}\hat{U}^{\dagger\eta}_{z_3}T^n\hat{U}^\eta_{z_3}\hat{U}^{\dagger\eta}_{z_2}\}
 -N_c {\rm Tr}\{\hat{U}^\eta_{z_1}\hat{U}^{\dagger\eta}_{z_2}\}]
\nonumber
 \end{eqnarray}
It is easy to see now that under the inversion (\ref{inversion})
the leading-order impact factor is invariant while the NLO impact factor is not because of the non-invariant logarithmic term
$\ln{\sigma s\over 4}\calz_3$. Since the original T-product of the currents in the l.h.s. of Eq. (\ref{ope1}) is conformal, it indicates that our operators  $\hat{U}^\eta$ with the ``rigid cutoff" (\ref{cutoff}) are not M\"obius invariant.  However,  if we expand the original T-product in composite conformal operators (\ref{confodipole}) instead, 
the resulting impact factor is conformally invariant: 
\begin{eqnarray}
&&\hspace{-5mm}
(x-y)^4 T\{\halo(x)\halo(y)\}~
=~{(x-y)^4\over \pi^2(N_c^2-1)}\!\int\! d^2 z_{1\perp}d^2 z_{2\perp}
{(x_\ast y_\ast)^{-2}\over \calz_1^2\calz_2^2}~[{\rm Tr}\{\hat{U}^\eta_{z_1}\hat{U}^{\dagger\eta}_{z_2}\}]^{\rm conf}
 \label{opeconfa}\\
&&\hspace{-5mm}
-~{\alpha_s(x-y)^4\over 2\pi^4(N_c^2-1)}\!\int\! d^2 z_{1\perp}d^2 z_{2\perp} d^2 z_3
{z_{12}^2(x_\ast y_\ast)^{-2}\over z_{13}^2z_{23}^2\calz_1^2\calz_2^2}
 \nonumber\\
&&\hspace{-5mm}
\times~
\Big(\ln{a\sigma^2 s^2z_{12}^2\over 16 z_{13}^2z_{23}^2}\Big[{(x-z_3)^2\over x_\ast}-{(y-z_3)^2\over y_\ast}\Big]^2-i\pi+2C\Big)
[ {\rm Tr}\{T^n\hat{U}^\eta_{z_1}\hat{U}^{\dagger\eta}_{z_3}T^n\hat{U}^\eta_{z_3}\hat{U}^{\dagger\eta}_{z_2}\}
 -N_c {\rm Tr}\{\hat{U}^\eta_{z_1}\hat{U}^{\dagger\eta}_{z_2}\}]\nonumber
 \end{eqnarray}
The arbitrary dimensional constant $a$ should be chosen in such a way that the impact factor in the r.h.s of Eq. (\ref{opeconf})
does not change under the rescaling (\ref{reggelimit}). The proper choice for our T-product is $a={ x_\ast y_\ast\over s^2(x-y)^2}$ so our final 
operator expansion takes the form 
\begin{eqnarray}
&&\hspace{-5mm}
(x-y)^4 T\{\halo(x)\halo(y)\}~
=~{(x-y)^4\over \pi^2(N_c^2-1)}\!\int\! d^2 z_1 d^2 z_2{(x_\ast y_\ast)^{-2}\over \calz_1^2\calz_2^2}~[{\rm Tr}\{\hat{U}^\eta_{z_1}\hat{U}^{\dagger\eta}_{z_2}\}]^{\rm conf}
 \label{opeconf}\\
&&\hspace{-5mm}
-~{\alpha_s(x-y)^4\over 2\pi^4(N_c^2-1)}\!\int\! d^2 z_1d^2 z_2 d^2 z_3
{z_{12}^2(x_\ast y_\ast)^{-2}\over z_{13}^2z_{23}^2\calz_1^2\calz_2^2}
\Big(\ln{ x_\ast y_\ast z_{12}^2e^{2\eta}\over 16(x-y)_\perp^2 z_{13}^2z_{23}^2}
\Big[{(x-z_3)^2\over x_\ast}-{(y-z_3)^2\over y_\ast}\Big]^2-i\pi+2C\Big)
 \nonumber\\
&&\hspace{65mm}
\times~
[ {\rm Tr}\{T^n\hat{U}^\eta_{z_1}\hat{U}^{\dagger\eta}_{z_3}T^n\hat{U}^\eta_{z_3}\hat{U}^{\dagger\eta}_{z_2}\}
 -N_c {\rm Tr}\{\hat{U}^\eta_{z_1}\hat{U}^{\dagger\eta}_{z_2}\}]\nonumber
 \end{eqnarray}
where the conformal dipole $[{\rm Tr}\{\hat{U}^\eta_{z_1}\hat{U}^{\dagger\eta}_{z_2}\}]^{\rm conf}$ is given by Eq. (\ref{confodipole}) with
$a={ x_\ast y_\ast\over s^2(x-y)^2}$.
Now it is evident that the impact factor in the r.h.s. of this equation is M\"obius invariant and does not scale with $\lambda$ so 
Eq. (\ref{confodipole1}) gives conformally invariant operator up to $\alpha_s^2$ order. In higher orders, one should expect the correction 
terms with more Wilson lines. This procedure of finding the dipole with conformally regularized rapidity divergence is analogous
 to the construction of the composite renormalized local operator by adding the appropriate counterterms order by order in perturbation theory.

%+++++++++++++++++++++++++++++++++++++++++++++++++++++++++++
\subsection{Leading-order evolution of the four-Wilson-line operator\label{sect:appenc}}

In this Appendix we derive the  evolution equation for the four-Wilson-line operator 
${\rm Tr}\{T^a\hat{U}^\eta_{z_1} \hat{U}^{\dagger\eta}_{z_3}T^a\hat{U}^\eta_{z_3} \hat{U}^{\dagger\eta}_{z_2}\}$ in the leading 
order in perturbation theory. As a first step, we rewrite the hierarchy equations of Ref. \cite{npb96} in the 
adjoint representation. It takes the form
\begin{eqnarray}
&&\hspace{-11mm}
{d\over d\eta}(\hat{U}^\eta_{z_1})_{ij}
(\hat{U}^\eta_{z_2})_{kl}~=~{\alpha_s\over 2\pi^2}
\!\int\! d^2z_3~{(z_{13},z_{23})\over z_{13}^2z_{23}^2}
\big[(T^a\hat{U}^\eta_{z_1})_{ij}(\hat{U}^\eta_{z_2}T^b)_{kl}+(\hat{U}^\eta_{z_1}T^b)_{ij}(T^a\hat{U}^\eta_{z_2})_{kl}\big]
(2\hat{U}^\eta_{z_3}-\hat{U}^\eta_{z_1}-\hat{U}^\eta_{z_2})^{ab} ,
\label{pairwise}
\end{eqnarray}
for the pair-wise interaction and
\begin{eqnarray}
{d\over d\eta}(\hat{U}^\eta_{z_1})_{ij}&=&
{\alpha_s\over \pi^2}
\int d^2z_3~{1\over z_{13}^2}
(T^a\hat{U}^\eta_{z_1}T^b)_{ij}(\hat{U}^\eta_{z_3}-\hat{U}^\eta_{z_1})^{ab}
\label{reggeiz}
\end{eqnarray}
for the self-interaction. Using the above equations we get
\begin{eqnarray}
&&\hspace{-1mm}
{d\over d\eta} [{\rm Tr}\{T^n\hat{U}^\eta_{z_1}\hat{U}^{\dagger\eta}_{z_3}T^n\hat{U}^\eta_{z_3}\hat{U}^{\dagger\eta}_{z_2}\}
-N_c{\rm Tr}\{\hat{U}^\eta_{z_1}\hat{U}^{\dagger\eta}_{z_2}\}] 
\nonumber\\
&&\hspace{-1mm}
=~{\alpha_s\over 2\pi^2}\!\int\! d^2z_4(\hat{U}^\eta_{z_3})^{aa'}
 \Big[{(z_{14},z_{34})\over z_{14}^2z_{34}^2}
 {\rm Tr}\{T^aT^b\hat{U}^\eta_{z_1}[T^{a'},T^{b'}]\hat{U}^{\dagger\eta}_{z_2}
 +[T^b,T^a]\hat{U}^\eta_{z_1}T^{b'}T^{a'}\hat{U}^{\dagger\eta}_{z_2}\}
 (2\hat{U}^\eta_{z_4}-\hat{U}^\eta_{z_1}-\hat{U}^\eta_{z_3})^{bb'}
 \nonumber\\
&&\hspace{-1mm}
+~{(z_{24},z_{34})\over z_{24}^2z_{34}^2}
 {\rm Tr}\{[T^a,T^b]\hat{U}^\eta_{z_1}T^{a'}T^{b'}\hat{U}^{\dagger\eta}_{z_2}
 + T^bT^a\hat{U}^\eta_{z_1}[T^{b'},T^{a'}]\hat{U}^{\dagger\eta}_{z_2}\}
 (2\hat{U}^\eta_{z_4}-\hat{U}^\eta_{z_2}-\hat{U}^\eta_{z_3})^{bb'}
 \nonumber\\
&&\hspace{-1mm}
-~{(z_{14},z_{24})\over z_{14}^2z_{24}^2}
 {\rm Tr}\{T^aT^b\hat{U}^\eta_{z_1}T^{a'}T^{b'}\hat{U}^{\dagger\eta}_{z_2}
 + T^bT^a\hat{U}^\eta_{z_1}T^{b'}T^{a'}\hat{U}^{\dagger\eta}_{z_2}\}
 (2\hat{U}^\eta_{z_4}-\hat{U}^\eta_{z_1}-\hat{U}^\eta_{z_2})^{bb'}
 \nonumber\\
&&\hspace{-1mm}
-~{1\over z_{34}^2}
 {\rm Tr}\{[T^a,T^b]\hat{U}^\eta_{z_1}[T^{a'},T^{b'}]\hat{U}^{\dagger\eta}_{z_2}\}
  (2\hat{U}^\eta_{z_4}-2\hat{U}^\eta_{z_3})^{bb'}
 \nonumber\\
&&\hspace{-1mm}
+~{1\over z_{14}^2} {\rm Tr}\{T^aT^b\hat{U}^\eta_{z_1}T^{b'}T^{a'}\hat{U}^{\dagger\eta}_{z_2}\} (2\hat{U}^\eta_{z_4}-2\hat{U}^\eta_{z_1})^{bb'}
 + {1\over z_{24}^2} {\rm Tr}\{T^bT^a\hat{U}^\eta_{z_1}T^{a'}T^{b'}\hat{U}^{\dagger\eta}_{z_2}\} (2\hat{U}^\eta_{z_4}-2\hat{U}^\eta_{z_2})^{bb'}
 \Big]
   \nonumber\\
&&\hspace{-1mm}
-~{~\alpha_sN_c\over \pi^2}\!\int\! d^2z_4{z_{12}^2\over z_{14}^2z_{24}^2}({\rm Tr}\{T^aU_{z_1}U_{z_4}^\dagger T^a U_{z_4}U_{z_2}^\dagger\}
-N_c{\rm Tr}\{U_{z_1}U_{z_2}^\dagger\})
\label{4wlinevolada}
 \end{eqnarray}
After some algebra, the r.h.s. of Eq. (\ref{4wlinevolada}) reduces to
\begin{eqnarray}
&&\hspace{-1mm}
{d\over d\eta} [{\rm Tr}\{T^n\hat{U}^\eta_{z_1}\hat{U}^{\dagger\eta}_{z_3}T^n\hat{U}^\eta_{z_3}\hat{U}^{\dagger\eta}_{z_2}\}
-N_c{\rm Tr}\{\hat{U}^\eta_{z_1}\hat{U}^{\dagger\eta}_{z_2}\}] 
  \nonumber\\
&&\hspace{-1mm}
 =~{\alpha_s\over 4\pi^2}\!\int\! d^2z_4(\hat{U}^\eta_{z_3})^{aa'}
 \Big[{z_{12}^2\over z_{14}^2z_{24}^2}
 {\rm Tr}\{T^aT^b\hat{U}^\eta_{z_1}T^{a'}T^{b'}\hat{U}^{\dagger\eta}_{z_2}
 + T^bT^a\hat{U}^\eta_{z_1}T^{b'}T^{a'}\hat{U}^{\dagger\eta}_{z_2}\}
 (2\hat{U}^\eta_{z_4}-\hat{U}^\eta_{z_1}-\hat{U}^\eta_{z_2})^{bb'} 
  \nonumber\\
&&\hspace{-1mm}
 -~{z_{13}^2\over z_{14}^2z_{34}^2}
 {\rm Tr}\{T^aT^b\hat{U}^\eta_{z_1}[T^{a'},T^{b'}]\hat{U}^{\dagger\eta}_{z_2}
 +[T^b,T^a]\hat{U}^\eta_{z_1}T^{b'}T^{a'}\hat{U}^{\dagger\eta}_{z_2}\}
 (2\hat{U}^\eta_{z_4}-\hat{U}^\eta_{z_1}-\hat{U}^\eta_{z_3})^{bb'}
 \nonumber\\
&&\hspace{-1mm}
-~{z_{23}^2\over z_{24}^2z_{34}^2}
 {\rm Tr}\{[T^a,T^b]\hat{U}^\eta_{z_1}T^{a'}T^{b'}\hat{U}^{\dagger\eta}_{z_2}
 + T^bT^a\hat{U}^\eta_{z_1}[T^{b'},T^{a'}]\hat{U}^{\dagger\eta}_{z_2}\}
 (2\hat{U}^\eta_{z_4}-\hat{U}^\eta_{z_2}-\hat{U}^\eta_{z_3})^{bb'}
\Big]
 \nonumber\\
&&\hspace{-1mm}
-~{\alpha_sN_c\over \pi^2}\!\int\! d^2z_4{z_{12}^2\over z_{14}^2z_{24}^2}
({\rm Tr}\{T^a\hat{U}^\eta_{z_1}\hat{U}^{\dagger\eta}_{z_4} T^a \hat{U}^\eta_{z_4}\hat{U}^{\dagger\eta}_{z_2}\}
-N_c{\rm Tr}\{U_{z_1}U_{z_2}^\dagger\})
\label{4wlinevolad}
 \end{eqnarray}
%

%++++++++++++++++++++++++++++++++++++++++++++++++++++++++++++++++++++++++
\subsection{Color traces\label{app:traces}}

In this section we rewrite the adjoint traces in our evolution equation (\ref{nlobksymconf}) in terms of the fundamental traces. 
The master formula for traces has the form (in this section we use the space-saving notations $U_i\equiv \hat{U}_{z_i}$)
\begin{eqnarray}
&&\hspace{-5mm}
 {\rm Tr}\{[T^a,T^b]U^\eta_1T^{a'}T^{b'}U^{\dagger}_2
 + T^bT^aU^\eta_{z_1}[T^{b'},T^{a'}]U^{\dagger}_{z_2}\}U_3^{aa'}U_4^{bb'}~
\label{trace}\\
&&\hspace{-5mm}
=~{1\over 2}\Big[-{\rm tr}\{U_1 U_2^\dagger\}{\rm tr}\{U_2 U_4^\dagger\}{\rm tr}\{U_4 U_3^\dagger\}{\rm tr}\{U_3 U_1^\dagger\}
-{\rm tr}\{U_1^\dagger U_2\}{\rm tr}\{U_2^\dagger U_4\}{\rm tr}\{U_4^\dagger U_3\}{\rm tr}\{U_3^\dagger U_1\}
\nonumber\\
&&\hspace{-5mm}
+~{\rm tr}\{U_1   U_2^\dagger\}{\rm tr}\{U_2 U_4^\dagger U_3 U_1^\dagger  U_4 U_3^\dagger\}
+~{\rm tr}\{U_1 U_3^\dagger\}{\rm tr}\{U_3  U_2^\dagger U_4 U_1^\dagger  U_2 U_4^\dagger\}
+{\rm tr}\{U_3 U_4^\dagger\}{\rm tr}\{U_4 U_2^\dagger U_1 U_3^\dagger U_2 U_1^\dagger\}
\nonumber\\
&&\hspace{-5mm}
+~{\rm tr}\{U_2 U_4^\dagger\}{\rm tr}\{U_4 U_1^\dagger U_3  U_2^\dagger U_1 U_3^\dagger\}
-{\rm tr}\{U_1  U_4^\dagger\}{\rm tr}\{U_4 U_2^\dagger U_3 U_1^\dagger  U_2 U_3^\dagger\}
-{\rm tr}\{U_2 U_3^\dagger\}{\rm tr}\{U_3 U_1^\dagger U_4 U_2^\dagger U_1  U_4^\dagger\}
\nonumber\\
&&\hspace{-5mm}
-~{\rm tr}\{U_1 U_2^\dagger U_4 U_3^\dagger\}{\rm tr}\{U_1^\dagger U_2 U_4^\dagger U_3\}          
+{\rm tr}\{U_1^\dagger U_2\}{\rm tr}\{U_2^\dagger U_4 U_3^\dagger U_1U_4^\dagger U_3\}
+{\rm tr}\{U_1^\dagger U_3\}{\rm tr}\{U_3^\dagger   U_2U_4^\dagger  U_1 U_2^\dagger U_4\}
\nonumber\\
&&\hspace{-5mm}
+~{\rm tr}\{U_3^\dagger U_4\}{\rm tr}\{U_4^\dagger U_2 U_1^\dagger U_3 U_2^\dagger U_1\}
+{\rm tr}\{U_2^\dagger U_4\}{\rm tr}\{U_4^\dagger U_1 U_3^\dagger  U_2 U_1^\dagger U_3\}
-{\rm tr}\{U_1^\dagger U_4\}{\rm tr}\{ U_4^\dagger U_2 U_3^\dagger U_1U_2^\dagger U_3\}
\nonumber\\
&&\hspace{-5mm}
-~{\rm tr}\{U_2^\dagger U_3\}{\rm tr}\{U_3^\dagger U_1U_4^\dagger U_2 U_1^\dagger  U_4\}
-{\rm tr}\{U_1^\dagger U_2 U_4^\dagger U_3\}{\rm tr}\{ U_1 U_2^\dagger U_4 U_3^\dagger\}\Big]
\nonumber
\end{eqnarray}
where ${\rm Tr}$ and ${\rm tr}$ stand for the trace in the adjoint and the fundamental representation, respectively. 
We will also need traces
\begin{eqnarray}
&&\hspace{-1mm} 
{\rm Tr}\{T^aU_1T^bU^\dagger_2\}U_3^{ab}~=~  {\rm Tr}\{T^aU_1U_3^\dagger T^bU_3U^\dagger_2\}
\nonumber\\
&&\hspace{-1mm}        
=~\half {\rm tr}\big[\{U_1U_2^\dagger\}{\rm tr}\{U_2U_3^\dagger\}{\rm tr}\{U_3U_1^\dagger\}
+{\rm tr}\{U_1U_3^\dagger\}{\rm tr}\{U_3U_2^\dagger\}{\rm tr}\{U_2U_1^\dagger\}
 -~{\rm tr}\{U_1U_2^\dagger U_3U_1^\dagger U_2U_3^\dagger\}
- {\rm tr}\{U_1U_3^\dagger U_2U_1^\dagger U_3U_2^\dagger\}\big],               
\nonumber\\
&&\hspace{-1mm}
{\rm Tr}\{[T^a,T^b]U_1T^{a'}T^{b'}U_2^\dagger+T^bT^aU_1[T^{b'},T^{a'}]U_2^\dagger\}  
U_3^{aa'}U_3^{bb'}
\nonumber\\
&&\hspace{-1mm}
=~ {\rm Tr}\{[T^a,T^b]U^\eta_1T^{a'}T^{b'}U^{\dagger}_2
 + T^bT^aU^\eta_2[T^{b'},T^{a'}]U^{\dagger}_2\}U_1^{aa'}U_3^{bb'}
  \nonumber\\
&&\hspace{-1mm}
=~ {\rm Tr}\{[T^a,T^b]U^\eta_1T^{a'}T^{b'}U^{\dagger}_2+ T^bT^aU^\eta_1[T^{b'},T^{a'}]U^{\dagger}_2\}U_3^{aa'}U_2^{bb'}
~=~-N_c{\rm Tr}\{T^aU_1U_3^\dagger T^aU_3 U_2^\dagger\},
 \nonumber\\
&&\hspace{-1mm}
 {\rm Tr}\{T^aT^bU_1T^{a'}T^{b'}U^{\dagger}_2 + T^bT^aU_1T^{b'}T^{a'}U_2^\dagger\}U_1^{aa'}U_3^{bb'}
  \nonumber\\
&&\hspace{-1mm}=~{\rm Tr}\{T^aT^bU_1T^{a'}T^{b'}U^{\dagger}_2 + T^bT^aU_1T^{b'}T^{a'}U_2^\dagger\}U_2^{aa'}U_3^{bb'}
 ~=~N_c{\rm Tr}\{T^aU_1U_3^\dagger T^aU_3 U_2^\dagger\}
\label{traces}
 \end{eqnarray}
which easily follow from Eq. (\ref{trace}).

Let us now find the master trace (\ref{trace}) in the two-gluon (BFKL) approximation. First, 
note that  in this approximation the trace of six Wilson lines reduces to one dipole.
Indeed,
\begin{eqnarray}
&&\hspace{-2mm}
{\rm tr}\{U_1U_3^\dagger U_4 U_2^\dagger U_3 U_4^\dagger\}~=~{\rm tr}\{(U_1U_3^\dagger -1)U_4(U_2^\dagger U_3 -1)U_4^\dagger\}+{\rm tr}\{(U_1 U_3^\dagger-1\}
+{\rm tr}\{(U_2^\dagger U_3-1\}+N_c^2
\nonumber\\
&&\hspace{-2mm} 
=~{\rm tr}\{(U_1U_3^\dagger -1)U_3(U_2^\dagger U_3 -1)U_3^\dagger\}+{\rm tr}\{(U_1 U_3^\dagger-1\}
+{\rm tr}\{(U_2^\dagger U_3-1\}+N_c^2
~=~{\rm tr}\{U_1 U_2^\dagger\}
\label{tracelin1}
\end{eqnarray}
where we replaced by $U_4$  by $U_3$  in the first term in the second line since it does not matter in the two-gluon approximation (both of them should be replaced by 1). 
All other six-Wilson-line terms in the r.h.s. of Eq. (\ref{trace}) can be similarly reduced to single-dipole terms.  

Using the same trick we can  reduce the trace of four Wilson lines to the sum of dipoles:
\begin{eqnarray}
&&\hspace{-2mm}
{\rm tr}\{U_1U_2^\dagger U_4  U_3^\dagger\}~=~{\rm tr}\{(U_1U_2^\dagger-1) (U_4  U_3^\dagger-1)\}+{\rm tr}\{U_1U_2^\dagger-1)\}+{\rm tr}\{U_4U_3^\dagger-1)\}+N_c^2
\label{tracelin2}\\
&&\hspace{-2mm} 
=~{\rm tr}\{(U_1U_2^\dagger-1)U_2U_4^\dagger(U_4  U_3^\dagger-1)\}+{\rm tr}\{U_1U_2^\dagger-1)\}+{\rm tr}\{U_4U_3^\dagger-1)\}+N_c^2
\nonumber\\
&&\hspace{-2mm} 
=~{\rm tr}\{(U_1U_2^\dagger-1)\}+{\rm tr}\{U_2U_4^\dagger-1\}+{\rm tr}\{U_4U_3^\dagger-1)\}+{\rm tr}\{U_1U_3^\dagger-1)\}-{\rm tr}\{U_1U_4^\dagger-1)\}-{\rm tr}\{U_2U_3^\dagger-1)\}
+N_c^2
\nonumber
\end{eqnarray}
where in the first term in the second line we have inserted $U_2U_4^\dagger$ since it does not matter in the two-gluon approximation.

Using formulas (\ref{tracelin1}) and (\ref{tracelin2}) it is easy to demonstrate that all the terms containing traces of four and six Wilson lines in the r.h.s. of the Eq. (\ref{trace}) 
sum to $2N_c^2$ in the two-gluon approximation so one gets
\begin{eqnarray}
&&\hspace{-5mm}
 {\rm Tr}\{[T^a,T^b]U^\eta_1T^{a'}T^{b'}U^{\dagger}_2
 + T^bT^aU^\eta_{z_1}[T^{b'},T^{a'}]U^{\dagger}_{z_2}\}U_3^{aa'}U_4^{bb'}~\stackrel{2-gluon}{=}~-N_c^2(N_c^2-1)-{1\over 2}N_c^3\big[{\rm tr}\{U_1 U_2^\dagger-1\}
\nonumber\\
&&\hspace{-5mm}
+~{\rm tr}\{U_1^\dagger U_2-1\}+{\rm tr}\{U_2 U_4^\dagger-1\}+{\rm tr}\{U_2^\dagger U_4-1\}+{\rm tr}\{U_4 U_3^\dagger-1\}+{\rm tr}\{U_4^\dagger U_3-1\}
+{\rm tr}\{U_3 U_1^\dagger-1\}+{\rm tr}\{U_3^\dagger U_1\}\big]
\nonumber\\
&&\hspace{-2mm} 
=~-N_c^2(N_c^2-1)+\half N_c^2(N_c^2-1)\big[\hat\calv(z_1,z_2)+\hat\calv(z_1,z_3)+\hat\calv(z_2,z_4)+\hat\calv(z_3,z_4)\big]
\label{tracelin}
\end{eqnarray}
where we use the definition (\ref{calv}). For completeness, let us present also $ {\rm Tr}\{T^aU_1U_3^\dagger T^bU_3U^\dagger_2\}$ in the two-gluon approximation. From the first line in Eq. (\ref{traces}) one easily obtains
\begin{eqnarray}
&&\hspace{-5mm}
 {\rm Tr}\{T^aU_1U_3^\dagger T^bU_3U^\dagger_2\}
~\stackrel{2-gluon}{=}~N_c(N_c^2-1)-{1\over 2}N_c(N_c^2-1)\big[\hat\calv(z_1,z_2)+\hat\calv(z_1,z_3)+\hat\calv(z_2,z_3)\big]
 \label{tracelintriv}
\end{eqnarray}
To describe the conformal operator (\ref{confoper4}) we need one more trace (which eventually drops out after the integration of the operator  (\ref{confoper4}) over 
$z_3$ with the weight ${z_{12}^2\over z_{13}^2z_{23}^2}$)
\begin{eqnarray}
&&\hspace{-5mm}
{\rm Tr}\{T^aT^bU_1T^{a'}T^{b'}U^\dagger_2\}U_3^{aa'}U_4^{bb'}~
=~~{1\over 4}
\Big[{\rm tr}\{U_1 U_3^\dagger\}{\rm tr}\{U_3 U_2^\dagger\}{\rm tr}\{U_2U_4^\dagger\}{\rm tr}\{U_4U_1^\dagger\}
\label{colorfla1}\\
&&\hspace{-5mm} 
+~{\rm tr}\{U_3 U_1^\dagger\}{\rm tr}\{U_2U_3 ^\dagger\}{\rm tr}\{U_4U_2^\dagger\}{\rm tr}\{U_4^\dagger U_1\}
+{\rm tr}\{U_3 U_1^\dagger U_4U_2^\dagger\}{\rm tr}\{ U_1U_4^\dagger U_2U_3^\dagger\}
+{\rm tr}\{U_2U_4^\dagger U_1 U_3^\dagger\}{\rm tr}\{U_1^\dagger  U_3 U_2^\dagger U_4\}
\nonumber\\
&&\hspace{-5mm} 
+~{\rm tr}\{U_4U_3^\dagger\}{\rm tr}\{U_3 U_1^\dagger U_2U_4^\dagger U_1  U_2^\dagger\}
+{\rm tr}\{U_1U_2^\dagger\}{\rm tr}\{U_4^\dagger U_3 U_1^\dagger U_4 U_3 ^\dagger U_2\}
+{\rm tr}\{U_4^\dagger U_3 \}{\rm tr}\{U_2U_1^\dagger U_4U_2^\dagger  U_1 U_3^\dagger\}
\nonumber\\
&&\hspace{-5mm} 
+~{\rm tr}\{U_1^\dagger U_2\}{\rm tr}\{U_3 U_4^\dagger U_1 U_3^\dagger U_4U_2^\dagger\}
-~{\rm tr}\{U_1 U_3^\dagger\}{\rm tr}\{U_2 U_1^\dagger U_4U_2^\dagger U_3 U_4^\dagger\}
-{\rm tr}\{U_1^\dagger U_4\}{\rm tr}\{U_1 U_3^\dagger U_2U_4^\dagger U_3 U_2^\dagger\}
\nonumber\\
&&\hspace{-5mm} 
-~{\rm tr}\{U_3 U_2^\dagger\}{\rm tr}\{U_1^\dagger  U_2U_4^\dagger U_1 U_3^\dagger U_4\}
-{\rm tr}\{U_2U_4^\dagger\}{\rm tr}\{U_1 U_2^\dagger U_3 U_1^\dagger U_4U_3^\dagger\}
-~{\rm tr}\{U_3 U_1^\dagger\}{\rm tr}\{U_4U_3^\dagger U_2U_4^\dagger U_1 U_2^\dagger \}
\nonumber\\
&&\hspace{-5mm} 
-~{\rm tr}\{U_4U_2^\dagger \}{\rm tr}\{U_1^\dagger U_3 U_4^\dagger U_1 U_3^\dagger U_2\}
-{\rm tr}\{U_2U_3 ^\dagger\}{\rm tr}\{U_1 U_4^\dagger U_3 U_1^\dagger U_4U_2^\dagger\}
- {\rm tr}\{U_4^\dagger U_1 \}{\rm tr}\{U_2U_3^\dagger U_4U_2^\dagger U_3 U_1^\dagger \}
\Big]
\nonumber
\end{eqnarray}
In the two-gluon approximation this yields
\begin{eqnarray}
&&\hspace{-5mm}
{\rm Tr}\{T^aT^bU_1T^{a'}T^{b'}U^\dagger_2\}U_3^{aa'}U_4^{bb'}~
\nonumber\\
&&\hspace{-5mm} 
=~\half N_c^2(N_c^2-1)-{1\over 4}N_c^2(N_c^2-1)\big[\hat\calv(z_1,z_3)+\hat\calv(z_2,z_3)+\hat\calv(z_1,z_4)+\hat\calv(z_2,z_4)\big]
\label{tracelin2g}
\end{eqnarray}
%

%++++++++++++++++++++++++++++++++++++++++++++++++++++++++++++++++++++
\subsection{Integrals}
In this section we  calculate two basic integrals which we use in this paper. The first one is
\begin{eqnarray}
&&\hspace{-6mm}
{z_{12}^2\over z_{13}^2  z_{23}^2}\!\int\! {d^2z_4\over\pi}\Big[
{z_{13}^2 \over z_{14}^2z_{34}^2 }+{ z_{23}^2\over  z_{24}^2z_{34}^2 }
-{z_{12}^2\over  z_{14}^2z_{24}^2}\Big]
\ln{z_{13}^2z_{24}^2\over z_{14}^2z_{23}^2}~=~4\zeta(3)[\delta(z_{23})-\delta(z_{13})]
\label{nashel}
\end{eqnarray}
The easiest way to prove this at $z_3\neq z_1,z_2$ is to set $z_2=0$ and make an inversion 
$x\rightarrow 1/\tilde{x}$ so the integral (\ref{nashel}) reduces to 
\begin{eqnarray}
&&\hspace{-6mm}
2(\tilde{z}_1-\tilde{z}_2)^2\!\int\! d^2\tilde{z}_4
{(\tilde{z}_1-\tilde{z}_3,\tilde{z}_1-\tilde{z}_4)\over (\tilde{z}_1-\tilde{z}_4)^2(\tilde{z}_3-\tilde{z}_4)^2 }
\ln{(\tilde{z}_1-\tilde{z}_3)^2\over (\tilde{z}_1-\tilde{z}_4)^2}~=~0
\nonumber
\end{eqnarray}
The $\delta$-function terms in the r.h.s. of Eq. (\ref{nashel}) can be restored from the formula
\begin{eqnarray}
&&\hspace{-6mm}
\int\! {d^2z_3d^2z_4\over\pi^2}\Big[
{(z_{12},z_{23})\over z_{34}^2 }\Big({1\over z_{13}^2z_{24}^2}+{1\over z_{23}^2z_{14}^2}\Big)
-{z_{12}^2(z_{12},z_{23})\over  z_{13}^2z_{24}^2z_{14}^2z_{23}^2}\Big]\ln{z_{13}^2z_{24}^2\over z_{23}^2z_{14}^2}
~=~4\zeta(3)
\label{zeta1}
\end{eqnarray}
which follows from the integral 
\begin{eqnarray}
&&\hspace{-11mm}
{1\over\pi}\!\int\! d^dzd^dz' \Big[{(x,z)\over (x-z)^2(z-z')^2{z'}^2}
+{(x,z)\over (x-z')^2(z-z')^2z^2}-{x^2(x,z)\over (x-z)^2z^2(x-z')^2{z'}^2}\Big]
\ln{(x-z)^2{z'}^2\over (x-z')^2z^2}~=~
\nonumber\\
&&\hspace{-11mm}
=~B\big({d\over 2},{d\over 2}-1\big)B\big(d-1,{d\over 2}-1\big)
{\Gamma(3-d)\over (x^2)^{2-d}}\Big\{
3\psi\big({d\over 2}-1\big)
-2\psi\big(d-2\big)+2\psi(1)-2\psi\big(2-{d\over 2}\big)
-~\psi\big({d\over 2}\big)
 \nonumber\\
&&\hspace{-11mm}
+~{4\over d-2}
{\Gamma\big({d\over 2}\big)\Gamma\big(3{d\over 2}-2\big)\Gamma^2\big(2-{d\over 2}\big)\over
\Gamma^2(d-1)\Gamma(3-d)}\Big\}
~\stackrel{d\rightarrow 2}{\rightarrow}~-4\zeta(3)
\nonumber
 \end{eqnarray}
The second integral is somewhat more tedious
\begin{eqnarray}
&&\hspace{-1mm}
\hspace{-5mm}                        
{z_{12}^2\over z_{13}^2z_{23}^2}\!\int\!{d^2z_4\over\pi}~
\Big[
{z_{23}^2\over z_{34}^2z_{24}^2}\ln{z_{12}^2z_{13}^2z_{34}^2 \over z_{14}^4z_{24}^2}  
+ {z_{13}^2\over z_{14}^2z_{34}^2}\ln{z_{13}^4z_{23}^2\over z_{12}^2z_{14}^2z_{34}^2}  
+{z_{12}^2\over z_{14}^2z_{24}^2}\ln{z_{14}^2z_{24}^2\over  z_{13}^2z_{23}^2} \Big]            
 \nonumber\\
&&\hspace{-1mm}
=~
2{z_{12}^2\over z_{13}^2z_{23}^2}\ln{z_{12}^2\over z_{13}^2}\ln{z_{12}^2\over z_{23}^2}  -8\zeta(3)\delta(z_{13})     
\label{integral1}
 \end{eqnarray}
First we prove this equation at $z_3\neq z_1,z_2$. To simplify the notations, we take $z_2=0$ an denote $z_1\equiv x, z_3\equiv z$ and $z_4\equiv z'$. 
Let us start from the first term in the square brackets in the l.h.s. of Eq. (\ref{integral1})
\begin{eqnarray}
&&\hspace{-11mm}
\!\int\!d^2z'~
{z^2\over (z-z')^2{z'}^2}\ln{x^2(x-z)^2(z-z')^2 \over  (x-z')^4{z'}^2}
\label{1stpart1}\\
&&\hspace{-11mm}   
=~-2\!\int\!d^2z'~\Big[{1\over (z-z')^2}-{1\over {z'}^2}\Big]\ln{(x-z')^2\over  x^2 }
-\ln{x^2\over (x-z)^2}\!\int\!d^2z'~{z^2\over (z-z')^2{z'}^2}
-4\!\int\!d^2z'~{(z,z-z')\over (z-z')^2{z'}^2}\ln{(x-z')^2\over  x^2 }
\nonumber
 \end{eqnarray}
We get
\begin{eqnarray}
&&\hspace{-11mm}
-2\!\int\!d^2z'~\Big[{1\over (z-z')^2}-{1\over {z'}^2}\Big]\ln{(x-z')^2\over  x^2 }
-\ln{x^2\over (x-z)^2}\!\int\!d^2z'~{z^2\over (z-z')^2{z'}^2}
\label{1stpart2}\\
&&\hspace{-11mm} 
=~\lim_{d\rightarrow 2}\Bigg(-2\!\int\!d^dz'~\Big[{1\over (z-z')^2}-{1\over {z'}^2}\Big]\ln{(x-z')^2\over  x^2 }
-\ln{x^2\over (x-z)^2}\!\int\!d^dz'~{z^2\over (z-z')^2{z'}^2}\Bigg)
\nonumber\\
&&\hspace{-11mm} 
=~\lim_{d\rightarrow 2}2\pi\Big[{\Gamma(1-{d\over 2})\over ((x-z)^2)^{1-{d\over 2}}}-{\Gamma(1-{d\over 2})\over (x^2)^{1-{d\over 2}}}
-{\Gamma(2-{d\over 2})\over (z^2)^{1-{d\over 2}}}\ln{x^2\over (x-z)^2}\Big]B\Big({d\over 2}, {d\over 2}-1\Big)
~=~-\pi\ln{(x-z)^2x^2\over z^4}\ln{(x-z)^2\over x^2}
\nonumber
 \end{eqnarray}
and
\begin{eqnarray}
&&\hspace{-11mm}
-4\!\int\!d^dz'~{(z,z-z')\over (z-z')^2{z'}^2}\ln{(x-z')^2\over  x^2 }~=~\lim_{d\rightarrow 2}4{\pi^{d/2}\over d-2}
{\Gamma(d/2)\over\Gamma(d-1)}\Gamma({d\over 2}-1)\Gamma(2-{d\over 2})
\Big[1+\Big({d\over 2}-1\Big)\ln z^2
\nonumber\\
&&\hspace{-12mm}
+~\Big({d\over 2}-1\Big)\ln z^2+
{1\over 2}\Big({d\over 2}-1\Big)^2\Big(\ln^2{z^2x^2\over(x-z)^2}
-\ln^2{x^4\over(x-z)^2}+\ln^2x^2\Big)
-1-\Big({d\over 2}-1\Big)\ln z^2-{1\over 2}\Big({d\over 2}-1\Big)^2\ln^2z^2\Big]
\nonumber\\
&&\hspace{-12mm}
=~2\pi\ln{x^2\over z^2}\ln{(x-z)^2\over x^2}
\label{fla105}
\end{eqnarray}
so
\begin{eqnarray}
&&\hspace{-11mm}
\!\int\!d^2z'~
{z^2\over (z-z')^2{z'}^2}\ln{x^2(x-z)^2 \over  (x-z')^4}
=~-\pi\ln^2{(x-z)^2\over x^2}
\label{1stpart}
 \end{eqnarray}
Similarly
\begin{eqnarray}
&&\hspace{-5mm}
\!\int\!d^2z'~
\Big[-{(x-z)^2\over (x-z')^2(z-z')^2}\ln{x^2(x-z')^2(z-z')^2 \over (x-z)^4z^2}  
+~{x^2\over (x-z')^2{z'}^2}\ln{(x-z')^2{z'}^2\over  (x-z)^2z^2} \Big]  
\nonumber\\
&&\hspace{-5mm}  
=~\lim_{d\rightarrow 2}2\pi{\Gamma\big(2-{d\over 2}\big)\Gamma^2\big({d\over 2}\big)\over  \big({d\over 2}-1\big)\Gamma(d-1)}
\Big\{|x-z|^{d-2}\ln{(x-z)^4z^2\over x^2}
-2|x-z|^{d-2}\Big[-{1\over d-2}+\ln(x-z)^2-\psi\big(2-{d\over 2}\big)+\psi\big({d\over 2}\big)
\nonumber\\
&&\hspace{-5mm}  
+~\psi(1)-\psi(d-1)\Big]+2|x|^{d-2}\Big[-{1\over d-2}+\ln x^2-\psi\big(2-{d\over 2}\big)+\psi\big({d\over 2}\big)+\psi(1)-\psi(d-1)\Big]
-|x|^{d-2}\ln(x-z)^2z^2\Big\}
\nonumber\\
&&\hspace{-5mm}  
=~\pi\ln^2{x^2\over(x-z)^2}+2\pi\ln{x^2\over z^2}\ln{x^2\over (x-z)^2}
\label{integral3}
 \end{eqnarray}
and therefore
\begin{eqnarray}
&&\hspace{-5mm}
\!\int\!d^2z'~
\Big[-{z^2\over (z-z')^2{z'}^2}\ln{(x-z')^4\over  x^2(x-z)^2}
+~{(x-z)^2\over (x-z')^2(z-z')^2}\ln{(x-z)^4z^2\over x^2(x-z')^2(z-z')^2 }  
+~{x^2\over (x-z')^2{z'}^2}\ln{(x-z')^2{z'}^2\over  (x-z)^2z^2} \Big]   
\nonumber\\
&&\hspace{-5mm}   
=~   2\pi\ln{x^2\over z^2}\ln{x^2\over (x-z)^2}
\label{pochti}
 \end{eqnarray}
However, it is easy to see that some sort of $\delta$-function contribution to the r.h.s. is necessary.
If we integrate the l.h.s. over $z$ with the weight ${x^2\over (x-z)^2z^2}$ we get zero because 
of the antisymmetry of the integrand with respect to $z\leftrightarrow  z'$. On the other hand,
the integral of the r.h.s. does not vanish because
\begin{eqnarray}
&&\hspace{-5mm}
\!\int\!d^2z~{x^2\over (x-z)^2z^2}\ln{x^2\over z^2}\ln{x^2\over (x-z)^2}~=~4\pi\zeta(3)
 \end{eqnarray}
To fix the coefficients in front of possible $\delta$-function contributions $\sim\delta(z)$ and/or $\delta(x-z)$ 
we calculate the integral of the l.h.s. of Eq. (\ref{pochti}) with the trial function ${(x,x-z)\over z^2(x-z)^2}$. We get
\begin{eqnarray}
&&\hspace{-5mm} 
{1\over\pi^2}\!\int\! d^2z d^2z' {(x,x-z)\over z^2(x-z)^2} 
\Big[{(x-z)^2\ln{ x^2z^2\over {z'}^4}\over (x-z')^2(z-z')^2}-{x^2\over (x-z')^2{z'}^2}\ln{z^2(x-z)^2\over {z'}^2(x-z')^2} 
+{z^2\over (z-z')^2{z'}^2}\ln{z^4(x-z)^2\over x^2{z'}^2(z-z')^2}\Big]
\label{vshiv3}\\
&&\hspace{-5mm}  
=~\lim_{d\rightarrow 2}B\big({d\over 2},{d\over 2}-1\big)  B\big(d-1,{d\over 2}-1\big) {\Gamma(3-d)\over (x^2)^{2-d}}   
\Big\{\psi(1)+\psi\big({d\over 2}\big)+\psi(3-d)+\psi\big(3{d\over 2}-2\big)-2\psi(d-1)-2\psi\big(2-{d\over 2}\big)\Big\}~
\nonumber\\
&&\hspace{-5mm}=~-4\zeta(3)
\nonumber
\end{eqnarray}
where $B(a,b)=\Gamma(a)\Gamma(b)/\Gamma(a+b)$. It is clear now that the
result for the integral in the l.h.s. of the formula (\ref{integral1}) should be as 
cited in the r.h.s. of Eq. (\ref{integral1}) - it satisfies both the Eq. (\ref{vshiv3}) and the requirement
that the integral of the l.h.s. with the trial function ${x^2\over z^2(x-z)^2}$ vanishes.

We will need one more integral which is obtained by antisymmetrization of Eq. (\ref{integral1}) with 
respect to $z_1\leftrightarrow z_2$ 
\begin{eqnarray}
&&\hspace{-1mm}
\hspace{-5mm}                        
{z_{12}^2\over z_{13}^2z_{23}^2}\!\int\!{d^2z_4\over\pi}~
\Big[{z_{13}^2\over z_{14}^2z_{34}^2}\ln{z_{12}^2z_{34}^2 \over z_{13}^2z_{24}^2}
-{z_{23}^2\over z_{34}^2z_{24}^2}\ln{z_{12}^2z_{34}^2 \over z_{14}^2z_{23}^2} 
\Big]~
=~4\zeta(3)[\delta(z_{13})- \delta(z_{23}) ]    
\label{integral2}
 \end{eqnarray}
%

%+++++++++++++++++++++++++++++++++++++++++++++++++++++++++
 
\end{document}